\documentclass[a4paper,11pt]{article}

\usepackage{jheppub} 

\usepackage[T1]{fontenc} 

\newtheorem{conjecture}{Conjecture}

\title{\boldmath A canonical purification for the entanglement wedge cross-section}

\author{Souvik Dutta}
\author{and Thomas Faulkner}

\affiliation[a]{Department of Physics, University of Illinois, Urbana-Champaign \\ 1110 W. Green St., Urbana IL 61801, USA.}

\abstract{In AdS/CFT we consider a class of bulk geometric quantities inside the entanglement wedge called reflected minimal surfaces. The areas of these surfaces are dual to the entanglement entropy associated to a canonical purification (the GNS state) that we dub the \emph{reflected entropy}. 
From the bulk point of view, we show that \emph{half} the area of the reflected minimal surface gives a reinterpretation of the notion of the entanglement wedge cross-section. 
We prove some general properties of the reflected entropy and
introduce a novel replica trick in CFTs for studying it. 
The duality is established using a recently introduced approach to holographic modular flow. We also consider an explicit holographic construction of the canonical purification, introduced by Engelhardt and Wall; the reflected minimal surfaces are simply RT surfaces in this new spacetime. We contrast our results with the entanglement of purification conjecture, and finally comment on the continuum limit where we find a relation to the split property: the reflected entropy computes the von Neumann entropy of a canonical splitting type-I factor introduced by Doplicher and Longo.  }

\begin{document} 
\maketitle
\flushbottom

\section{Introduction}

The main purpose of this paper is to study a quantum information quantity, which we will refer to as the \emph{reflected entropy}, that computes the area of the entanglement wedge cross section in holographic theories. The geometric description of the cross section, denoted $E_W$, was given in \cite{Takayanagi:2017knl,Nguyen:2017yqw} as the area of the minimal surface that bipartitions the entanglement wedge region. In these papers, a dual boundary description was suggested in terms of the entanglement of purification \cite{terhal2002entanglement}.\footnote{ For further developments related to this original duality conjecture see \cite{Bao:2017nhh,Tamaoka:2018ned,Hirai:2018jwy,Umemoto:2018jpc,Bao:2018gck,Espindola:2018ozt,Bao:2018fso,Bao:2018pvs,Caputa:2018xuf,kudler2018entanglement,BabaeiVelni:2019pkw,Guo:2019pfl}.} The entanglement of purification is in general hard to compute since it involves a minimization over the space of all possible purifications.  In this paper, we will give evidence that there is a much simpler dual for the cross section. 

The \emph{reflected entropy} is defined for a bipartite quantum system $AB$ and a mixed state $\rho_{AB}$. For simplicity we will mainly work with finite dimensional Hilbert spaces and we will discuss the more general case later. There is a simple and canonical purification of the state $\rho_{AB}$ in a doubled Hilbert space:
\begin{equation}
\left| \sqrt{\rho_{AB}} \right> \in {\rm End}(\mathcal{H}_A) \otimes {\rm End}(\mathcal{H}_B)
= \left( \mathcal{H}_A  \otimes \mathcal{H}_A^\star \right) \otimes  \left( \mathcal{H}_B  \otimes \mathcal{H}_B^\star \right) \equiv \mathcal{H}_{A A^\star B B^\star}
\label{can}
\end{equation}
where the space of linear maps/matrices ${\rm End}(\mathcal{H}_A)$, acting on $\mathcal{H}_A$, forms a Hilbert space with the trace inner product $\left< \sigma_A \right| \left. \sigma_A' \right> = {\rm Tr}_A \sigma_A^\dagger \sigma_A'$. This Hilbert space is isomorphic to  $ \mathcal{H}_A  \otimes \mathcal{H}_A^\star $. Similar definitions hold for $A$ replaced $B$ or $AB$.

It is not hard to show that:
\begin{equation}
{\rm Tr}_{\mathcal{H}_A^\star \otimes \mathcal{H}_B^\star} \left| \sqrt{\rho_{AB}} \right> \left< \sqrt{\rho_{AB}} \right| = \rho_{AB}
\label{traceab}
\end{equation}
and so this does represent a genuine purification.\footnote{In fact this doubled Hilbert space arises from the GNS representation of the matrix algebra acting on the original Hilbert space. We discuss this further in Appendix~\ref{app:gns}.}  We then define the reflected entropy as:
\begin{equation}
S_R(A:B) \equiv S(A A^\star)_{\sqrt{\rho_{AB}}} = S_{vN}(\rho_{AA^\star})
\end{equation}
where $\rho_{AA^\star}$ is the reduced density matrix after tracing over $\mathcal{H}_B \otimes \mathcal{H}_B^\star$. We will present evidence for the following duality:\footnote{Another simple proposal has been made for $E_W$ in \cite{Tamaoka:2018ned} using the so called odd entropy. The reflected entropy is distinct from their proposal. For example $S_R$ can never be negative, while the quantity studied in \cite{Tamaoka:2018ned} can be negative. } 
\begin{equation}
\label{ewdual}
S_R(A:B) = 2 E_W(A:B) +  \ldots 
\end{equation}
where $E_W$ is the area of the entanglement wedge cross section divided by $4 G_N$. This is the leading term in an expansion of $G_N$, and we will discuss the first quantum correction in Section~\ref{sec:qc}.

We will present several approaches to proving \eqref{ewdual}. We will find the spacetime dual of $\sqrt{\rho_{AB}}$ using the tools developed in \cite{Engelhardt:2018kcs,Engelhardt:2017aux}. The reflected entropy are simple RT\cite{Ryu:2006bv}/HRT\cite{Hubeny:2007xt} surfaces in this new spacetime. We will also use a newly developed approach to modular flow in holographic theories \cite{Faulkner:2018faa,Chen:2018rgz}. 

Our results should be compared to the original entanglement of purification conjecture \cite{Takayanagi:2017knl,Nguyen:2017yqw}. The entanglement of purification can be defined as:
\begin{equation}
\label{defep}
E_{p}(A:B) = \mathop{{\rm inf}}_{U} S_{vN}(\rho_{A A^\star}^U)
\end{equation}
where 
\begin{equation}
\rho_{A A^\star }^U = {\rm Tr}_{B^\star B} \left| \sqrt{\rho_{AB}} U \right>\left< \right. \! \sqrt{\rho_{AB}} U \! \left.\right|,
\end{equation}
for some unitary matrix $U$ which acts on $\mathcal{H}_{AB}^\star$, which here is represented by the right action on matrices. Note that to define $E_p$ we should really have allowed the purifying system $\mathcal{H}_{AB}^\star$ to vary in size, although for the eventual application to QFT the relevant Hilbert spaces will be infinite dimensional and so this is not a real concern.

Notice the important factor of $1/2$ in \eqref{ewdual}.  For example, on comparing $E_p$ to $S_R$ we only know that $E_p \leq S_R(A:B)$, and this does not immediately contradict the original conjectured duality $E_p = E_W = S_R/2$. It is however easy to find states where $E_p \neq \frac{1}{2}  S_R$, which means that the original duality, when interpreted as the boundary theory statement that $E_p = \frac{1}{2}S_R$, can only possibly be true for a special class of states. This class of states may include the states of a holographic theory associated to a classical bulk geometry. Indeed, we view this as a reasonable conjecture, given the discussion developed in \cite{Takayanagi:2017knl}. However, the complexity of finding the entanglement of purification does make this later conjecture a harder task to prove. 

\begin{figure}[h!]
\centering 
\includegraphics[width=.49\textwidth]{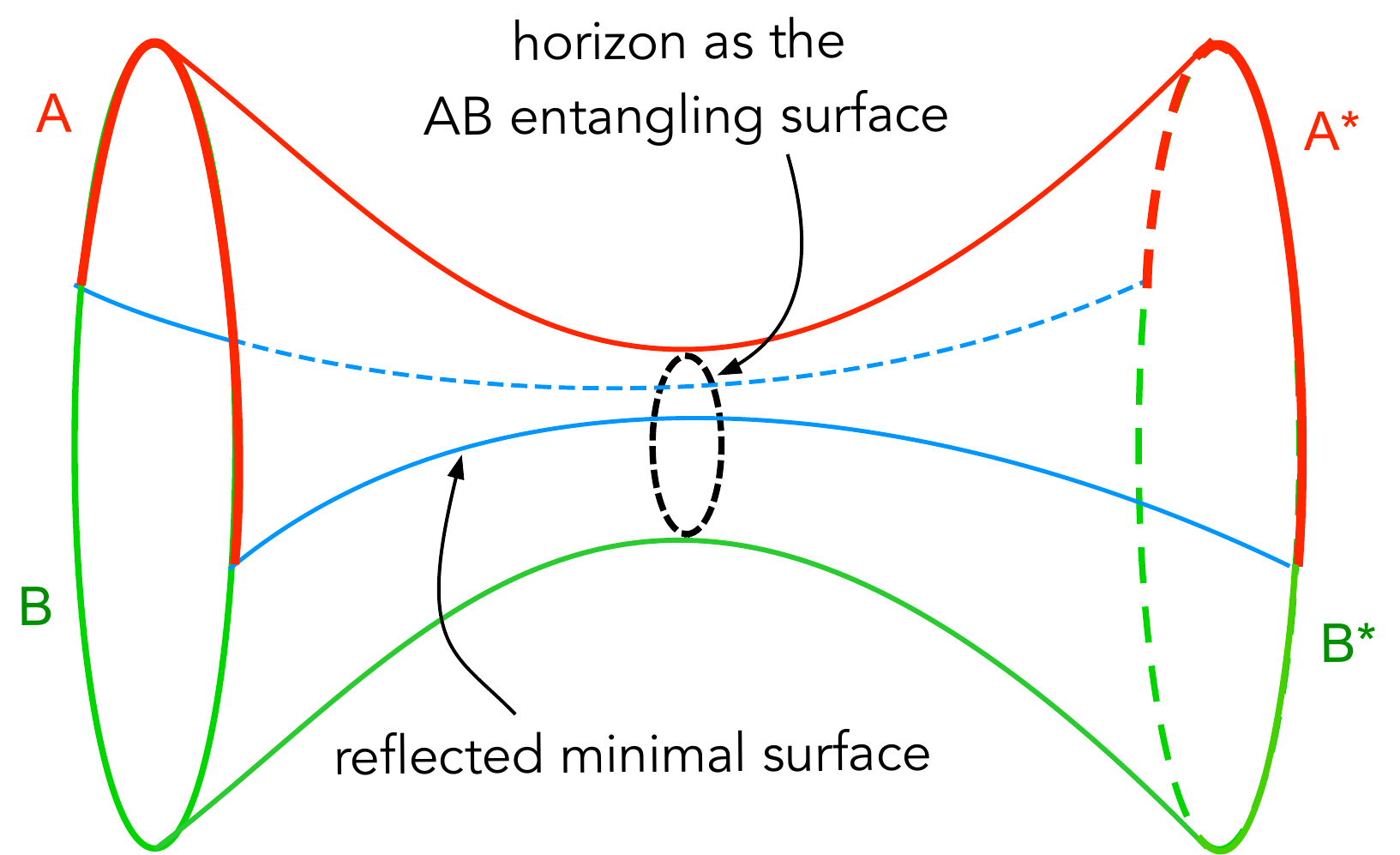}
\caption{ \label{bh:worm} When $AB$ is the full CFT Hilbert space, the Gibbs state is canonically purified by a thermofield double state in $\mathcal H_{L}\otimes \mathcal H_R $, which is dual to a two-sided Schwarzschild black hole geometry. The horizon is the entangling surface for $AB$. For this bipartition $A:B$ of the left CFT, the minimal cross-section of its entanglement wedge is $1/2$ the the area of RT surface (shown in blue) for $AA^\star$, that passes through the horizon. We will generalize this picture to other regions $AB$ and states in holographic theories. }
\end{figure}

We can motivate this conjecture by an analogy with the thermofield double state, whose dual is the maximally extended AdS Schwarzschild black hole, shown in Figure~\ref{bh:worm}.  In this case, we should take the region $AB$ to be the entire left CFT and the original state $\rho_{AB} = e^{ - \beta H}/Z$ is simply the Gibbs state.  Then the purification $\left| \sqrt{\rho_{AB}} \right>$ exactly corresponds to the thermofield double state in $\mathcal{H}_{AB} \otimes \mathcal{H}_{AB}^\star$. The entanglement wedge of the left CFT is the exterior region of the black hole and there is a anti-unitary CPT operator  $J$ that exchanges this wedge with the right wedge. An entangling surface anchored between $A$ and $A^\star = J A J$ then exactly divides the entanglement wedge at the minimal cross section for $AB$. This is shown in blue in Figure~\ref{bh:worm}. The factor of $2$ in \eqref{ewdual} accounts for the double counting of both the wedges.

In the more general case, we will develop a picture where one takes an arbitrary entanglement wedge and glues it to the CPT conjugate of itself, using the Engelhardt-Wall procedure \cite{Engelhardt:2018kcs,Engelhardt:2017aux}. Symmetric entangling surfaces in this new spacetime compute $S(AA^\star)$, and also map out twice the EW cross section. 

We note that, like the entanglement of purification $E_p$, the reflected entropy $S_R$ is a measure of correlations between $A$ and $B$: both quantum and classical. Although unlike $E_p$ \cite{terhal2002entanglement}, we do not have an operational interpretation for this quantity that establishes this. If anything, $S_R$ is only singled out for its simplicity.

We will end this paper with a new conjecture, which aims to identify the dual of $E_W$ directly in the continuum limit where a tensor factorization cannot be used to describe sub-regions of the QFT. The best approach here is algebraic \cite{haag2012local}, and the relevant algebras $\mathcal{A}_{A}, \mathcal{A}_B$ in QFT, associated to causally complete sub-regions of spacetime $\mathcal{D}(A),\mathcal{D}(B)$, are von Neumann algebras that contain type-III$_1$ factors. We point out a very suggestive correspondence to the split property in QFT \cite{buchholz1974product,doplicher1984standard}. The split property is an extra condition that is sometimes included in the axioms of algebraic QFT. Physically, the split property requires that algebras associated to regions which are space-like separated are independent, such that it is possible to prepare states independently in each region. Roughly speaking, it should hold for theories with well defined thermodynamic properties at high temperatures. In this paper, we will simply assume that the split property applies to the relevant QFTs with holographic duals, when quantized on the spatial manifolds of interest.\footnote{There is a large literature studying this question, see for example: \cite{buchholz1986causal,buchholz1990nuclear,buchholz1987universal}. See \cite{Harlow:2018tng} for a recent discussion of this assumption in a holographic context.} The split property is equivalent to the following more abstract formulation: there exist a type-I factor $\mathcal{N}$ splitting the algebras:
\begin{equation}
\label{split}
\mathcal{A}_A \subset \mathcal{N} \subset \mathcal{A}_B'.
\end{equation} 
We work here with the case that the two spacetime regions $\mathcal{D}(A), \mathcal{D}(B)$  are spacelike separated. 
$\mathcal{A}_B'$ is the commutant of the algebra associated to $B$, of which $\mathcal{A}_A$ is a subalgebra. 
Actually, there could be many $ \mathcal{N}$ satisfying \eqref{split}; it however turns out that given a state $\psi$ that is cyclic and separating for the algebras $\mathcal{A}_{A},\mathcal{A}_{B},\mathcal{A}_{AB}$, there
is a \emph{canonical} type-I factor  $\mathcal{N}_\psi$. This was constructed in \cite{doplicher1984standard}, and can be defined as $\mathcal{N}_\psi = \mathcal{A}_A \vee \tilde{J} \mathcal{A}_A \tilde{J}$ where $\tilde{J} = J_{\psi;AB}$ is the modular conjugation operator defined using Tomita-Takesaki theory associated to $\psi$ and the algebra $\mathcal{A}_{AB}$. We will identify $\tilde{J}$ (up to conjugation by a fixed unitary) with the anti-unitary CPT operator $J$, that exchanges the two entanglement wedges of the dual spacetime associated to the canonical purification.

Note that the cross section $E_W$ is actually UV finite when $A$ and $B$ are spacelike separated since the relevant surface does not reach the boundary of AdS. We thus conjecture that this finite area is computing the entropy of this type-I factor. 
\begin{conjecture}
\label{conj}
The continuum limit of the reflected entropy/entanglement wedge cross section duality for holographic QFTs becomes:
\begin{equation}
 S(\mathcal{N}_\psi)_\psi =  2 E_W(A: B)  + \mathcal{O}(G_N^{0})
\end{equation}
where $S(\mathcal{M})_\psi$ is the von Neumann entropy of the density matrix representing $\psi$ on a type-I factor $\mathcal{M}$. 
\end{conjecture}

We will give evidence for this correspondence by relating $ S(\mathcal{N}_\psi)_\psi $ to the reflected entropy above. Beyond holography, our results also give a replica/path integral method for computing quantities associated to the split property. There are several interesting applications of this conjecture, including a geometric picture for a regulator of entanglement in QFT that has been suggested previously \cite{schroer2007localization,schroer2010localization,narnhofer1994entropy,narnhofer2002entanglement,Narnhofer:2011zz,Otani:2017pmn}. This regulator shares similarities to the mutual information regulator discussed in \cite{Casini:2015woa}, but has the advantage that the regulated entropy is still an entropy.

Note that the reflected entropy in the thermofield double case has been studied previously in the literature \cite{Morrison:2012iz,Shenker:2013pqa,Hosur:2015ylk}. Indeed, whenever $\rho_{AB}$ is a state associated to a local modular Hamiltonian, i.e. for half space cuts of a QFT \cite{Bisognano:1976za} and the various generalizations \cite{Casini:2011kv,Cardy:2016fqc},  then the result we claim follows immediately from symmetry as well as the original Ryu-Takayangi (RT) formula \cite{Ryu:2006bv}. In this paper, we are claiming a generalization that works for other regions and states with holographic descriptions. $S_R$ is also a kind of operator entanglement entropy \cite{Hosur:2015ylk,zhou2017operator,dubail2017entanglement} which is here associated to the operator $\sqrt{\rho_{AB}}$.  The operator space entanglement entropy studied in \cite{zhou2017operator} is associated to the operator $\rho_{AB}/\sqrt{ \rm Tr \rho_{AB}^2}$, and the replica computations in that paper have some relations to those presented here, as we discuss in Section~\ref{sec:replica}.  

When we make the final connection between the reflected entropy and minimal surfaces in a holographic dual, we will assume the original state is prepared via a Euclidean path integral and that we are interested in sub-regions $A,B,AB$, which live on a constant time slice that represents a moment of time reflection symmetry. This situation allows for an easy Wick rotation to real times. It also means that we will be dealing with minimal RT like surfaces and not their dynamical generalization - the HRT \cite{Hubeny:2007xt} surfaces. We do this only to simplify the discussion - there is no reason to expect that our results do not work in the dynamical setting. Indeed, the entanglement wedge cross section certainly has a dynamical generalization, and so does the reflected entropy. So we continue to conjecture their duality. 

The plan of this paper is as follows. In Section~\ref{sec:refent}, we will study properties of the reflected entropy,  setting up some notation and then proving some basic inequalities involving $S_{AA^\star}$.  We also introduce a natural generalization involving regions that are not symmetric across the purification. In Section~\ref{sec:refmin}, we discuss the holographic dual that we call reflected minimal surfaces. In the symmetric case, we show that these surfaces give nothing more than a reinterpretation of the entanglement wedge cross section. The non-symmetric generalizations can also be computed in holography using reflected minimal surfaces.
In Section~\ref{sec:replica}, we discuss a novel replica trick that computes $S_R$, and in Section~\ref{sec:duality}, we use this and other approaches to establish the holographic dictionary for the reflected entropy, including its quantum corrections.  In Section~\ref{sec:comp}, we compute the reflected entropy in some simple quantum systems. In Section~\ref{sec:split}, we expand on our discussion of the relation to the split property.
In Appendix~\ref{app:ineq}, we attempt to prove that $S_R$ satisfies the more non-trivial inequalities that were
proven geometrically for $E_W$ in \cite{Takayanagi:2017knl,Nguyen:2017yqw}. These inequalities were used to conjecture the duality with $E_p$. We will manage to show that the reflected entropy satisfies some of these inequalities, but not all. We will however give independent evidence that all of the inequalities should be satisfied in holographic states.

\section{Reflected entropy}
\label{sec:refent}

In this section we summarize some properties of reflected entropy in the case of finite dimensional Hilbert spaces.
Consider the canonical purification in \eqref{can} but now as a pure density matrix on the doubled Hilbert space:
\begin{equation}
\rho_{A A^\star B B^\star} = \left| \sqrt{\rho_{AB}}\right> \left< \sqrt{\rho_{AB}} \right|
\end{equation}
where it is natural to use the same symbol for this new density matrix since tracing out $A^\star B^\star$ gives back the original density matrix \eqref{traceab}.
We will now trace over various sub-systems. The reflected entropy, $S_R$, is defined as:
\begin{equation}
S_R(A:B) \equiv S(AA^\star)_{\sqrt{\rho_{AB}}} = S(BB^\star)
\end{equation}
and purity of the underlying state shows that the definition is symmetric under the exchange $A \leftrightarrow B$. 
Here, $S(X)_\psi$ refers to the von Neumann entropy of $\rho_X = {\rm Tr}_{X^c} \left| \psi \right> \left< \psi \right|$ where $X^c$ is the complement region. We will sometimes suppress the $\psi$ subscript where the state should be understood.

We will often work in the Hilbert space ${\rm End}(\mathcal{H})$, which makes it clear that the results are canonical and not dependent on any particular choice of basis. However, it is sometimes convenient to pass to a more standard Hilbert space description. There is an isometric isomorphism between the Hilbert space ${\rm End}( \mathcal{H})
= \mathcal{H} \otimes \mathcal{H}^\star$, and the doubled Hilbert space $\mathcal{H}_L \otimes \mathcal{H}_R$, that we will denote by $K$. It is defined in a particular basis $\left| i \right>$ of $\mathcal{H}$ via:
\begin{equation}
\label{defK}
K \left| \sigma_{ij} \right> \equiv \left| i \right> \otimes \left| j \right>
\end{equation}
where we have defined a basis for ${\rm End} (\mathcal{H})$:
\begin{equation}
\label{sigbasis}
\sigma_{ij} = \left| i \right> \left< j \right|.
\end{equation}
The map in \eqref{defK} sends $\mathcal{H}^\star$ to the right copy $\mathcal{H}_R$. 
Note that the definition of $K$ depends on the choice of basis. For example, in a different basis $\left| \tilde{i} \right>
= U^\dagger \left| i \right>$, but keeping the definition of $K$ fixed, 
\begin{equation}
\label{kdef}
K \left| \right. \! \sigma_{\tilde{i} \tilde{j}} \! \left. \right> = \left| \tilde{i} \right> \otimes \left( U^T U \left| \tilde{j} \right> \right),
\end{equation}
where the transpose is taken in the original basis. However, it turns out that as long as we pick the $\{ \left| i \right> \}$ compatible with the various tensor factorizations of $\mathcal{H}$ the entropies defined below are unaffected. In particular we will require $\left| i_{AB} \right> = \left| i_A \right>  \otimes \left| i_B \right>$ such that $K_{AB} = K_A \otimes K_B$ where $K_{X}$ is the isomorphism defined for ${\rm End} (\mathcal{H}_X)$.

The canonical purification maps to:
\begin{equation}
\left| \Psi_\rho \right> = K_{AB} \left| \sqrt{\rho_{AB}} \right>
= \left( \sqrt{\rho_{AB}} \otimes 1  \right)  \left| \Phi^+_{AB} \right>
= \left(1 \otimes  \sqrt{\rho_{AB}}^T \right)  \left| \Phi^+_{AB} \right>
\end{equation}
where the maximally entangled state is  $\left| \Phi^+ \right> = \sum_i \left| i \right> \otimes \left| i \right>$. 
Reducing to the right factor we have:
\begin{equation}
{\rm Tr}_{A_L B_L} \left| \Psi_\rho \right> \left< \Psi_\rho \right|
= \rho_{AB}^T,
\end{equation}
and since the transpose acts in a factorized bases we conclude that $S(A) = S(A^\star)$ and $S(B) = S(B^\star)$. More generally, the entropy reduced to a sub-region contained only inside the reflected region $(AB)^\star$ is equal to the entropy of the equivalent unreflected sub-region. 

The reflected entropy is a measure of correlations between $A$ and $B$. Consider an entangled pure state where a simple calculation (using for example \eqref{explicit}) gives:
\begin{equation}
K_A \rho_{AA^{\star}} \left(K_A\right)^{-1} = \rho_A \otimes \rho_A^T,
\end{equation}
such that:
\begin{equation}
\label{sppure}
{\rm pure\,\,\,state:} \quad  S_R(A:B) = 2 S(A).
\end{equation}
At the opposite extreme, consider the factorized density matrix:
\begin{equation}
\label{factstate}
\rho_{AB} = \rho_A \otimes \rho_B,
\end{equation}
which has zero mutual information $I(A:B) = 0$. Since the state
$\left| \sqrt{\rho_A } \right> \otimes \left| \sqrt{\rho_B} \right>$ is clearly factorized between
$AA^\star$ and $BB^\star$, the entropy vanishes:
\begin{equation}
\label{spfact}
{\rm factorized\,\,\,state:}  \qquad S_R(A :B )= 0.
\end{equation}
However classical correlations also contribute to $S_R$ and it is easy to derive a bound:
\begin{equation}
\label{srmi}
S_R(A:B) \geq I(A:B)
\end{equation}
which follows from considering strong sub-additivity in the following form:
\begin{equation}
S(AA^\star) - S(A^\star) + S(AB) - S(B) \geq 0
\end{equation}
and using the fact that $S(A^\star) = S(A)$. We can also realize this by writing $S_R$ as a relative entropy:
\begin{equation}
\label{reldef}
S_R = S_{\rm rel}( \rho_{AA^\star B} |  \rho_{A A^\star} \otimes \rho_B)
\end{equation}
and applying monotonicity to tracing out $A^\star$.

Furthermore positivity of mutual information gives us:
\begin{equation}
I(A:A^\star) = 2 S(A) - S(AA^\star ) \geq 0 \qquad I(B:B^\star) = 2 S(B) - S(AA^\star) \geq 0
\end{equation}
from which we conclude that:
\begin{equation}
\label{sr2s}
S_R(A:B) \leq 2 \, {\rm min}\left\{ S(A),S(B) \right\}
\end{equation}
Together we have the following bounds on the reflected entropy:
\begin{equation}
\label{miss}
I(A:B) \leq S_R(A:B) \leq 2 \, {\rm min} \{ S(A), S(B) \},
\end{equation}
which can also be used to explain the special cases considered above: \eqref{sppure} and \eqref{spfact}. The bounds in \eqref{miss} are also satisfied by $2 E_p$ as was discussed in \cite{Takayanagi:2017knl,Nguyen:2017yqw}. Note that \eqref{miss} leads directly to other entropy relations that are thus also satisfied by $2 E_p$. For example the polygamy inequality for a tripartite pure state:
\begin{equation}
S_R(A:B) + S_R(A:C) \geq S_R(A:BC)
\end{equation}
and the fixed value in states saturating the Araki-Lieb inequality \cite{Nguyen:2017yqw}:
\begin{equation}
|S(A) - S(B)| = S(AB) \implies S_R = 2 {\rm min}(S(A),S(B))
\end{equation}

Now consider some simple separable states. 
These are mixtures of factorized states such as \eqref{factstate}. We will not attempt to make a general statement about separable states, however if we consider a slight simplification:
\begin{equation}
\rho_{AB} = \sum_k p_k \rho_A^k \otimes \rho_B^k
\end{equation}
where we take $\rho_B^k$ to be proportional to projectors with orthogonal support:
\begin{equation}
\rho^k_B  \rho^{k'}_B \propto \delta^{kk'}
\end{equation}
and no requirement on $\rho_A^k$. Then upon tracing over $B B^\star$ we find that $\rho_{AA^\star}$
becomes an ensemble of pure states $\{ p_k, \big| \sqrt{\rho^k_A} \big> \}$. If the $\rho_A^k$ are further proportional to projectors with orthogonal support then we have:
\begin{equation}
{\rm classical\,\,\,mixture} \qquad S_R(A:B) = H(\{ p_k\})
\end{equation}
where $H$ is the Shannon entropy of the classical probability distribution. This differs from twice the entanglement of purification which has $2 E_p(A:B) = 2 H(\{p_k\})$ as shown in \cite{Nguyen:2017yqw}. So we conclude, unsurprisingly, that $E_p \neq S_R/2$. 

We have however shown that many inequalities satisfied by $E_p/2$ are also satisfied by $S_R$. This was one of the main pieces of evidence presented in \cite{Takayanagi:2017knl,Nguyen:2017yqw} for the duality with the entanglement wedge cross section. Examining the list of agreeing inequalities considered in \cite{Takayanagi:2017knl,Nguyen:2017yqw} we are missing two:
\begin{equation}
\label{miss1}
?? \qquad S_R(A:BC) \geq I(A:B) + I(A:C) \qquad ??
\end{equation}
and monotonicity:
\begin{equation}
\label{miss2}
?? \qquad S_R(A:BC) \geq S_R(A:B) \qquad ??
\end{equation}
Indeed it is easy to find counterexamples to \eqref{miss1}. For example the classically mixed state:
\begin{equation}
\sum_k p_k \left| k_A \right> \left< k_A \right| \otimes  \left| k_B \right> \left< k_B \right| \otimes  \left| k_C \right> \left< k_C \right|
\end{equation}
However given the monogamy of mutual information inequality $I(A:BC) \geq I(A:B) + I(A:C)$, which is known to be true for geometric states of a theory with a classical gravity dual \cite{Hayden:2011ag}, then \eqref{miss1} follows from \eqref{srmi}. So at least in holographic theories we can derive \eqref{miss1}.

We unsuccessfully attempt to derive \eqref{miss2} in Appendix~\ref{app:ineq}. We do come close by proving a Renyi version of this inequality, where the von Neumann entropy is replaced by the Renyi entropy $S_n$ for \emph{integer} $n \geq 1$. We understand this as follows. The holographic version of \eqref{miss2} for the entanglement wedge cross section $E_W/2$ was related to entanglement wedge nesting \cite{Wall:2012uf} in \cite{Takayanagi:2017knl}. Entanglement wedge nesting has consequences for correlation functions involving modular flow that was studied in \cite{Balakrishnan:2017bjg,Faulkner:2018faa}. The Renyi version of the reflected entropy can be thought of as such a correlation function \cite{Chen:2018rgz}. The bound we derive in  Appendix~\ref{app:ineq} follows from analyticity and unitarity of these correlators. However for the case at hand we must apply these arguments to correlation functions of twist/swap operators of an $n$-replicated theory that compute the Renyi version of $S_R$. This only makes sense for integer $n$ and even after applying the usual $n$-analytic continuation trick we cannot guarantee the bounds continue to hold away from the integers. However in AdS/CFT such twist correlation functions can be continued in $n$ in a simple way such that at leading order in $G_N$ they give the same answer as correlation functions of heavy probe operators in a single copy of the theory (these are higher dimensional defect operators in $d>2$.)
Thus the $n$ analytic continuation of the inequality should still hold at this order in $G_N$. 
These ideas were inspired by those in \cite{casini2010entropy,headrick2014general}.  
Note that we have also not found any counterexamples to \eqref{miss2}. 

Another inequality that was proven for $E_W$ (but not generally for $E_p$) in \cite{Takayanagi:2017knl} is strong superaddativity. For the reflected entropy it is the statement that:
\begin{equation}
\label{miss3}
?? \qquad S_R(A_1 A_2 :B_1 B_2) \geq S_R(A_1:B_1) + S_R(A_2:B_2) \qquad ??
\end{equation}
It is again easy to find counterexamples to this using classically correlated states. However with some basic assumptions about holographic like states we sketch a proof of this inequality in Appendix~\ref{app:ineq}. 

We conclude that even without going into the detailed derivation of later sections, we can claim there is another plausible candidate duality for $E_W/2$ in $S_R$.

\subsection{Conditional mutual information}

Now that we have a canonical purification $\sqrt{ \rho_{AB}}$ there are various entropy quantities that one can define. For example, consider splitting the reference system in a non-symmetric way $A^\star B^\star = C^\star D^\star$ compared to the original $AB$ and then computing $S(B C^\star)$. We will demonstrate how to compute this more general quantity in holography momentarily. 

A natural quantity to study given a purification is the conditional mutual information (CMI) defined in this case as:
\begin{align}
\label{cr}
C_R(A:B) &\equiv I(A:B|A^\star) =  I(A:B|B^\star) =  S(AA^\star) + S(BA^\star) - S(A) - S(B) \\
& = S_R(A:B) - I(B:A^\star)
\end{align}
For example this is the quantity that goes into the squashed entanglement \cite{christandl2004squashed} - which is found after minimizing over possible extensions like $A^\star$. 
We see the appearance of $S(BA^\star)$ which is in the more general class of reflected entropy quantities that are not symmetric.  Since the conditional mutual information is positive by SSA we find the inequalities:
\begin{equation}\label{ineq1}
0 \leq C_R \leq S_R \leq 2 {\rm min}\{ S(A),S(B) \}
\end{equation}

We will give evidence that in holographic theories with geometric states $I(B:A^\star) = 0$ such that $S_R = C_R$. At least this is true to leading order in the $G_N$ expansion. Our proposal will still distinguish these quantities at sub-leading order in $G_N$.  

Consider also the tripartite information:
\begin{equation}
I_3(A:B:A^\star) = I(A:B) - C_R(A:B)
\end{equation}
then it is interesting to observe that for classical holographic states the statement that $C_R = S_R$ implies that $I_3 < 0$, via \eqref{miss}.  It is well known that the tripartite information in holographic states is non-positive \cite{Hayden:2011ag}. See for example \cite{Hosur:2015ylk} and \cite{ding2016conditional} where similar quantities to $I_3$ and $C_R$ were studied.  In these papers a unitary operator $U_{AB}$, rather than a density matrix, was re-interpreted as a state on ${\rm End}(\mathcal{H}_{AB})$.

\section{Reflected minimal surfaces}

\label{sec:refmin}

In this section we will introduce the gravitational dual for the reflected entropy, as well as its non-symmetric counterpart.  Consider the entanglement wedge of the boundary region $AB$. For simplicity, we will restrict to the case where we have a bulk time slice/Cauchy surface $\mathcal{S}$ with a time reflection symmetry, such that all the RT surfaces lie on $\mathcal{S}$ and are thus anchored to the boundary of $\mathcal{S}$ at the boundary of AdS. The entanglement wedge is a spatial region of $ r(AB) \subset \mathcal{S}$ that lies between $AB$ on the boundary and the minimal surface $m(AB)$, such that $\partial r(AB) = AB \cup m(AB)$. Note that usually the entanglement wedge refers to the \emph{spacetime} region given by the bulk domain of dependence of $r(AB)$: $\mathcal{E}(AB) = \mathcal{D}(r(AB))$, and then $r(AB)$ is sometimes referred to as the homology region. In the time-symmetric case, we can blur this distinction since one can be determined from the other; e.g. $r(AB) =   \mathcal{E}(AB) \cap \mathcal{S}$. The area of $m(AB)$ divided by $4 G_N$ computes the entanglement entropy, as conjectured in \cite{Ryu:2006bv} and proven in \cite{Lewkowycz:2013nqa}.

Consider the following construction. Take two copies of the wedge region $r(AB)$ denoting the second by $r^\star(AB)$ and sew these together by gluing the boundaries of the two regions along the minimal/RT surfaces $m(AB)$ and $m^\star(AB)$. The resulting euclidean manifold, denoted  $rr^\star(AB) = r(AB) \cup r^\star(AB)$, can be visualized by taking the two wedges on top of each other and folding them together along the RT surface. This construction then has the property that the $m(AB)$ RT surface behaves like a mirror - smooth curves reflect off the RT surface, while passing from one wedge into the other wedge region. While the resulting space is a smooth manifold, the natural metric induced from $r(AB)$ can have a discontinuous first derivative. 
We do not view this singularity as an issue for several reasons. On the one hand, in Section~\ref{sec:duality}, we will see that it arises from the replica trick and is on par with the conical singularities of the LM construction \cite{Lewkowycz:2013nqa}. We will also later embed this glued manifold into a real-time solution of Einstein's equations, following \cite{Engelhardt:2018kcs}. Here these singularities pass to discontinuities in the shear across $m(AB)$ which result in impulsive gravitational shockwaves that nevertheless solve Einstein's equations. 

Using $rr^\star(AB)$ we conjecture that the entropy $S(AA^\star)$ is computed as the area of a co-dimension-$1$ minimal surface $m(AA^\star)$ that lives inside $rr^\star(AB)$ ending on the boundary at $ \partial (AA^\star)$, and where the surface passes through $m(AB)$ via the smooth reflection rule. There is a homology condition on the folded space, where we define a new entanglement wedge region $r(AA^\star)$ in such a way that $\partial r(AA^\star) = m(AA^\star) \cup  AA^\star$. Then we conjecture that:
\begin{equation}
S_R(A:B) 
= \frac{ {\rm Area}[m(AA^\star)] }{4 G_N} + \ldots,
\end{equation}
and an appropriate regularization of the area should be understood if the minimal surface is boundary anchored.
We call $m(AA^\star)$ the reflected minimal surface.

To make sense of this prescription one should think of the boundary of the glued wedge $\partial (rr^\star)$ as a boundary time slice for the CFT, where the new doubled CFT Hilbert space $\mathcal{H}_{AA^\star BB^\star}$ resides. The glued bulk geometry is then describing the relevant state in this Hilbert space.\footnote{In QFT the Hilbert space never factorizes like this, so at this point we are assuming some natural UV regulator. We will  confront the continuum description directly in Section~\ref{sec:split}.} 
Note that when $\partial A \cap \partial B = 0$, the region $AA^\star$ of the CFT that is on the boundary of $rr^\star(AB)$ must be boundary-less: $\partial( AA^\star) = 0$. In this case, the homology condition is everything, and will result in a non-zero $S(AA^\star)$ only if $AA^\star$ is not contractible inside of  $rr^\star(AB)$. 
As we will discuss in Section~\ref{sec:split}, the bulk geometry in this case is a wormhole and the non-contractible cycle wraps the throat of this wormhole. 

Since the two boundary regions $A,A^\star$ are related by a symmetry that exchanges the glued wedges we expect that the minimization procedure for $m(AA^\star)$ will generally result in a surface that is symmetric under this exchange.  We can then give the following alternative prescription for finding $m(AA^\star)$:

Consider splitting the minimal surface $m(AB)$ into two regions, $m(AB) = \Gamma_A \cup \Gamma_B$ and then finding the minimal surface $m(\Gamma_A A)$ inside $r(AB)$ that ends on $\partial\left( \Gamma_A A\right)$. It is easy to see that minimizing the area of $m(\Gamma_A  A)$ over all possible such splittings, constructs half of $m(AA^\star)$. Specifically, defining the minimal splitting with $\Gamma_A = \Gamma_A^{\rm min}$, we have:
\begin{equation}
m(AA^\star) = m(\Gamma_A^{\rm min}  A) \cup m(\Gamma_{A^\star}^{\rm min}  A^\star)
\end{equation}
 where the later region is simply the mirror of $m(\Gamma_A^{\rm min}  A)$ on the other wedge.  Note that locally, the surface $m(\Gamma_A^{\rm min} A)$ can be found by minimizing over the intersection surface $\gamma = \partial \Gamma_A$, a co-dimension $1$ portion of  $m(AB)$.  Locally minimizing $\gamma$ also results in a surface $m(AA^\star)$, that intersects $m(AB)$ orthogonally.

\begin{figure}[h!]
\centering 
\includegraphics[width=.45\textwidth]{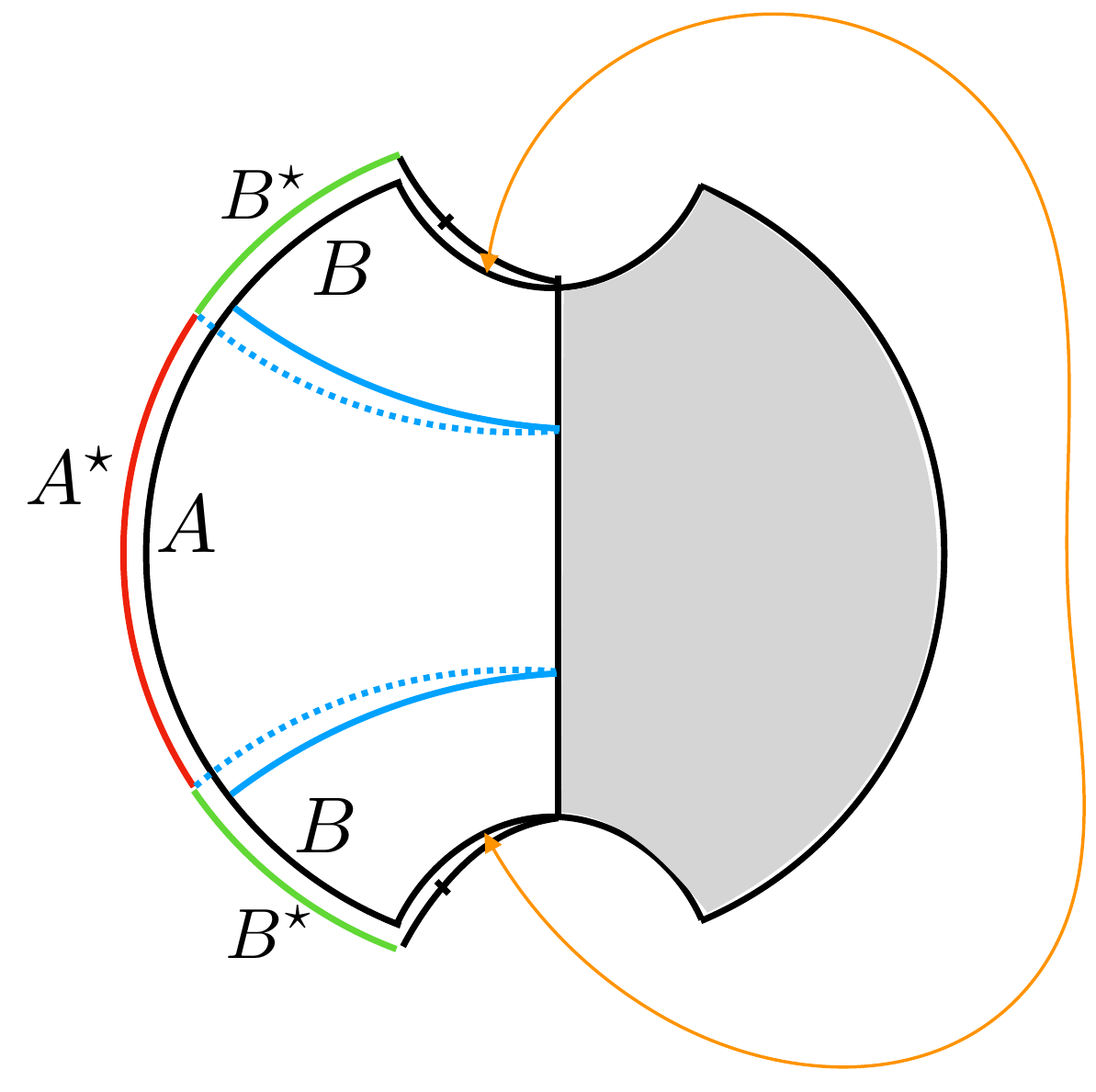}
\hfill
\includegraphics[width=.45\textwidth]{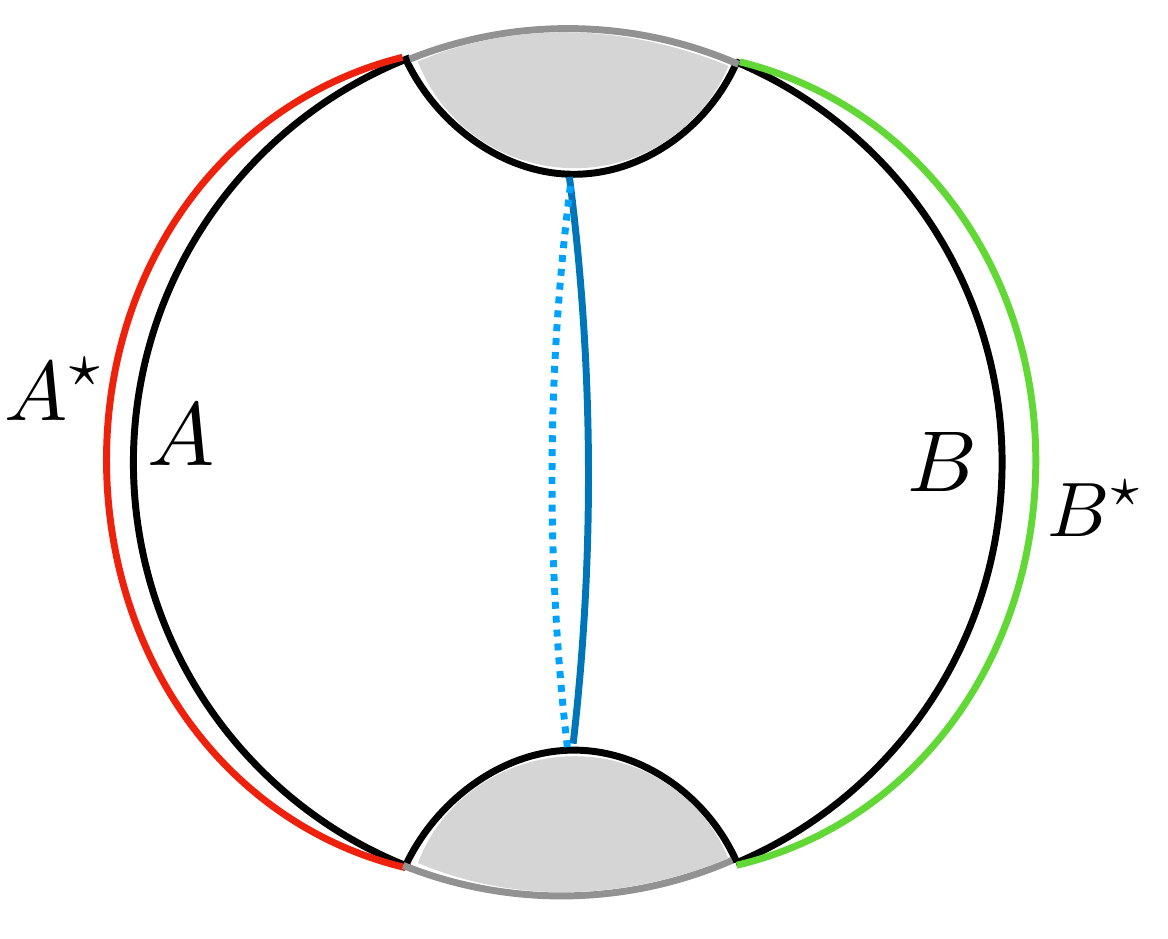}
\caption{ \label{fig:sr} Some example computations of the reflected minimal surfaces shown here in blue. The \emph{left} case involves the holographic thermofield double which is dual to the maximal extension of the BTZ black hole. Here $AB$ is the entire $S^1$ boundary region of the left copy of $\mathcal{H}_{CFT}^L$.  The right Hilbert space $\mathcal{H}_{CFT}^R$ is naturally drawn on the right hand side of this figure but we choose to place it on top of the other space folding the wormhole along the $AB$ entangling surface, which in this case is the horizon of the black hole.   The reflected minimal surfaces, in this case are trivially correct, as can be seen by unfolding.  On the \emph{right} we show the case that was important in the original conjectures of \cite{Takayanagi:2017knl} where the CFT is in the ground state and is cut into three regions $A$, $B$ and $(AB)^c$. The entanglement wedge in this case is not disconnected and shown in white. We have sketched the doubled space $rr^\star(AB)$ which in this case has the topology of a cylinder - similar to the wormhole slice of the eternal black hole. The reflected minimal surface wraps the horizon of this wormhole.  }
\end{figure}

\begin{figure}[h!]
\centering 
\includegraphics[width=.45\textwidth]{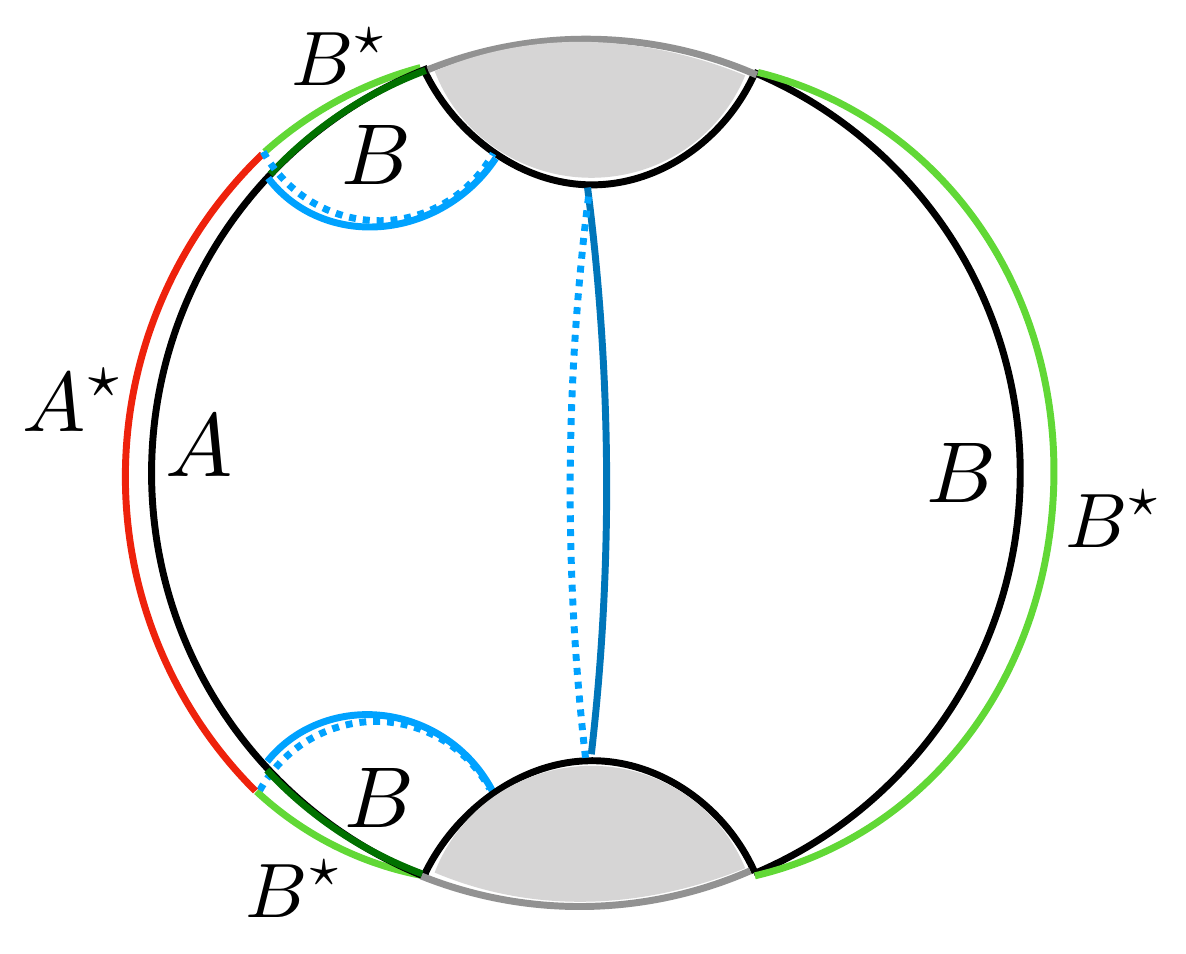}
\hfill
\includegraphics[width=.45\textwidth]{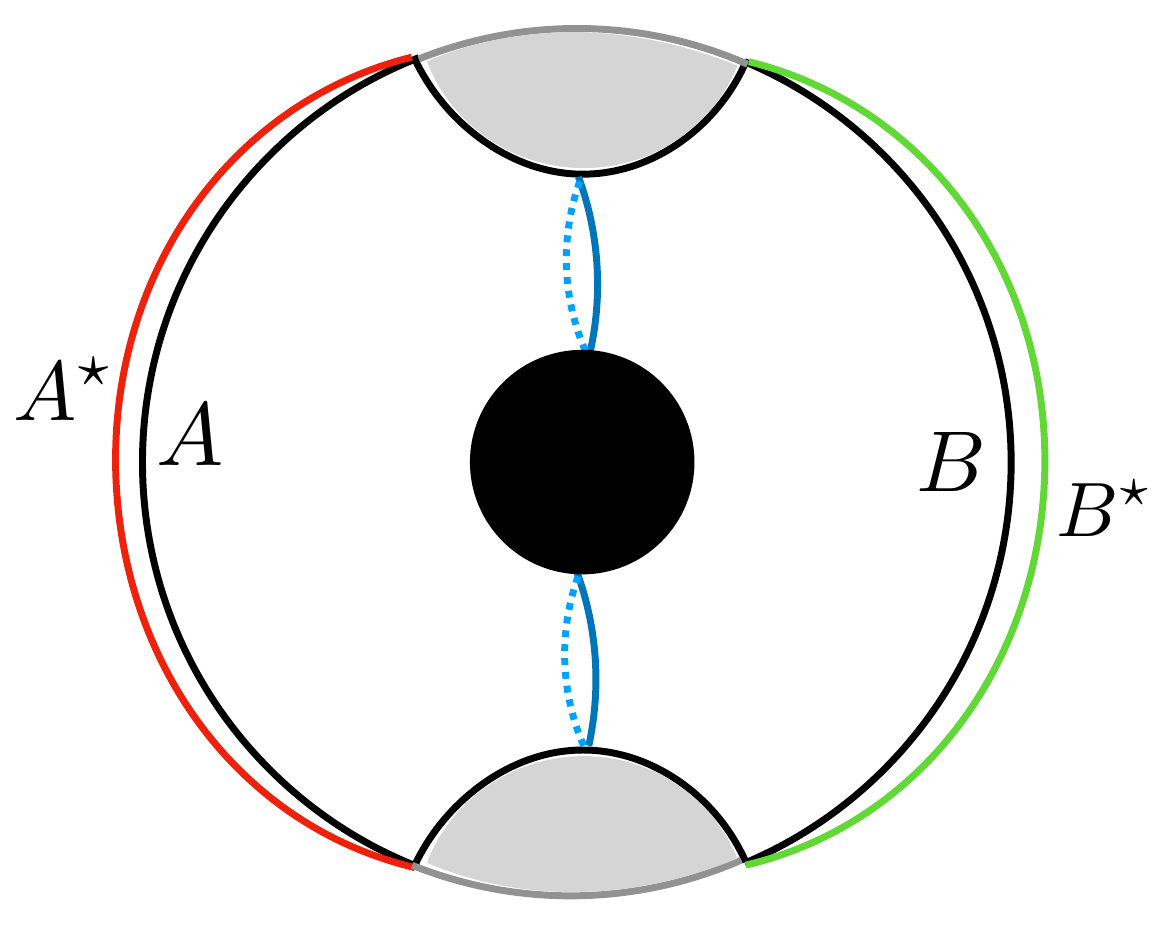}
\caption{ \label{fig:sr2} Some more examples of reflected minimal surface. On the \emph{left} we show a case where $(\partial A) \cap (\partial B) \neq 0$ and which is a small deformation of the right panel in Figure~\ref{fig:sr}. The \emph{right} figure demonstrates what happens in the presence of a mixed state/black hole. The black hole horizon in this case acts also like a mirror. }
\end{figure}

Of course, the surface $ m(\Gamma_A^{\rm min}  A)$ is exactly the entanglement wedge cross section surface defined in \cite{Takayanagi:2017knl}. It is constructed in exactly the same way. Thus we find the relation:
\begin{equation}
S_R(A:B) = 2 E_W(A:B) = 2\frac{ {\rm Area}[m(\Gamma_A^{\rm min} A)] }{4 G_N} + \ldots
\end{equation}
We give several examples in Figure~\ref{fig:sr}-\ref{fig:sr2} which show the re-interpretation of the cross section as the reflected minimal surface.
While the pictures of the cross section were mostly previously discussed for the case where $\partial A \cup \partial B = 0$, the definition also works in the overlapping case, where the interpretation in terms of the reflected minimal surface is more transparent, since in this case $m(AA^\star)$ is boundary anchored.

The reader might complain that this is rather a trivial reinterpretation of the cross section - why do we even bother to describe it? There are several reasons. Firstly, it points to a natural generalization, where we split $A^\star B^\star = C^\star D^\star$ and compute something like:
\begin{equation}
S(AC^\star)_{\sqrt{\rho_{AB}}} =  \frac{ {\rm Area}[m(AC^\star)] }{4 G_N} + \ldots
\end{equation}
where $m(AC^\star)$ is a minimal surface in $rr^\star(AB)$ that ends on $AC^\star$.

We will also see that this re-interpretation allows us to identify the boundary duality with the reflected entropy. In particular, recently \cite{Faulkner:2018faa} it was shown that correlation functions for heavy probe operators, and involving the imaginary modular flow operator $\Delta_{AB}^{1/2}$ (equivalently $J=J_{AB}$), are computed by reflected geodesics at the RT that look very similar to the description of $m(AA^\star)$. The only difference being that the later may be a higher dimensional object for boundary dimensions $d>2$. As discussed in a related proposal for modular flow  \cite{Chen:2018rgz}, entangling surfaces themselves give nice examples of such heavy probe operators, and so combining the ideas of \cite{Faulkner:2018faa} and \cite{Chen:2018rgz} will give us one way to arrive at the proposed duality.  

\subsection{Conditional mutual information}

We consider here the conditional mutual information $C_R(A:B)$ defined in \eqref{cr}. Here we must confront the computation of $S(A^\star B)$. This is found via the area of a minimal surface $m(A B^\star)$ defined to live inside of $rr^\star(AB)$ and end on $\partial (A B^\star)$ with the appropriate homology condition inside $rr^\star(AB)$. In fact, we can give a holographic proof that such a minimal surface is always $m(AB^\star) = m(A) \cup m(B^\star)$ and the two regions do not talk to each other across the reflection surface $m(AB)$. In the local case, for example, for the thermofield double purification, this was already observed in \cite{Morrison:2012iz}. In particular they found that the RT surface for the union of two regions, one on the left and one on the right section of the wormhole, factorizes like this if the boundary regions have no overlap after identifying the two boundaries under the CPT reflection that exchanges the two entangled CFTs. 

\begin{figure}[h!]
\centering 
\includegraphics[width=.44\textwidth]{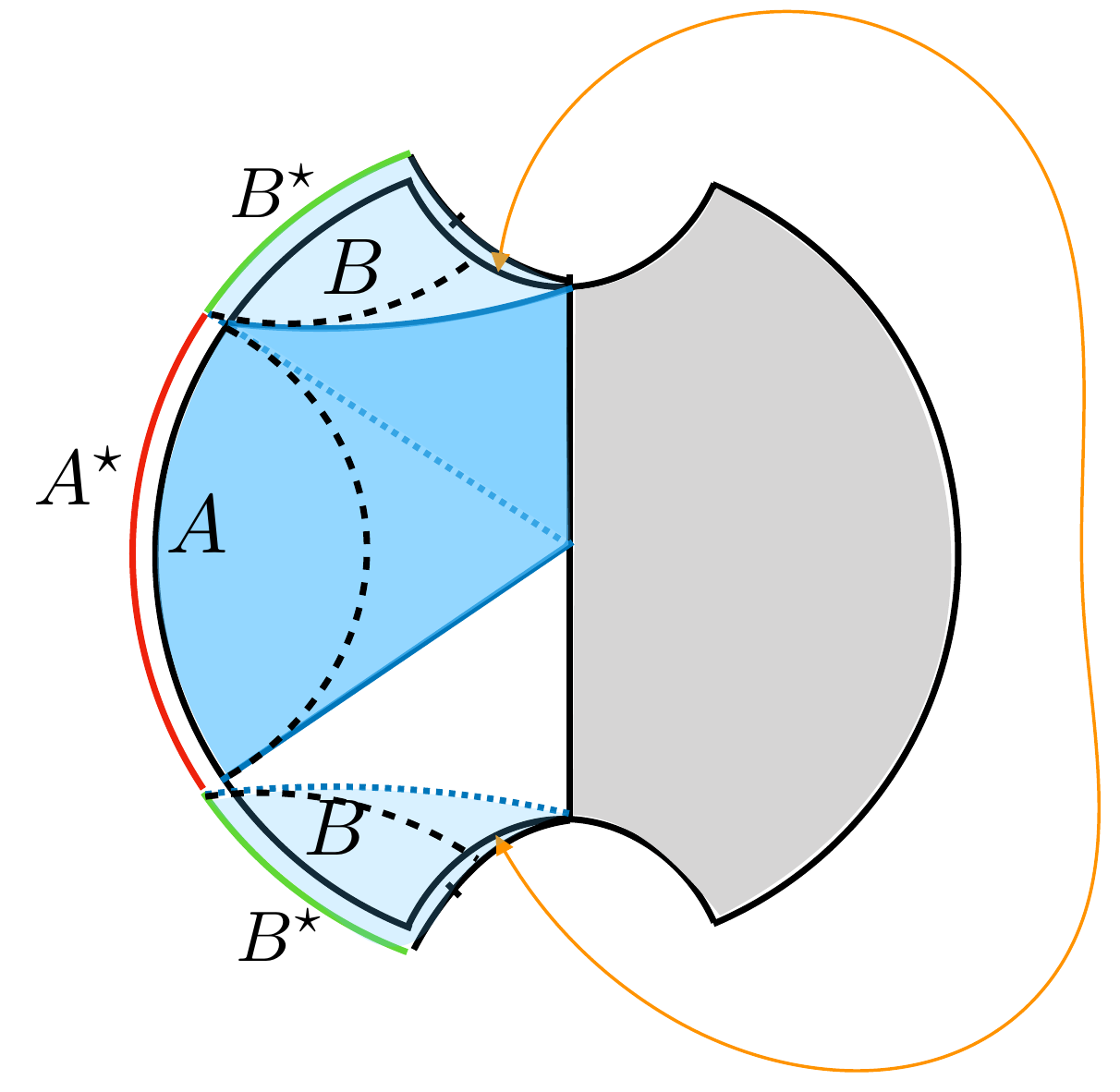}
\hfill
\includegraphics[width=.44\textwidth]{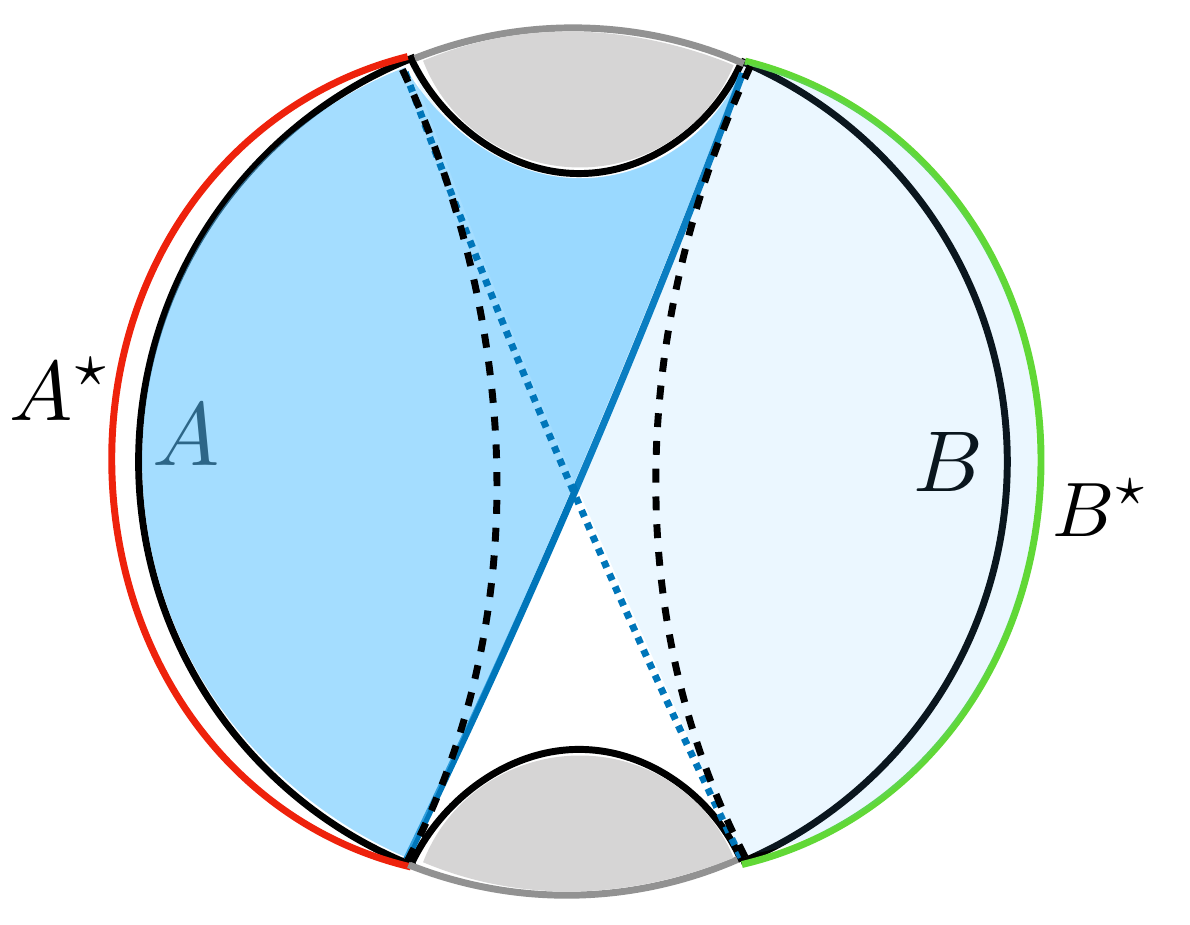}
\caption{\label{fig:cr} These figures demonstrate the fact that $I(A,B^\star) = 0$.  The \emph{left} figure follows the representation of the thermofield double discussed in Figure~\ref{fig:sr}. The dominant configuration follows the black dashed curves, while the blue curves (and associated homology region) has a larger area.  Only the blue curves pass through the AB entangling surface. When projecting the blue curves onto a single wedge region $r(AB)$ one can cut and join these curves so they form surfaces with the same boundary condition as the $A,B$ minimal surfaces. They thus have larger area.
This means the minimal entangling surfaces are always disconnected such that the mutual information vanishes. On the \emph{right} we have shown the equivalent picture in the non-local case. }
\end{figure}

\begin{figure}[h!]
\centering 
\includegraphics[width=.43\textwidth]{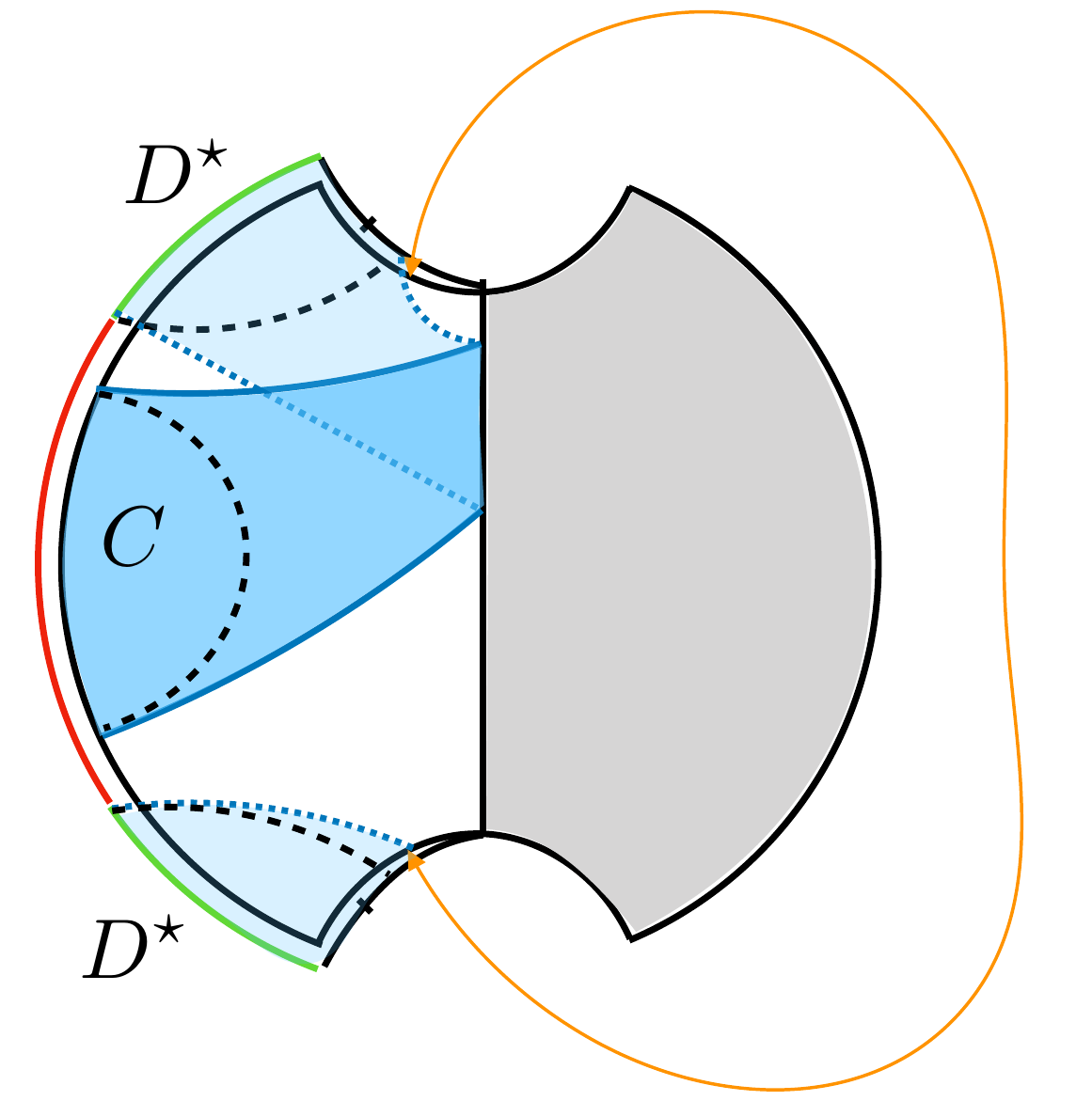}
\hfill
\includegraphics[width=.43\textwidth]{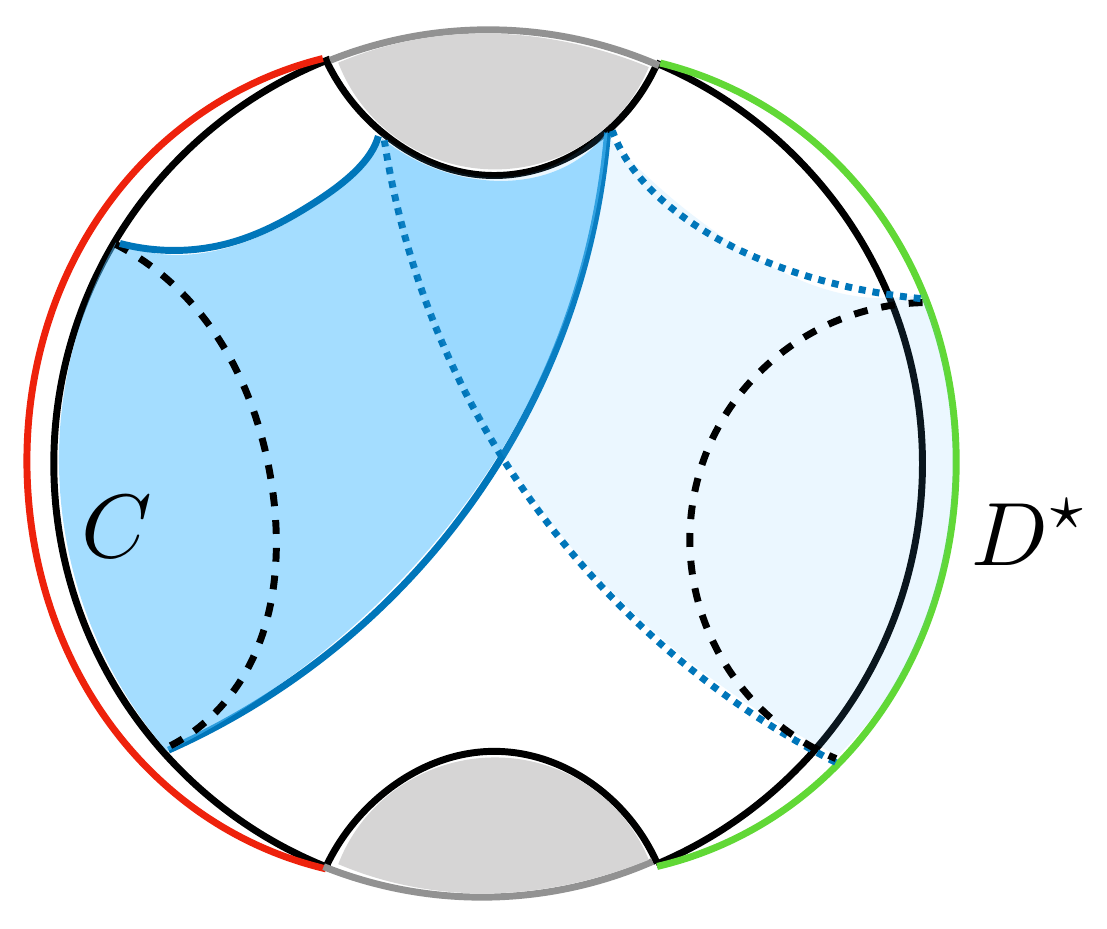}
\caption{Keeping the doubled entanglement wedge $rr^\star(AB)$ fixed we consider reduced sub-regions $C,D^\star$ and their associated minimal surface. We consider only the case where on the original copy $C\cap D = 0$. We can give the same argument as in Fig~\ref{fig:cr} to show that $I(C,D^\star) =0$. \label{fig:cr2}}
\end{figure}

Inspired by this observation, the general rule we would like to establish is as follows. If $C \cap D = 0$ where both $C,D \subset AB$ then for the reflected regions inside of $rr^\star(AB)$  the entropies satisfy $S(CD^\star) = S(C) + S(D^\star) = S(C) + S(D)$ to leading order in $G_N$. Note that here we are imagining holding fixed the state $\left| \sqrt{\rho_{AB}} \right>$ and partial tracing to the subregion $CD^\star$. 
We can prove this geometrically both for the local case and the non-local case, see Figure~\ref{fig:cr}-\ref{fig:cr2}  for the argument.  Since $A \cap B = 0$ (taking the spatial regions to be open subsets) we conclude that $I(A,B^\star)$ vanishes and that:
\begin{equation}
\label{srcr}
S_R(A:B) = C_R(A:B) + \mathcal{O}(G_N^0)
\end{equation}
where it is certain that these quantities will be distinguished by their quantum corrections. 

One important issue that remains is to generalize \eqref{srcr} to real times. While the definition of $S_R(A:B)$ generalizes simply to real times, it is no longer so simple to compute $S(A B^\star)$. In this case the extremal surface inside the doubled spacetime that we construct in Section~\ref{sec:duality} will move into the past or future of the $m(AB)$ HRT surface where we do not give an explicit bulk solution and so it is harder to compute the area of this extremal surface. This happens for $S(AB^\star)$ and not $S(AA^\star)$ because the former boundary regions are not symmetric under the $J$ operator. The best we might hope to do for $S(AB^\star)$ is to give some bound along the lines discussed in \cite{Chen:2018rgz} and using such a bound it might still be possible to show \eqref{srcr} in the dynamical setting, but we will leave this to future work. 

\subsection{Quantum corrections}
\label{sec:qc}

\begin{figure}[h!]
\centering 
\includegraphics[width=.43\textwidth]{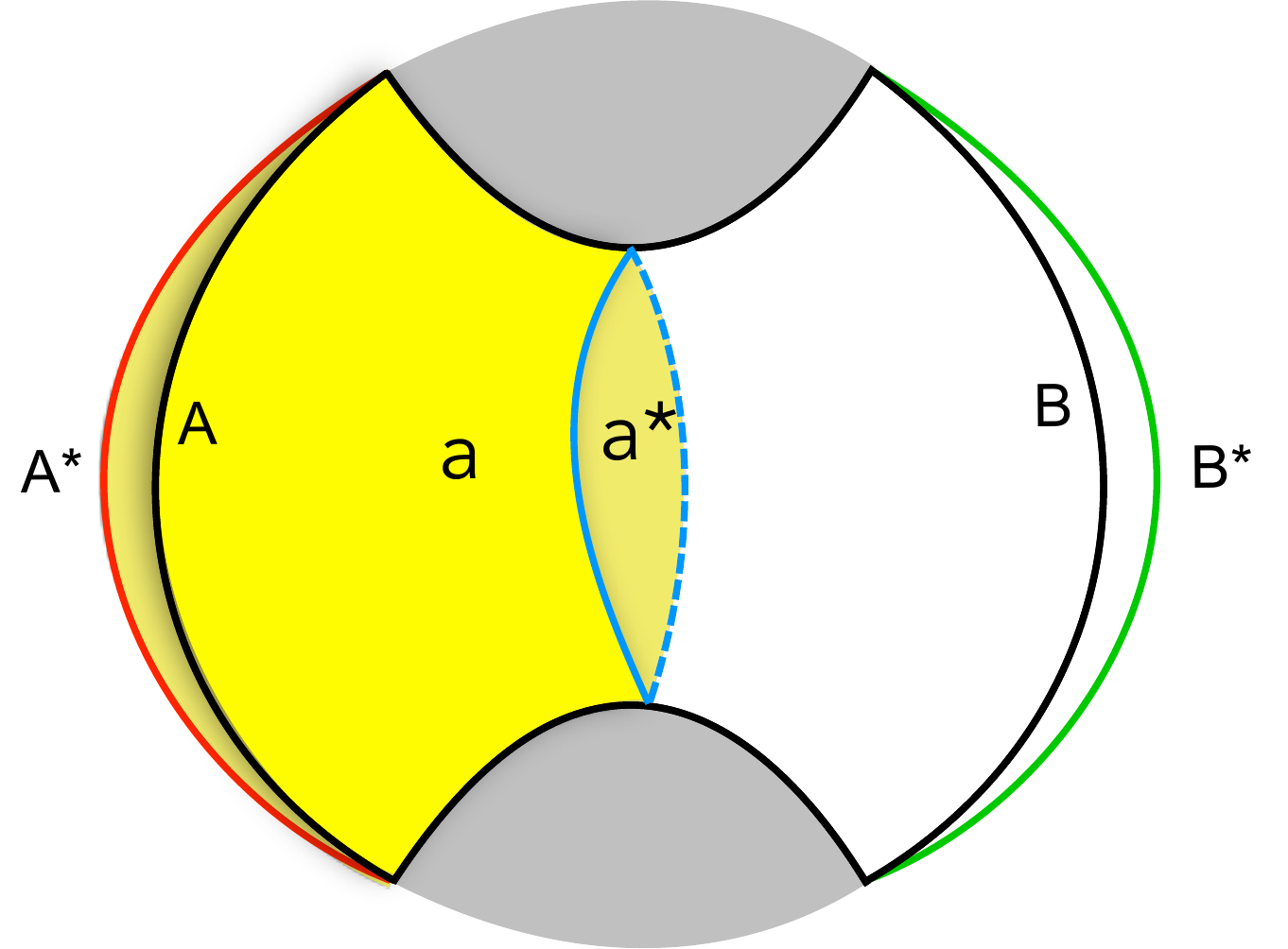}
\hspace{10 mm}
\includegraphics[width=.41\textwidth]{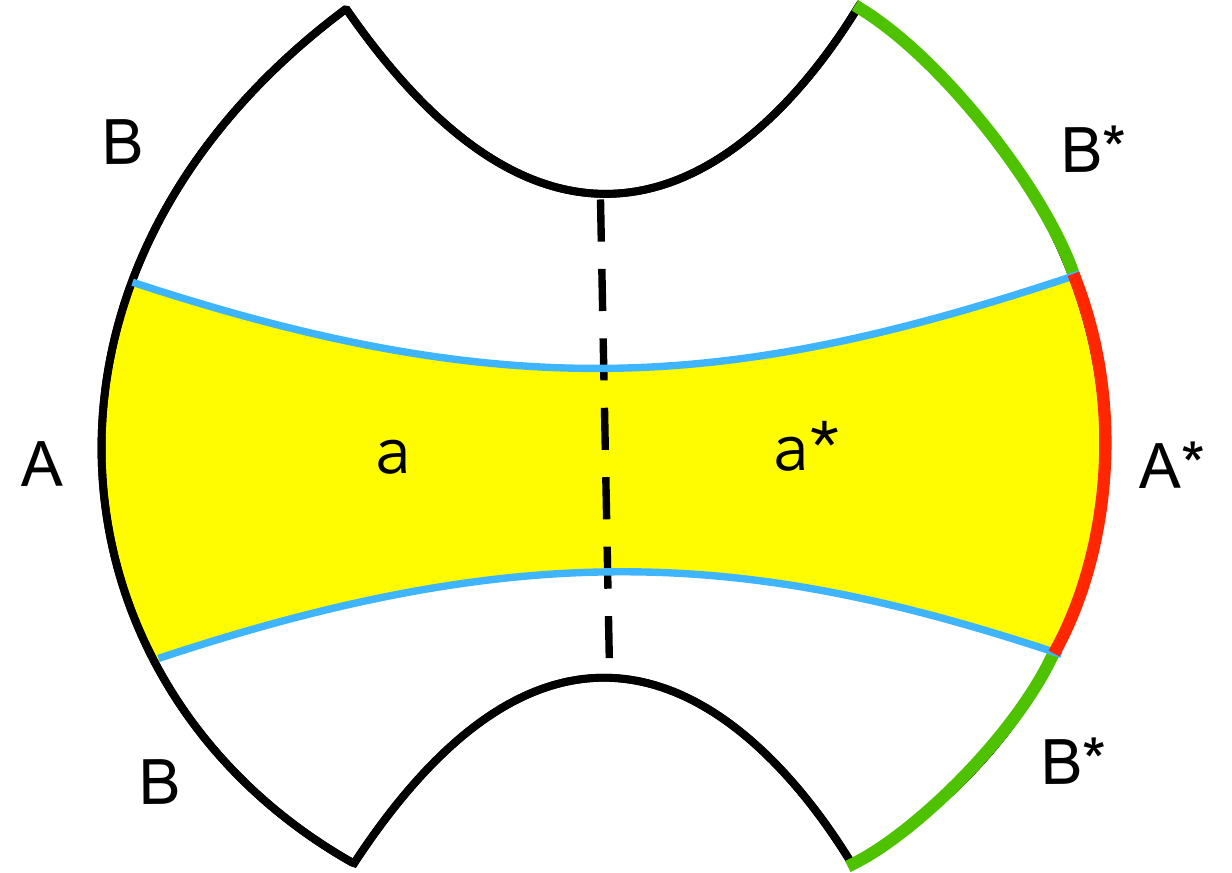}
\caption{The entanglement wedge $r(AA^\star) = aa^\star$ (shown in yellow), for the cases $\partial A \cap \partial B = 0$ (\emph{left}) and $\partial A \cap \partial B \neq 0$ (\emph{right}). The quantum correction $S_R^{\rm bulk}(a:b)$ comes from the bulk entanglement between this region and $bb^\star$ (shown in white), across the reflected minimal surfaces (blue).\label{srquant}}
\end{figure}

Since the conjectured dual for $E_W$ is now precise and computable, it is easy to
guess the quantum corrections following \cite{Faulkner:2013ana,Jafferis:2015del}:
\begin{equation}
\label{srquantum}
S_R(A:B) =  \frac{  \left< \widehat{\mathcal{A}}[m(AA^\star)] \right>_{\sqrt{\rho_{ab}^{\rm bulk}}} }{4 G_N} + 
S_R^{\rm bulk}(a:b) + \mathcal{O}(G_N)
\end{equation}
where $r(AB)$ is split into two regions $a$, $b$ by the cross section surface $\Gamma^{\rm min}_A= \partial a \cap \partial b$, and the entropy on the RHS is the reflected entropy for bulk QFT reduced to $ab$ with the associated mixed state in that region. The area operator $\widehat{\mathcal{A}}$ computes the classical area plus any quantum corrections due to bulk quantum fluctuations that arise from the canonically purified (GNS) state $\sqrt{\rho_{ab}^{\rm bulk}}$. 
More explicitly we have $r(AA^\star) = a a^\star$ and:
\begin{equation}
\label{srquantum2}
S_R^{\rm bulk}(a:b) = S^{\rm bulk}(aa^\star)_{\sqrt{\rho_{ab}^{\rm bulk}}}
\end{equation} 
See Figure~\ref{srquant}.

There are several subtleties with \eqref{srquantum} which we do not explore in detail. Firstly, the \emph{location} of $m(AB)$ is subject to $G_N$ corrections. These would show up in \eqref{srquantum} as an $\mathcal{O}(G_N)$ change in expectation value of the area operator arising at the intersection surface of the $m(AA^\star)$ and $m(AB)$ RT surfaces. This is a rather delicate $\mathcal{O}(1)$ contribution to $S_R$. 
In the quantum theory the entanglement wedge should be determined by a quantum extremal surface \cite{Engelhardt:2014gca}. Thus we should use the quantum corrected location of the surface $m(AB)$, including the $\mathcal{O}(G_N)$ corrections, to construct the doubled space used in our expression \eqref{srquantum}. We will give some evidence that this is the correct prescription in Section~\ref{sec:duality}. 

Secondly, we must be careful about constructing the pure quantum state $\sqrt{\rho_{ab}^{\rm bulk}}$ for gauge fields and metric fluctuations. As with the difficulties of defining entanglement entropy in these cases, in order to construct the localized Hilbert space on $ab$, we need to include central elements/edge modes \cite{Casini:2013rba,Donnelly:2016auv}
: operators that commute with all operators in $ab$ and $(ab)^\star$. This is likely necessary here also, although we leave the details to future work. 

\section{Replica trick for reflected entropy}
\label{sec:replica}

In this section, we will develop a replica trick for computing reflected entropy that lets us, in principle, work with arbitrary CFTs. As usual the most power will come for 2d CFTs \cite{Holzhey:1994we,calabrese2004entanglement,Headrick:2010zt}, however our approach also works in higher dimensions. 

We point out that this section 
can be understood as giving a CFT prescription for computing correlations (entropies or otherwise) in the GNS representation for states of a QFT that might be useful for other purposes.

\subsection{Some formalism}

Working with the Hilbert space ${\rm End}(\mathcal{H})$, there are several natural operators. We have the left and right action:
\begin{equation}
\label{lraction}
L_M \left| \sqrt{\rho} \right> = \left| M \sqrt{\rho} \right> \qquad
R_M \left| \sqrt{\rho} \right> = \left|  \sqrt{\rho} M\right>
\end{equation}
From this we can construct the Tomita-Takesaki modular operators associated to the state $\left| \sqrt{\rho} \right>$ and the algebra of operators given by the left action $L_M$:
\footnote{We do not label the modular operators by the state that defines them - we hope this will be clear from the context.}
\begin{equation}
\label{moddef}
\Delta = L_{\rho} R_{\rho}^{-1}\,, \qquad
J \left| M \right>  = \big| M^\dagger \big>
\end{equation}
which together satisfy the defining equation $ J \Delta^{1/2} L_M \left| \sqrt{\rho} \right> = (L_M)^\dagger  \left| \sqrt{\rho} \right> $. See for example \cite{Witten:2018lha} for a review of modular operators. 
We have taken the state to be cyclic and separating, which means that $\rho_{AB}$ should be invertible.  We will assume this here. This is not a necessary assumption, just one that makes our life a little easier, since we do not need to keep track of the support of various modular/density operators.  

In order to represent the purified state in a path integral language, we need to apply a generalization of the replica trick. We will replicate twice by taking $n \times m$ copies of the theory. The first replication relates simply to a replacement of the canonically purified state by:
\begin{equation}
\label{psim}
\left| \psi_m \right> = \frac{1}{ \sqrt{ {\rm Tr} \rho_{AB}^m }} \left| \rho_{AB}^{m/2} \right>
\end{equation}
which is normalized. We will be able to describe this state via a path integral for $m \in 2 \mathbb{Z}^+$.\footnote{We will later analytically continue $m$ away from even integers. This is reminiscent of entanglement negativity computations in QFT \cite{calabrese2012entanglement}.}
 Finally, we would like to compute the entanglement entropy of this state. Instead, we compute the Renyi entropy for integer $n$:
\begin{equation}
S_n(AA^\star)_{\psi_m} = \frac{1}{n-1} \ln {\rm Tr} \left( \rho^{(m)}_{AA^\star} \right)^n, 
\end{equation}
where 
\begin{equation}
\label{rhoaa}
\rho^{(m)}_{AA^\star} \equiv {\rm Tr}_{BB^\star} \left| \psi_m \right> \left< \psi_m \right| \equiv\frac{1}{{\rm Tr} \rho_{AB}^m } {\rm Tr}_{BB^\star}  \left| \rho_{AB}^{m/2} \right>  \left< \rho_{AB}^{m/2} \right|.
\end{equation}
We can easily extract $S_n$ from the following un-normalized ``partition functions'':
\begin{equation}
Z_{n,m} \equiv {\rm Tr}_{AA^\star}  \left( {\rm Tr}_{BB^\star}  \left| \rho_{AB}^{m/2} \right>  \left< \rho_{AB}^{m/2} \right| \right)^n
\end{equation}
which satisfy $Z_{1,m} = {\rm Tr} \rho_{AB}^m$ such that:
\begin{equation}
S_n(AA^\star)_{\psi_m} = \frac{1}{n-1} \ln \frac{Z_{n,m}}{ \left(Z_{1,m}\right)^n}.
\end{equation}
In the next subsection, we will give a CFT prescription for computing $Z_{n,m}$ for even $m$ and integer $n$. 

One way to compute $Z_{n,m}$, involves using cyclic swap operators, which are precursors to twist operators in QFT. We consider $n$ copies of the Hilbert space $\left( {\rm End}(\mathcal{H}_{AB}) \right)^{\otimes n}$. In this new Hilbert space we can define unitary cyclic-swap operator for the factor $\mathcal{H}_A$ via:
\begin{equation}
\label{aswap}
\mathbb{S}_{n}(A) \left| \right. \! \sigma^A_{i_1 i_1'}\!\left. \right> \otimes  \left| \right. \! \sigma^A_{i_2 i_2'}  \!\left. \right> \otimes \ldots \otimes
 \big| \sigma^A_{i_n i_n'} \big> 
 \equiv  \left| \right.\! \sigma^A_{i_2 i_1'} \! \left. \right> \otimes  \left| \right. \! \sigma^A_{i_3 i_2'} \! \left. \right> \otimes \ldots \otimes
 \big| \sigma^A_{i_1 i_n'} \big> 
\end{equation}
and which acts trivially on ${\rm End}(\mathcal{H}_B)$. We are working in the basis \eqref{sigbasis}. 
We also define a similar swap operator $\mathbb{S}_{n}(A^\star)$, which permutes the primed indices instead of the unprimed ones  but in the same direction as \eqref{aswap}.

These definitions can be used to compute:
\begin{equation}
\label{twistopexp}
Z_{n,m} = \left(  \big< \rho_{AB}^{m/2} \big|\right)^{\otimes n}  \, \mathbb{S}_{n}(A) \mathbb{S}_{n}(A^\star)\left( \big| \rho_{AB}^{m/2} \big>\right)^{\otimes n} \equiv  \left(\big< \rho_{AB}^{m/2} \big| \right)^{\otimes n} \,  \mathbb{S}_{n}(AA^\star) \left( \big| \rho_{AB}^{m/2} \big>\right)^{\otimes n}.
\end{equation}
Notice that $\mathbb{S}_n(A)$ ($\mathbb{S}_n(A^\star)$) only involves an action on the left (right) indices so it must satisfy:
\begin{equation}
\mathbb{S}_n(A) = L_{\Sigma_n(A)} \qquad \mathbb{S}_n(A^\star) = R_{\Sigma_n(A)^\dagger}
\end{equation}
for an operator $\Sigma_n(A)$ acting on $\mathcal{H}_{A}^{\otimes n}$. This allows us to write:
\begin{equation}
\label{znmtrace}
Z_{n,m} = {\rm Tr} (\rho_{AB}^{\otimes n})^{m/2} \Sigma_n(A) (\rho_{AB}^{\otimes n})^{m/2} \Sigma_n(A)^\dagger
\end{equation}
Let us now consider the Tomita-Takesaki modular operators for the replicated theory. 
By noting that:
\begin{equation}
\mathbb{S}_{n}(A)^\dagger \left| 1_{AB} \right>^{\otimes n} 
 = \mathbb{S}_{n}(A^\star) \left| 1_{AB} \right>^{\otimes n} 
\end{equation}
we can show that:
\begin{equation}
\left( \Delta_{AB}^{m/2}\right)^{\otimes n}
\mathbb{S}_{n}(A)^\dagger \left| \rho_{AB}^{m/2}\right>^{\otimes n} = \mathbb{S}_n(A^\star) \left| \rho_{AB}^{m/2}\right>^{\otimes n}.
\end{equation}
which for $m=1$ is simply telling us that the Tomita-Takesaki mirror operator of the twist operator on $A$ is the twist operator on the factor $A^\star$. So we can also write the partition functions as:
\begin{equation}
\label{dhalf}
Z_{n,m} =  \left< \rho_{AB}^{m/2} \right|^{\otimes n} \mathbb{S}_{n}(A)  \left( \Delta_{AB}^{\otimes n}\right)^{m/2}
\mathbb{S}_{n}(A)^\dagger \left| \rho_{AB}^{m/2}\right>^{\otimes n}
\end{equation}
Correlators such as these were studied in \cite{Faulkner:2018faa} as a method for computing correlation functions involving $\Delta^{1/2}$ via an analytic continuation of $m \rightarrow 1$.

\subsection{Description in terms of twist operators}

We next work out the path integral expression for $Z_{n,m}$. We will go through similar manipulations as in the previous subsection, but here our pictures will involve euclidean path integrals. The basic answer can already be inferred from \eqref{znmtrace}, which involves a correlation function of $n$-fold twist operators for the regions $A$ and $A^\star$ on the Euclidean branched manifold $\mathcal{M}_m$. The manifold $\mathcal{M}_m$ can be used to compute the $m^\text{th}$ Renyi entropy via the partition function/path integral on this manifold. It can also be used to compute  correlation functions such as ${\rm Tr} \rho_{AB}^m \mathcal{O}_1 \mathcal{O}_2 \ldots $ by inserting other operators into the path integral. The correlation functions here are then associated to twist operators for an $n$-fold tensor product of the CFT placed on $\mathcal{M}_m$. 
 
To draw pictures in this section we will consider $AB = $ two disjoint intervals in a 2d CFT.  However the same ideas work in a more general setting. We also specialize to the case where the Euclidean path integrals have a time-reflection symmetry about a fixed time slice where the QFT Hilbert space lives. This is a $\mathbb{Z}_2$ symmetry on a single replica  but on $\mathcal{M}_m$ it gets mixed with the replica symmetry $\mathbb{Z}_m$ which is enhanced to the dihedral group $D_m$. 

\begin{figure}[h!]
\centering 
\includegraphics[width=.9\textwidth]{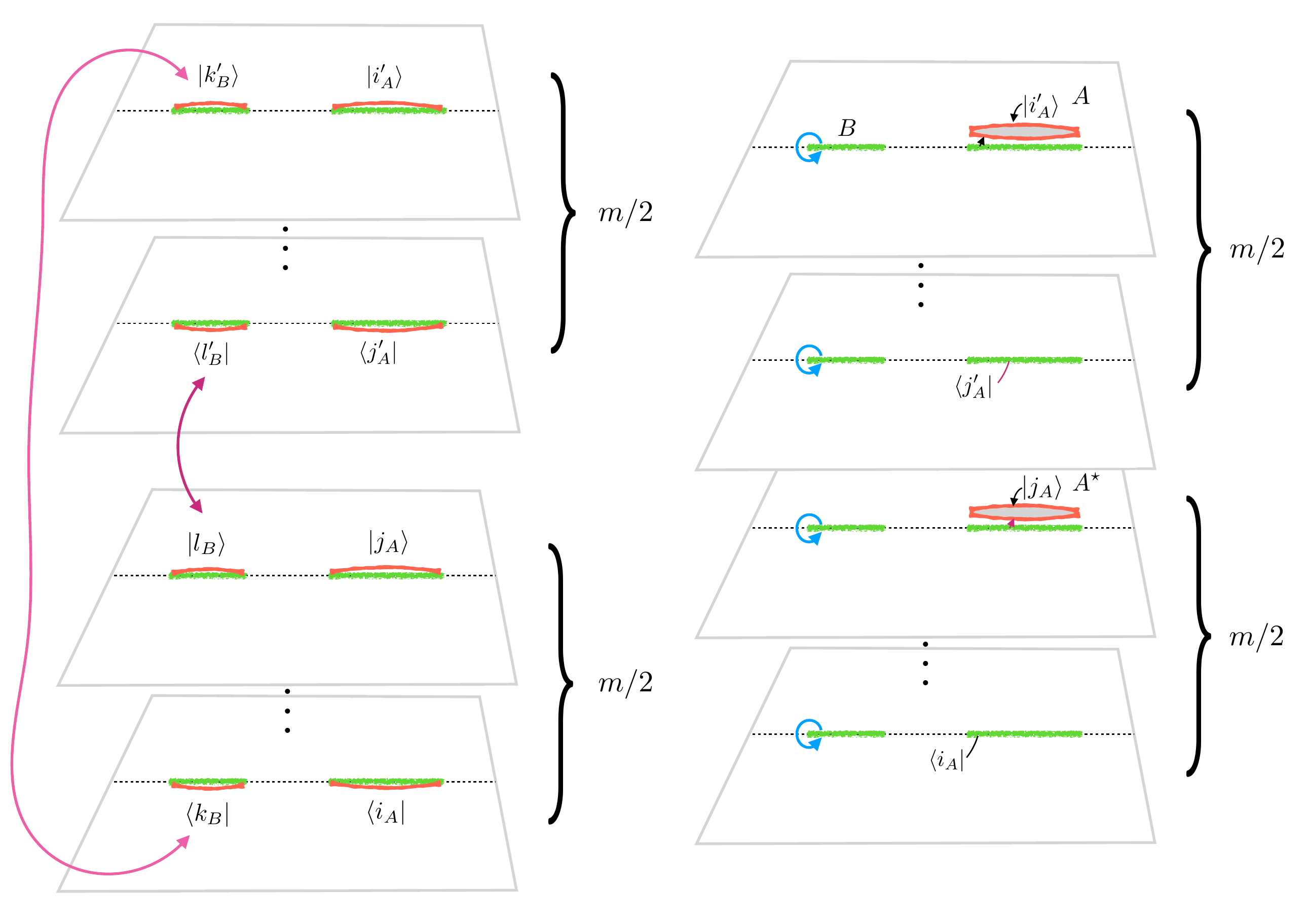}
\caption{ \label{showfig} (\emph{left}) Path integral that defines the reduced density \eqref{rhoaa} matrix of the state \eqref{psim} for the factors $AA^\star$. The arrows indicate gluings and $i_A',j_A',j_A,i_A$ are left free.  Several gluings, involving replicas in the $\ldots$ have not been shown, but these happen on both the $A$ and $B$ cuts in a hopefully obvious way. 
There are $m$ replicas and $m$ must be even. (\emph{right}) A simple deformation of the left picture making clear that the result is a path integral on $\mathcal{M}_m$ with cuts on the zeroth and $m/2$'th replica.
Our conventions are such that moving from the top of $A,B$ at Euclidean time $\tau = \epsilon$ to the bottom of $A,B$ at $\tau=-\epsilon$ decreases the replica index and shifts upwards.}
\end{figure}

We start by giving a path integral representation for the state $\psi_m$ with
 $m$= even. Consider the matrix elements of the un-normalized state:
\begin{equation}
\label{me}
 \left< \sigma^B_{kl} \otimes \sigma_{ij}^A \right| \left. \rho_{AB}^{m/2} \right>
\end{equation}
These are pictured on the lower half of the left part of Figure~\ref{showfig}.  Note that we are using standard methods from \cite{calabrese2004entanglement} to construct these path integrals.  Including the hermitian conjugate of \eqref{me} which is shown at the top half of the left part of Figure~\ref{showfig}, and tracing over $BB^\star$ results in
\begin{equation}
\label{redmat}
\left< \sigma_{i j}^A \right| {\rm Tr}_{BB^\star} \left( \left|\rho_{AB}^{m/2} \right> \left< \rho_{AB}^{m/2} \right| \right) 
\left| \sigma_{i' j'}^A \right> 
\end{equation}
which is depicted via identifications/gluings in the left part of Figure~\ref{showfig}. In the right part of  Figure~\ref{showfig} we have deformed the free $A$ cuts slightly  to make it clear that the external states $i,i',j,j'$ are inserted along two cuts through the replicated manifold $\mathcal{M}_m$.

Thes cuts are inserted on the zeroth sheet and the $(m/2)^\text{th}$ sheet. To compute the $n^\text{th}$ power of the (un-normalized) reduced density matrix in \eqref{redmat} we should apply a further $n$-fold replication of the CFT living on $\mathcal{M}_m$ and insert twist operators along the two cuts $A^{(0)}$ and $A^{(m/2)}$ where the superscript tells us which replica the operator is located on.  The twist operators have opposite orientation on the two different cuts. We can write this as:
\begin{equation}
Z_{n,m} = \left<\Sigma_n(A^{(0)}) \Sigma_n(A^{(m/2)})^\dagger \right>_{\left(\mathcal{M}_m\right)^{\otimes n}}
\end{equation}
This expression should be compared to \eqref{znmtrace}. Up to global considerations, the twist operator $\Sigma_n(A)$ can be considered as a co-dimension $2$ operator living at $\partial A$ which then in two dimensions becomes two local twist operators $\Sigma_n(A) = \sigma_{n}(z_L) \sigma_{-n}(z_R)$ where $\partial A = z_L \cup z_R$. Thus the end result, in 2d, is a twist operator four point function on $\mathcal{M}_m$ for the product theory $CFT^{\otimes n}$.

Consider an alternative description where we take the CFT to have  $n \times m$ replicas which each now live on flat Euclidean space. Consider the symmetric permutation group $S_{nm}$ that acts on these replicas. We can label each replica with $(\nu,\mu) \in (\mathbb{Z}_n,\mathbb{Z}_m)$ on which we can define the action of some special group elements of $S_{nm}$:
\begin{align}
\tau_m^{(k)} (\nu,\mu) = (\nu,\mu+\delta_{\nu,k})  \qquad \tau_n^{(q)}(\nu,\mu) =  (\nu+\delta_{\mu,q},\mu)
\end{align}
The first one is an $m$-cyclic permutation on a fixed $n$-replica and the second switches is the same with $n$ and $m$ swtiched. 
The full $m$-cyclic permutation on the $n$ replicas is
simply $g_m = \prod_{k=0}^{n-1} \tau_m^{(k)}$ and similarly for $g_n$. These obey the group relations:
\begin{equation}
g_m \tau_n^{(k)} = \tau_n^{(k+1)} g_m
\end{equation}

\begin{figure}[h!]
\centering 
\includegraphics[width=.9\textwidth]{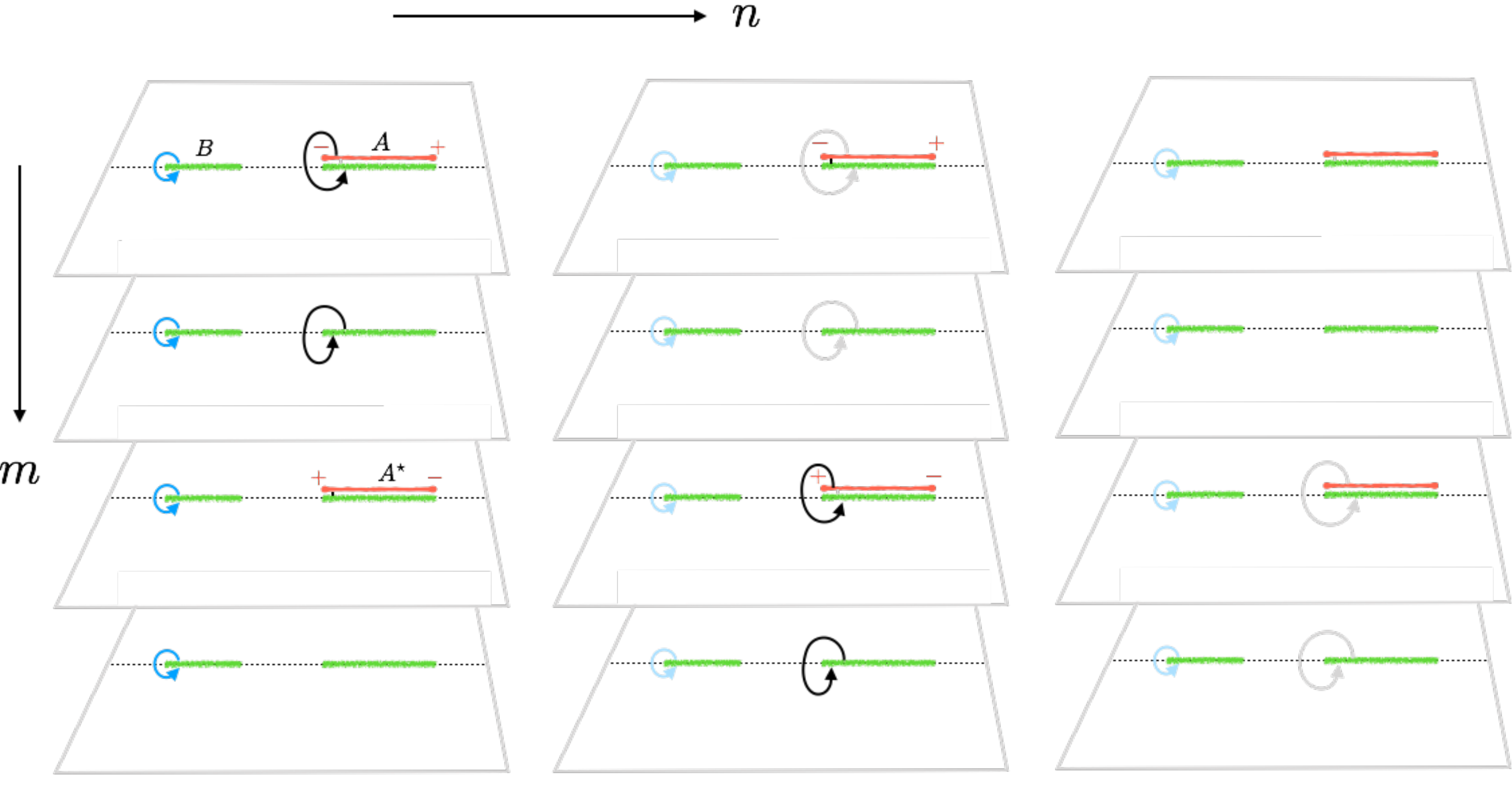}
\caption{ \label{fig:allreplica} Including the $n$-fold replication of the space $\mathcal{M}_m$ we arrive at the above picture (shown here with $m=4$ and $n=3$.)  Some simple curves on this space are depicted. The red cuts move you in the horizontal direction and the green in the vertical direction. }
\end{figure}

Now fix a replica $(\nu,\mu)$ and imagine moving along a curve inside this replica, passing from the bottom of the cut $A^-$ to above the cut $A^+$ staying close to the boundary $\partial A$. 
In doing this we will pass to another replica $g_A(\nu,\mu)$ which is determined by the path integral that computes $Z_{n,m}$ as shown in Figure~\ref{fig:allreplica}.
Here $g_A \in S_{mn}$. We can define a similar group element $g_B$ moving from $B^-$ to $B^+$.  From the path integral picture we can read off the respective group elements: \footnote{We can also describe these using cycles for the symmetric group. This allows us to compare to the work of \cite{dubail2017entanglement} which coincides with our discussion for $m=2$. Denoting a single replica that we previously labelled $(\nu,\mu)\in(\mathbb{Z}_n,\mathbb{Z}_m)$ by a number $1+\nu + n \mu$ which runs from $1$ to $nm$ we have:
\begin{equation}
g_B = \prod_{k=1}^{n} (k,k+n,\ldots, k+n(m-1))
\end{equation}
 that is a product of $m$-cycles on each $n$ replica. Also:
 \begin{align}
 g_A &=\prod_{k=1}^{n} (k ,k+n,\ldots,k+n(m/2-1),k+1+nm/2, \ldots, k+1+n(m-1)) \\ 
 g_B g_A^{-1} &= (1,2, \ldots n) (n (m/2+1),n(m/2+1)-1, \ldots, nm/2+1)
 \end{align}
 where the numbers above are all defined mod $nm$. 
} 
\begin{align}
\partial A \, &: \quad g_A = (\tau_n^{(0)})^{-1}  \tau_{n}^{(m/2)}   g_m  \\
\partial B\, &: \quad g_B = g_m
\end{align}
These group elements define quasi co-dimension 2 twist operators at $\partial A$ (or $\partial B$) that we will denote by $\Sigma_{g_A}(A)$ (or $\Sigma_{g_B}(B)$). These operators are not quite co-dimension 2 because they still remember (in a homological sense) the region $A$ that defines them. If we were to orbifold or gauge the tensor product theory by $S_{nm}$ or some appropriate sub-group, we could make these into genuine co-dimension 2 operators. In this case, the twist operators would be labelled by group conjugacy classes since the gauging identifies different twist operators that are conjugate. We will not do this gauging here.  It is certainly possible to work in the orbifold theory but this requires introducing other objects/complications \cite{Balakrishnan:2017bjg}. 

Note that the two group elements are conjugate:
\begin{equation}
g_{A} = \gamma^{-1} g_B \gamma \qquad \gamma = \prod_{k=0}^{m/2-1} \tau_n^{(k)}.
\end{equation}
Using these twist operators we can write:
\begin{equation}
Z_{n,m} = \left< \Sigma_{g_B}(B) \Sigma_{g_A}(A) \right>_{CFT^{\otimes nm}}. 
\end{equation}
When $n=1$, we reproduce:
\begin{equation}
Z_{1,m} = \left< \Sigma_{g_m}(A) \Sigma_{g_m}(B) \right>_{CFT^{\otimes m}} \propto {\rm Tr} \rho_{AB}^m
\end{equation}
up to normalization, such that the reflected entropy is finally computed via:
\begin{equation}
S_n(AA^\star)_{ \psi_m}  
= \frac{1}{1-n} \ln \frac{\left<  \Sigma_{g_B}(B) \Sigma_{g_A}(A) \right>_{CFT^{\otimes nm}} }{\left( \left<   \Sigma_{g_m}(A) \Sigma_{g_m}(B) \right>_{CFT^{\otimes m}}\right)^n}
\end{equation}

Note that until now, we have concentrated on the case where $\partial A \cap \partial B = 0$. If this is not the case then we can give the following prescription which follows from the above discussion via a simple deformation argument:
\begin{equation}
Z_{n,m} = \left< \Sigma_{  g_A g_B^{-1} }(A^+) \Sigma_{g_B} (AB)  \right> 
=  \left<  \Sigma_{ \tau_{n}^{(m/2)} }(A^+) \Sigma_{\tau_n^{(0)}}(A^-)^\dagger \Sigma_{g_m} (AB) \right>  \qquad 
\end{equation}
where we insert the $A$ twist operators slightly above the $AB$ operators in the Euclidean time ordering sense, and the Hermitian conjugation involves a reflection about the fixed time slice. For example: 
\begin{equation}
\Sigma_{\tau_n^{(0)}}(A^-)^\dagger =  \Sigma_{(\tau_n^{(0)})^{-1}}(A^+)
\end{equation}
This also allows us to define the generalized version where we split $A^\star B^\star = C^\star D^\star$ and compute:
\begin{equation}
S_n(AC^\star)_{ \psi_m}  
= \frac{1}{1-n} \ln \frac{\left< \Sigma_{ \tau_{n}^{(m/2)} }(A^+) \Sigma_{\tau_n^{(0)}}(C^-)^\dagger \Sigma_{g_m} (AB) \right>_{CFT^{\otimes nm}} }{\left( \left<   \Sigma_{g_m}(AB) \right>_{CFT^{\otimes m}}\right)^n}
\end{equation}
It should be possible to get a handle on the higher ($d>2$) dimensional version of reflected entropy in certain OPE limits where the regions $A,B$ are far separated. This is the limit where it is possible to compute the mutual information in higher dimensions \cite{Cardy:2013nua,Agon:2015ftl} for any CFT.  We will now switch to studying $2d$ CFTs, where we can get a fairly good handle on reflected entropy for general CFTs . 

Let us return to the case where $\partial A \cap \partial B = 0$. In 2d CFTs, the twist operators above become two (quasi) local twist operators $\sigma_g$ at the end of each interval.
We take the intervals $A=[a_1,a_2]$ and $B=[b_1,b_2]$ with $a_1 < a_2 < b_1 < b_2 $
\begin{equation}
\label{4pt}
S_n(AA^\star)_{\psi_m} = \frac{1}{1-n} \ln \frac{ \left<  \sigma_{g_A}(a_1) \sigma_{g_A^{-1}}(a_2)  \sigma_{g_B}(b_1) \sigma_{g_B^{-1}}(b_2) \right>_{CFT^{\otimes mn}}}{\left( \left<  \sigma_{g_m}(a_1) \sigma_{g_m^{-1}}(a_2)  \sigma_{g_m}(b_1) \sigma_{g_m^{-1}}(b_2) \right>_{CFT^{\otimes m}}\right)^n }
\end{equation}
There are some subtleties
because we have chosen to not gauge the permutation group. For example, this is not really a correlation function of four local operators that, would otherwise have been single valued upon moving the operators around each other in Euclidean. There are some topological string operators that should be attached between the boundaries of the $A$ and $B$ interval (that we have suppressed above).  The operator product expansion must similarly be treated with care.
In particular, the fusion rule for the OPE is:
\begin{equation}
\label{fusion}
\sigma_{g_A^{-1}} \sigma_{g_B} \rightarrow \sigma_{ g_{B} g_{A}^{-1} } + \ldots
\end{equation}
and does not include the unit operator despite the fact that the two group elements $g_A$ and $g_B$ are conjugate. The twist operator on the right hand side of \eqref{fusion} can be worked out by examining Figure~\ref{fig:allreplica}, as the boundaries of the intervals $\partial A$ and $\partial B$ come close to each other, by following a path that circles both the operators in a clockwise direction.

We can calculate the holomorphic weights/dimensions of these twist operators:
\begin{equation}
\label{dims}
h_{g_B} = h_{g_A^{-1}} =  \frac{c n(m^2-1)}{24 m} \qquad h_{g_{B} g_A^{-1}} =  \frac{2 c (n^2-1)}{24 n}.
\end{equation}
The first two operators have the same dimension because the corresponding group elements are conjugate to each other. Since $g_B$ involves an $m$-twist operator on each of the $n$ replicas the dimension is $n \times$ the standard twist operator dimension $h_m = c m(1-m^{-2})/24$.
The other dimension in \eqref{dims} follows by noting that $ g_{B} g_{A}^{-1}=   (\tau_{n}^{(m/2)} )^{-1} \tau_n^{(0)} $ corresponds to two $n$-twist operators on two of the $m$ sheets. 

Despite not having genuine local operators in \eqref{4pt}, we can still compute the correlation function using conformal blocks, since the OPE is still well defined, with the caveat about the fusion rules discussed above. Indeed, the conformal blocks themselves do not correspond to single valued correlation functions.

Let us define the cross ratio:
\begin{equation}
x = \frac{(a_2 - b_1)(b_2 - a_1)}{ (b_1 - a_1)(a_2 - b_2)}, 
\end{equation}
and consider first, the limit where $x \rightarrow 0$. For any CFT,  the leading term in the OPE will come from \eqref{fusion} and we find:
\begin{equation}
\label{opeexp}
S_n(AA^\star)_{\psi_m}
= - \frac{  c (n+1)}{6 n} \ln x +  \frac{2 \ln C_{n,m}}{1-n} + \ldots
\end{equation}
where $C_{n,m}$ is the OPE coefficient for the operator fusion in \eqref{fusion} (where we normalize the twist operators in the standard way.) In fact, it is possible to use the methods of \cite{Lunin:2000yv} to compute this OPE coefficient by unwrapping the twist operators in the three point function so that it reduces to a trivial CFT partition function on a zero-genus covering space.  We will give a shortcut method to get to the answer in Appendix~\ref{app:ope}. Both methods give:
\begin{equation}
\label{opec}
C_{n,m} = \left( 2m \right)^{- 4 h_n} \qquad h_n = \frac{c}{24} \left(n-\frac{1}{n} \right),
\end{equation}
such that:
\begin{equation}
S_n(AA^\star)_{\psi_m}
= - \frac{  c (n+1)}{3 n} \ln \frac{\sqrt{x}}{2m}  +  \ldots.
\end{equation}

In a special class of ``holographic'' like CFTs, we can actually do much better. Such theories arise from a family of CFTs in which we can take the large $c$ limit and that also have a sufficiently sparse spectrum of low dimension (compared to $\# c$) operators. In this case, correlators such as those in \eqref{4pt} are expected to be determined by a single Virasoro block for some finite window of $x$ at leading order in $1/c$ \cite{Hartman:2013mia}. In this case, we should consider the block corresponding to exchange of the twist operator $\sigma_{g_B g_A^{-1}}$.  For $m \approx 1$ and $n \approx 1$ all the operator dimensions are small and we can apply the known analytic form for the  Virasoro block:
\begin{equation}
\ln F = - 4 h \log(x) + 2 h_p \log\left( \frac{1+ \sqrt{1-x}}{2\sqrt{x}} \right)
\end{equation}
where $h$ is the weight of the external operators (all equal) and $h_p$ is the weight of the exchanged operator. 
Here for the numerator in \eqref{4pt} we should take $h = \frac{ cn(m^2-1)}{24 m} \,,
h_p = \frac{ 2c (n^2-1)}{24 n}$ and for the denominator we have $h = \frac{ c(m^2-1)}{24 m}, \, h_p =0$ since the vacuum exchange is allowed in this case. Adding these two contributions, we find the terms proportional to the external weights ($h$) cancel leaving just the internal contribution: 
\begin{equation}
\lim_{m,n \rightarrow 1}S_n(AA^\star)_{\psi_m} \approx \frac{2c}{3} \log\left(\frac{1+ \sqrt{1-x}}{\sqrt{x}} \right) + \mathcal{O}(c^0)
\end{equation}
where we have used the OPE coefficient \eqref{opec}. Comparing to the bulk calculation of the cross section using $AdS_3$, which is given in Eq.~20 of \cite{Caputa:2018xuf}, and using $E_W = S(AA^\star)/2$, we find exact agreement.\footnote{The cross ratio they define is related to ours via $x=1/(z+1)$.} 

It should be possible to compute $S_n(AA^\star)_{\psi_m}$ at finite $m$ and $n$ for such holographic theories. This can be done in two related ways. The conformal blocks above are in principle calculable at finite weight using the ``monodromy method'' \cite{Hartman:2013mia}.  Secondly one can simply try to compute the action of the dominant solution of 3d gravity whose boundary is the non-trivial surface implied by the twist operators in \eqref{4pt}. Assuming a simple handlebody solution should make this problem tractable \cite{Faulkner:2013yia}. It would be interesting to compute the Renyi entropy of the reflected reduced density matrix in this way in order to get a handle on a consistently regulated version of the  entanglement spectrum \cite{calabresees}. We leave further exploration of this to the future. 

\section{Establishing the duality}
\label{sec:duality}

We will now give several methods for proving the duality between the area of reflected minimal surfaces $2 E_W$ and the reflected entropy $S_R$. 

\subsection{Entangling surfaces as mirrors}

The first idea is to consider the expression \eqref{dhalf}, with $m=1$, in the light of the work in \cite{Faulkner:2018faa}. This paper gave a prescription, that we can apply here, for computing correlation functions such as $\left< \mathcal{O} \Delta^{1/2}_{AB} \mathcal{O}^\dagger \right> $ where $\mathcal{O}$ is a \emph{heavy probe operators}. Note that \eqref{dhalf} is effectively of this form in the limit $n \rightarrow 1$, since the entangling surfaces are well known to become such heavy probe operators under this limit. Indeed, while \cite{Faulkner:2018faa} worked exclusively with local operators, a related discussion \cite{Chen:2018rgz} worked with entangling surfaces. In this case $\mathcal{O}$ is replaced by a higher dimensional defect operator such as the twist operators that compute entanglement.  So indeed we can simply apply the prescription of  \cite{Faulkner:2018faa} to the case at hand. 

These papers more generally gave a prescription for computing $\left< \mathcal{O}(x) \Delta^{-is/2\pi}_{AB} \mathcal{O}^\dagger(y) \right>$ for a subset of parameter values: $(x,y,s)$. In particular, the dominant bulk saddle that computes this correlation function, needs to correspond to a geodesic that passes from $x$ through to $m(AB)$, where it is deflected via a local boost of  rapidity $s$, before passing back to $y$. For a random set of parameters $(x,y,s)$ this boost condition will not be satisfied. However one can argue that there is a co-dimension $1$ surface in parameter space where this is satisfied and where the modular flow correlator is computable. Importantly for us it was observed in \cite{Faulkner:2018faa} that one way to always satisfy this condition is to set $x=y$ and $s = i\pi$ corresponding to a local Euclidean rotation where the geodesic reflects back on itself. 
Several methods for proving this prescription were put forward and we will rehash some of these in the next subsection for the case in point.\footnote{There is a nice method put forward in \cite{Chen:2018rgz} that works in boundary $d=2$ dimensions and that we will not end up using. This involves the modular zero modes defined in \cite{Faulkner:2017vdd} as an integral of the flowed operator over all modular time $s$.
In the heavy probe limit this $s$-integral, inside a correlation function, has a saddle point when the local boost condition is satisfied.  Equating this to the a saddle point evaluation of the bulk description of this zero mode operator (a smearing of the bulk operator over $m(AB)$)  \cite{Faulkner:2017vdd} one arrives at the prescription. This proof was worked out for real modular flows, however it likely works for imaginary flows also where the saddle point of the $s$-integral would now be at complex $s$. We leave further investigation of this method to future work. }

The prescription above clearly leads to the reflected entangling surfaces discussed in Section~\ref{sec:refmin} where the $\Delta^{1/2}_{AB}$ acts as a $\pi$ rotation reflecting $m(AA^\star)$ exactly when it intersects the $AB$ entangling surface. 
Note that the minimal surface that computes the $AA^\star$ entanglement is constrained to remain entirely inside the $AB$ entanglement wedge due to this reflection condition.  The relationship to \cite{Faulkner:2018faa} is most clearly seen for the situation depicted in the left figure of Fig~\ref{fig:sr2} where $\partial A \cap \partial B \neq 0$, since the $AA^\star$ entangling surface is boundary anchored and reflects off the $AB$ entangling surface, folding back to the boundary just like in the examples given in \cite{Faulkner:2018faa}. The homology condition, and thus the need to include the non-boundary anchored reflected minimal surface that wraps around the entanglement wedge, is less clear but plausible. Since this condition will naturally arise in our other approaches to this duality, we will not give further evidence for it here.

\subsection{Replica path integral and Engelhardt-Wall}

We now turn to a replica path integral proof, which turns out to be an interesting extension of the Lewkowycz-Maldacena method \cite{Lewkowycz:2013nqa}. This method follows Section~4.1 of \cite{Faulkner:2018faa} which was used to justify the mirror prescription discussed in the subsection above. 
We will also connect this method to a spacetime gluing procedure discussed by Engelhardt-Wall \cite{Engelhardt:2017aux,Engelhardt:2018kcs}, and in some sense this will give a path integral justification for one minor aspect of their work (specifically Footnote~15 of that paper.) The construction here works by firstly finding a bulk dual of the purification $\left| \rho_{AB}^{1/2} \right>$, and then using this spacetime to find entanglement entropies with the usual RT formula.\footnote{We thank Matt Headrick and Aron Wall for comments that led us to think about our construction in this way.}

\begin{figure}[h!]
\centering 
\includegraphics[width=.32\textwidth]{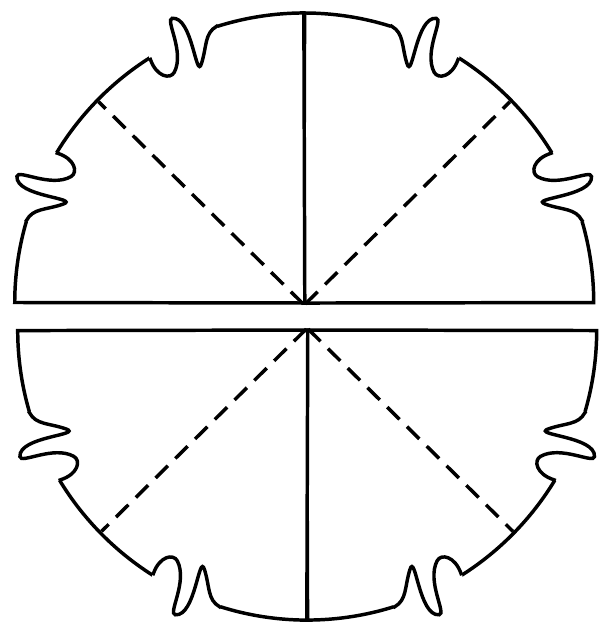} \hspace{1.5cm}
\includegraphics[width=.44\textwidth]{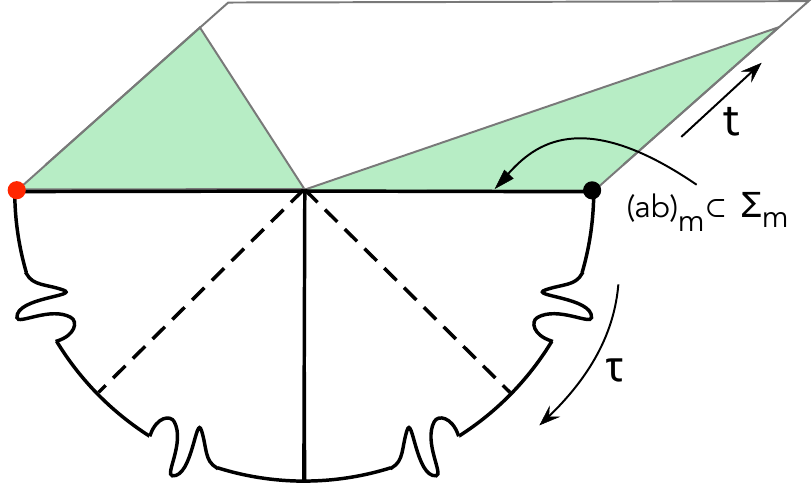}
\caption{ \emph{Left:} The LM bulk replica picture for $m$ even. Each wedge is a bulk region that is associated with a copy of the replicated boundary region $AB$. The solid and dashed lines end at the boundary on a fixed time slice that is a fixed point of the time reflection symmetry on each replica. There is an  enhanced dihedral symmetry and we have cut the bulk into two along the moment of time reflection symmetry governed by $\tau$ defined in the text. \emph{Right:} Continuing into real time based on this time reflection symmetry in the standard way. 
\label{LMconst}}
\end{figure}

We start by reviewing the construction of the bulk spacetime $\mathcal{B}_m$ that is used to compute the standard Renyi entropies in AdS/CFT \cite{Faulkner:2013yia,Lewkowycz:2013nqa}. This Euclidean spacetime solves Einstein's equations with a negative cosmological constant with a boundary condition that it must approach $\mathcal{M}_m$ near the conformal boundary. The regularized on-shell gravitation action computes the $m^\text{th}$ Renyi entropy of $\rho_{AB}$. We will assume that the least action configuration maintains the $\mathbb{Z}_m$ replica symmetry of the boundary manifold $\mathcal{M}_m $.
The bulk spacetime is visualized in the left part of Figure~\ref{LMconst} where only a two dimensional slice is shown. The boundary of this slice is  then a one dimensional slice through $\mathcal{M}_m$. The angular direction on the boundary passes around the $AB$ entangling surface $m$ times. The central point is located in the bulk at the fixed point of the $\mathbb{Z}_m$ replica symmetry of $\mathcal{B}_m$. This point becomes the RT surface as one sends $m \rightarrow 1$. The appropriate analytic continuation involves constructing the quotient $\mathcal{B}_m/\mathbb{Z}_m$ which can then be continued away from integer $m$ by maintaining a conical opening angle of $2\pi/m$ at the center, solving Einstein's equations away form this point and imposing that the spacetime approaches the boundary $\mathcal{M}_1 = \mathcal{M}_m/\mathbb{Z}_m$.

In Figure~\ref{LMconst} we have also shown an enhanced replica symmetry which is the dihedral group $D_m$ and arises from the assumption of a time reflection symmetry on a single copy of the boundary theory. This group is generated by a clockwise rotation $g$ between the replicas and a reflection $\tau$ through one of the axes, which we take to be the horizontal axis of Figure~\ref{LMconst}. Note that $g^m =1$,  $\tau^2 = 1$ and $\tau g \tau = g^{-1}$. 
The dashed radial curves in the figure end at the boundary region $(AB)^c$ (the complement of $AB$) and the solid radial curves end at $AB$. 

We consider $\tau$ as a $\mathbb{Z}_2$ moment of time reflection symmetry that we can then use to analytically continue the gravitational solution $\mathcal{B}_m$ into real times. Define the fixed point surface of $\tau$ to be $\Sigma_m$. This surface is a Cauchy surface for the analytically continued spacetime. 
We can use the Hamiltonian formulation of classical GR to then evolve the data on $\Sigma_m$ forward into real times. This produces a picture such as in the right of Figure~\ref{LMconst}.  We need only specify the induced metric $\gamma_{ij}^{(m)}$  and the lapse function $N^{(m)}$ on $\Sigma_m$ since the extrinsic curvature and the shift vector vanish there by the $\tau$-symmetry. Additionally, in the quantum theory, the bulk Euclidean path integral in the lower plane of Figure~\ref{LMconst} up to $\Sigma_m$ will produce a state that is then analogous to the path integral construction of the thermofield double/Hartle-Hawking state in the Euclidean black hole geometry \cite{Maldacena:2001kr}.
In more detail, we can see that this will produce the bulk state:
\begin{equation}
\label{bulkm}
\left| \psi_m^{\rm bulk} \right> = \frac{1}{\sqrt{Z_m}} \left| \left(\rho_{(ab)_m}^{(m)}\right)^{\frac{m}{2}} \right>\,, \qquad
Z_m = {\rm Tr} \left(\rho_{(ab)_m}^{(m)}\right)^{m}
 \end{equation}
 where $\rho_{(ab)_m}^{(m)}$ is the reduced density matrix produced via the bulk Euclidean path integral on a single replica section of $\mathcal{B}_m$. We pick a replica that lives between two solid radial lines of Figure~\ref{LMconst}. These solid curves represent the ``entanglement wedge'' region $r_{m}(AB) \equiv (ab)_m$; i.e. the homology region between the $\mathbb{Z}_m$ fixed point and the $AB$ region on the boundary of a fixed replica. The bulk density matrix is defined on the $(ab)_m$ Hilbert space which can then be doubled in the usual way. The pure state \eqref{bulkm} is defined in this doubled Hilbert space.

It should be now clear that the lower half of bulk Euclidean solution $\mathcal{B}_m$ and its continuation into real time, along with the associated bulk state on $\Sigma_m$ implied by the path integral, should be interpreted as the holographic dual of $\left| \psi_m \right> \propto \left| \rho_{AB}^{m/2} \right>$ up to normalization. 
We can thus use this bulk solution to compute entanglement entropies of this new state in the usual way. In particular note that $\Sigma_m = rr_m^{\star}(AB) = (ab)_m(a b)^\star_m$  - the doubled entanglement wedge region. The boundary of $\Sigma_m$ is then simply $AA^\star BB^\star$. We will split $(ab)_m = a_m \cup b_m \subset \Sigma_m$ such that $\partial a_m = A \cup \left( \Gamma_A^{\rm min} \right)_m$ and $\partial b_m = B \cup \left( \Gamma_A^{\rm min} \right)_m$. Here $\left( \Gamma_A^{\rm min} \right)_m$ is the minimal cross section of the $m$-dependent entanglement wedge region. This is the $m$ generalization of the cross section defined in \cite{Takayanagi:2017knl}. 

Due to the time reflection symmetry $\tau$ we can then simply apply the RT formula to compute the entanglement entropy as minimal surfaces on $\Sigma_m$. These surfaces are clearly a generalization of the reflected minimal surfaces of Section~\ref{sec:refmin}, that now depend on the even integer $m$. For example: 
\begin{equation}
m_m(AA^\star) = 
\left( \Gamma_A^{\rm min} \right)_m \cup \left( \Gamma_A^{\rm min} \right)_m^\star
\end{equation} 
is the minimal surface on $\Sigma_m$ that ends on $AA^\star$.
The area of these surfaces will then compute, for example:
\begin{equation}
\label{mverclass}
\lim_{n \rightarrow 1} S_n(AA^\star)_{\psi_m} = \frac{{\rm Area}[ m_m(AA^\star) ]}{4 G_N} + \mathcal{O}(G_N^0)
\end{equation}
Note that in the case where $\partial A \cup \partial B = 0$ (the main case of interest in \cite{Takayanagi:2017knl}) the two reflected regions $A$ and $A^\star$ share a common boundary. In this case the surface $m_m(AA^\star)$ is not boundary anchored and wraps some non-trivial horizon homologous to $AA^\star$. 
However the surface $AA^\star$ passes right through the conical singularity of $\mathcal{M}_m$, and one might worry that one cannot directly apply the homology condition in this case. Indeed there seems to be significant possibility for UV issues here. One can approach this situation by defining a slightly cutoff $A^\epsilon \subset A$ while maintaining $A^\epsilon \tilde{B} = AB$ (such that $B \subset \tilde{B}$) in which case the bulk entangling surfaces look like the left panel of Figure~\ref{fig:sr2}. The new reflected minimal surface never sees the boundary conical singularity which gets smoothed moving into the bulk \cite{Faulkner:2013yia}. However now since the $A^\epsilon A^{\epsilon\star}$ is boundary anchored, the entropy itself is divergent, and it is hard to take the limit $\epsilon \rightarrow 0$ for this reason. We can easily fix this problem by applying a kind of mutual information regulator for entanglement \cite{Casini:2015woa} to the case at hand:
\begin{equation}
 S(AA^\star)_{\psi_m} = \lim_{\epsilon \rightarrow 0} \frac{1}{2} I\left(A^\epsilon A^{\epsilon\star}: B^\epsilon B^{\epsilon\star}\right)_{\psi_m}
\end{equation}
where $B^\epsilon \subset B$ and where we compute the right hand side using \eqref{mverclass} for different subregions of  $AB A^\star B^\star$. The limit is smooth and gives the expected prediction for \eqref{mverclass} in the non boundary anchored case. 

We can also include the first quantum correction to \eqref{mverclass}:
\begin{equation}
\label{mver}
\lim_{n \rightarrow 1} S_n(AA^\star)_{\psi_m} =   \frac{  \left< \widehat{\mathcal{A}}[m_m(AA^\star)] \right>_{\psi_m^{\rm bulk}}}{4 G_N} + 
 S^{\rm bulk}(a_ma_m^{\star})_{\psi_m^{\rm bulk}}+ \mathcal{O}(G_N)
\end{equation}
This is the $m$ version of \eqref{srquantum}. Of course actually computing this quantity, even the classical piece, is very difficult since we don't in general have an explicit construction of $\mathcal{M}_m$. 

Note that there is another important $\mathbb{Z}_2$ symmetry generated by
\begin{equation}
j =  \tau g^{m/2}
\label{littlej}
\end{equation}
and this acts on $\Sigma_m$ to exchange the regions $a_m,b_m$ with their mirrors $a_m^\star,b_m^\star$.  This becomes an anti-unitary CPT symmetry $J$ acting on the Hilbert space. For example, if we consider an RT surface homologous to $AA^\star$, then it will maintain this $J$ reflection symmetry in such a way that the minimal cross section of one wedge is exchanged with the cross section of the mirror. Thus it will always meet the $\mathbb{Z}_m$ fixed point (or $AB$ ``RT surface'') perpendicularly. 

\begin{figure}[h!]
\centering 
\includegraphics[width=.52\textwidth]{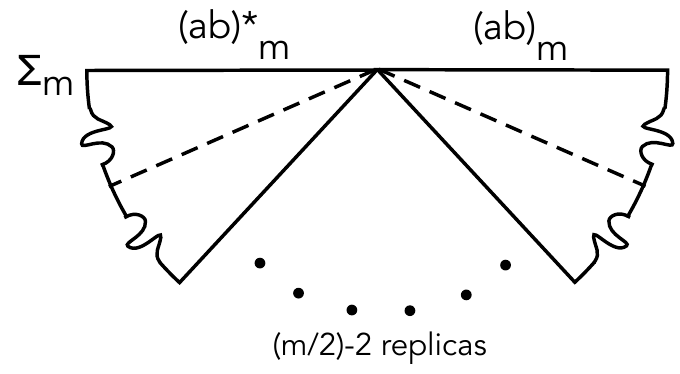}
\caption{ \label{LMconst2}  The ``$m/2$'' replicas placed below the entanglement wedges $(ab)_m$ and $(ab)^\star_m$ }
\end{figure}

The discussion is so far only for positive even $m$. We now endeavor to give an analytic continuation in $m$ along the lines of \cite{Lewkowycz:2013nqa,Faulkner:2018faa}. The quotient $\hat{B}_m = \mathcal{M}_m/\mathbb{Z}_m$ can be continued away from even $m$.\footnote{It is interesting to consider the analytic continuation from odd $m$. These solutions also have a $\mathbb{Z}_2$ symmetry that we can treat as a moment of time reflection. However the Cauchy surface associated to this has a boundary equal to the union of $AB$ and its complement $(AB)^c$. Thus the continuation of these saddles to $m=1$ just gives back the starting point - the original state of the CFT on $AB \cup (AB)^c$.} The region $(ab)_m$ is again a co-dimension one ``homology region'' of $\hat{B}_m$ that lives between the conical singularity and the boundary region $AB$. 
We can now define the reduced density matrix $\rho_{(ab)_m}^{(m)}$ for $m \notin 2\mathbb {Z}^+$ as a bulk path integral with open boundary conditions above and below the surface $(ab)_{(m)}$ on $\hat{B}_m$.  
Similarly, we can extract the intrinsic/induced metric $\gamma_{ij}^{(m)}$ and lapse $N^{(m)}$ on $(ab)_m$ from $\hat{B}_m$. We can then construct the initial data on $\Sigma_m$ by gluing two copies of the homology region,
$\Sigma_m = (ab)_m \cup (ab)^{\star}_m$,
together along the location of the conical singularity and while maintain the $j$ symmetry that exchanges these two copies.  We then have ``$m/2$'' replicas placed below these two spatial regions (see Figure~\ref{LMconst2}). 
We need not have a geometric picture of these ``$m/2$'' replicas since we only need data local to $\Sigma_m$ to specify the real time evolution. This is true at least classically. Furthermore for the quantum case we only need to know the state on $\Sigma_m$, and while we don't have a geometric bulk path integral that computes this state, we can simply use the obvious analytic continuation of \eqref{bulkm} based on $\rho_{(ab)_m}^{(m)}$. 

We thus have all the ingredients to continue \eqref{mver} in $m$ away from even integers. The limit $m \rightarrow 1$ then lands as on \eqref{srquantum} as we had wanted to show, where the bulk state is as expected:
\begin{equation}
\label{m1state}
\lim_{m \rightarrow 1} \left| \left(\rho_{(ab)_m}^{(m)}  \right)^{m/2} \right> =  \left| \sqrt{ \rho^{\rm bulk}_{ab} } \right>
\end{equation}

We know from the discussion in \cite{Dong:2017xht} that the conical surface of $\hat{B}_m$ becomes the quantum extremal surface \cite{Engelhardt:2014gca} as $m\rightarrow 1$. Thus the gluing for $m=1$ will take place here and so the reflected entangling surface $m(AA^\star)$ should be taken to reflect off of this quantum extremal surface surface. For even integer $m$ the quantum state $\psi_m^{\rm bulk}$ is actually well defined, for bulk gauge fields and the graviton, as is - simply by the bulk path integral. However the analytic continuation to non integer $m$ for \eqref{m1state} requires us to confront the meaning of the reduced density matrix of sub-regions for such gauge fields. This issue was already predicted below \eqref{srquantum2}. 

We note that our gluing procedure is very much like a Euclidean version of the CPT gluing that was applied as a final step of the construction in \cite{Engelhardt:2018kcs}. Indeed these are the same construction as $m \rightarrow 1$, where we the CPT symmetry of \cite{Engelhardt:2018kcs} is the $J$ symmetry here.
Thus we can by-pass entirely the discussion of the $m$-replicas above and simply take the real spacetime entanglement wedge:
\begin{equation}
\mathcal{E}_{AB} = \mathcal{D}(r(AB)) =  \mathcal{D}(ab)
\end{equation}
and apply the co-dimension two gluing procedure worked out in \cite{Engelhardt:2018kcs} to glue $\mathcal{E}_{AB}$ to it's CPT conjugate $\mathcal{E}_{AB}^\star$. One can then solve Einstein's equations into the future and past of the HRT surface $m(AB)$ to directly construct the gravitational dual of $\left| \rho_{AB}^{1/2} \right>$. Considering the relation to the GNS construction that we discuss in Appendix~\ref{app:gns} this then demonstrates the claim in Footnore~15 of that paper. Our results also show that the quantum corrections to the CPT gluing in  \cite{Engelhardt:2018kcs} involve picking the bulk state to be the GNS state of the bulk theory $\left| \sqrt{\rho_{ab}^{\rm bulk}} \right>$.

Note that there are several boundary conditions one has to satisfy in order to glue two spacetimes together along a co-dimension $2$ surface \cite{Engelhardt:2018kcs}. The most important is the requirement of vanishing expansions in both null directions. The expansion picks up a minus sign under $CPT$ so the gluing would otherwise result in a discontinuity that makes it hard to solve Einstein's equations. These conditions are all satisfied if one glues an entanglement wedge to its CPT conjugate along an arbitrary extremal surface \cite{Engelhardt:2018kcs}. However including quantum corrections, the quantum extremal surface does not quite have vanishing expansion and so the CPT gluing seems to result in a jump discontinuity in the expansion. This discontinuity would need to be sourced by a matter shock wave.
Indeed in simple situations \cite{Ceyhan:2018zfg,cheap} one can check that the bulk GNS state $\left| \sqrt{\rho_{ab}^{\rm bulk}} \right>$ actually has a delta function shock wave in the stress energy tensor at this surface and this shock will exactly make up for the discontinuity in the expansion. We leave a general discussion of this to future work.

\section{Simple computations of reflected entropy}
\label{sec:comp}

\begin{figure}[h!]
\centering 
\includegraphics[scale=.62]{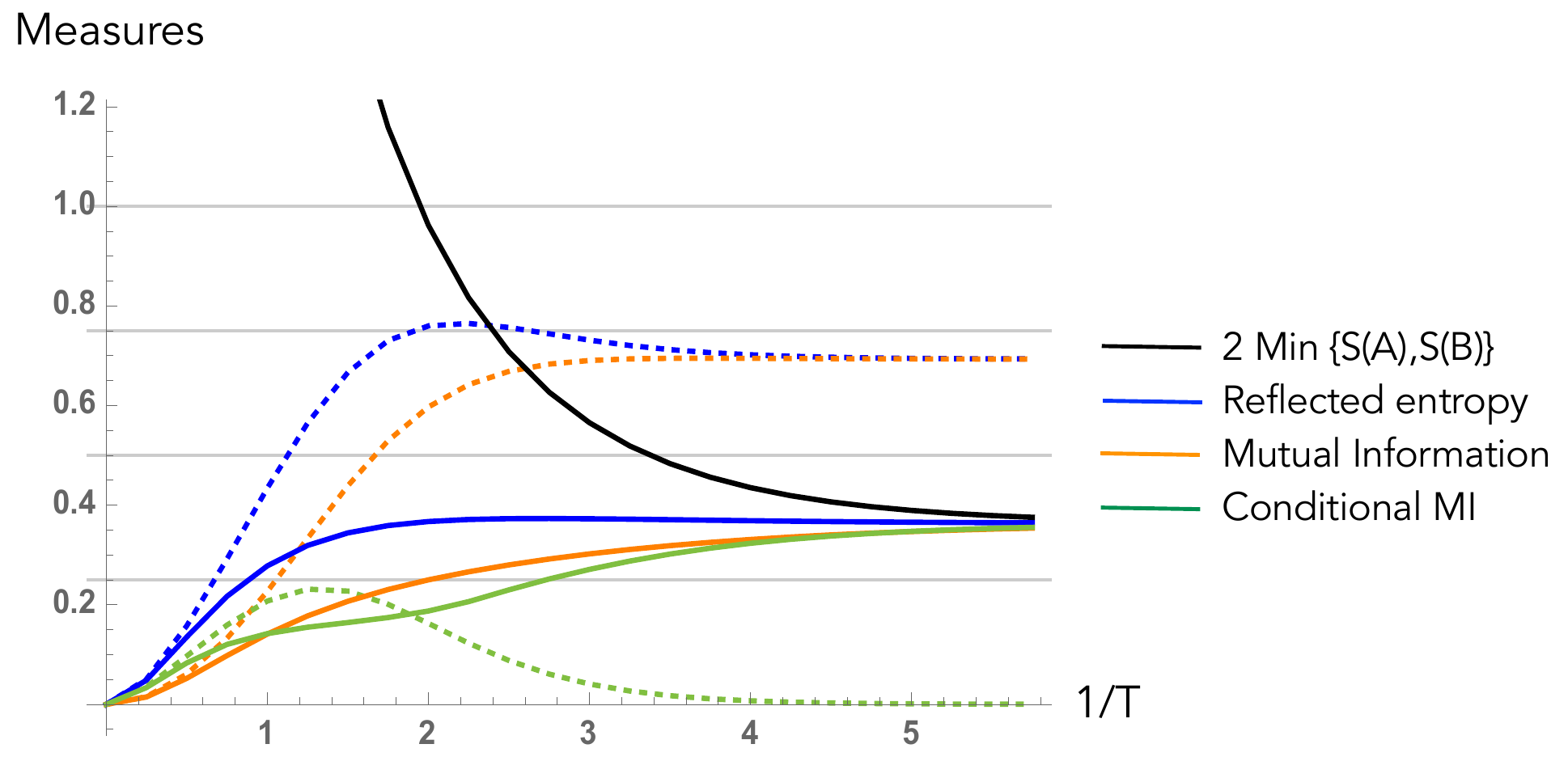}
\caption{ \label{ising-all} For the Ising model with $N=8$, the temperature dependence of the reflected entropy (blue) is shown, in comparison to the mutual information (orange) and the conditional mutual information (green). We consider $g=0.01$ and $g=2$, shown by dots and solid respectively.}
\end{figure}

In this section, we shall verify various properties of the reflected entropy, $S_R(A:B)$ for some sample states associated to eigenstates of local Hamiltonians. Consider the XY spin chain with $N$ spins, described by the Hamiltonian
\begin{equation}\label{XYM}
H_\text{XY} = -\frac{1}{2}\sum_{k=1}^{N}\left[ \left(\frac{1+\gamma}{2}\right)\hat\sigma^x_k \hat\sigma^x_{k+1}+ \left(\frac{1-\gamma}{2}\right)\hat\sigma^y_k \hat\sigma^y_{k+1}\right] -\frac{g}{2} \sum_{k=1}^N \hat\sigma^z_k.
\end{equation}
on which we impose periodic boundary conditions, $\hat\sigma^\mu_{N+1} = \hat\sigma^\mu_1$. The parameter $g$ is an external magnetic field, and $\gamma$ measures anisotropy in the $x$ and $y$ directions $-$ when $\gamma=1$, \eqref{XYM} describes the transverse-field Ising model. In the simplest case, we can pick the mixed state $\rho_{AB}$ to be the thermal Gibbs density matrix, $= \exp(-\beta H_\text{XY})$. Figure~\ref{ising-all} shows for the Ising model, that the reflected entropy, $S_R(A:B)$, obeys the inequality relations \eqref{miss} and \eqref{ineq1} with other entropy quantities. We also computed the conditional mutual information $C_R(A:B)$ which was defined in \eqref{cr}.
In the high temperature limit, the entanglement between $AB$ and $A^\star B^\star$ reduces to a product of Bell pairs between mirrored regions. Therefore, $AA^\star$ is not entangled with $BB^\star$ and $S_R(A:B)$ vanishes in the limit. By a similar argument, both $I(A:B)$ and $I(A:B^\star)$ also vanish in this limit.

For low temperatures, we consider separately the couplings $|g|<1$ and $|g|>1$. For $g\gg1$, there is a unique ground state. As discussed in \eqref{sppure}, $I(A:B) = 2S(A) = S_R(A:B)$ in this limit, which is exhibited by the solid curves in Fig \ref{ising-all}. However for $g\to0$, the ground state incurs an approximate double degeneracy, $|\Omega_\pm\rangle = \prod_{i} |\pm\rangle$, where  $\sigma_x |\pm\rangle = (\pm1)|\pm\rangle$, etc. Then the limit $T\to 0$ projects $\rho_{AB}$ not onto a single ground state, but onto the mixed state in $|\Omega_+\rangle$ and $|\Omega_-\rangle$. Therefore, $I(A:B^\star) \neq 0$ which explains the split between the dotted curves in Figure~\ref{ising-all} (see \eqref{cr}.)

\begin{figure}[h!]
\centering 
\includegraphics[scale=.57]{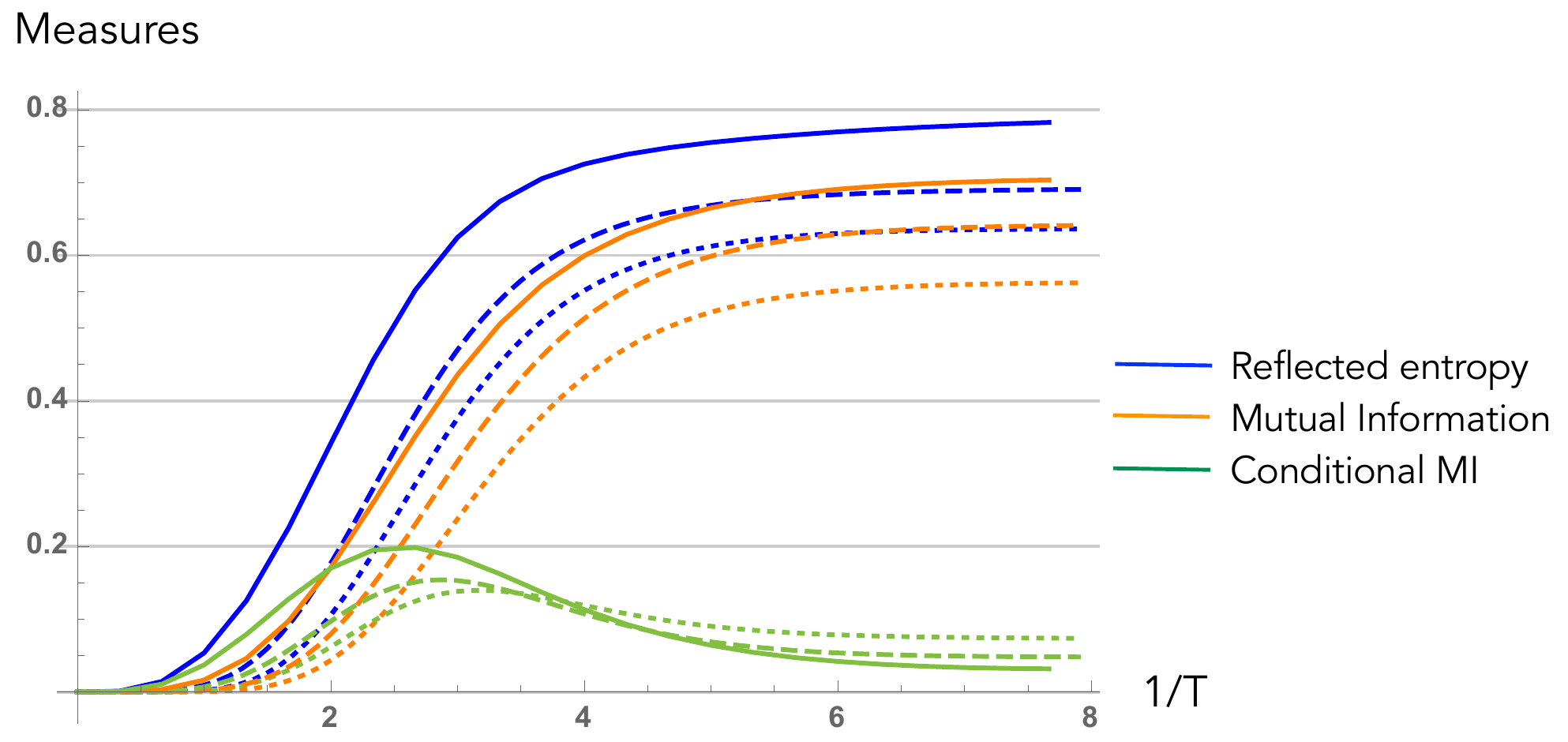}
\caption{ \label{xy-all} For the XY model, with $\{g=0, N=8, \gamma=1/2\}$, the temperature dependence of the reflected entropy (blue) is shown, in comparison with the mutual information (orange) and the conditional mutual information (green). We consider various lengths $\ell=2,4,6$ of the subsystem $C$, shown by solid, dashes and dots respectively.} 
\end{figure}

\begin{figure}[h!]
\centering 
\includegraphics[scale=.57]{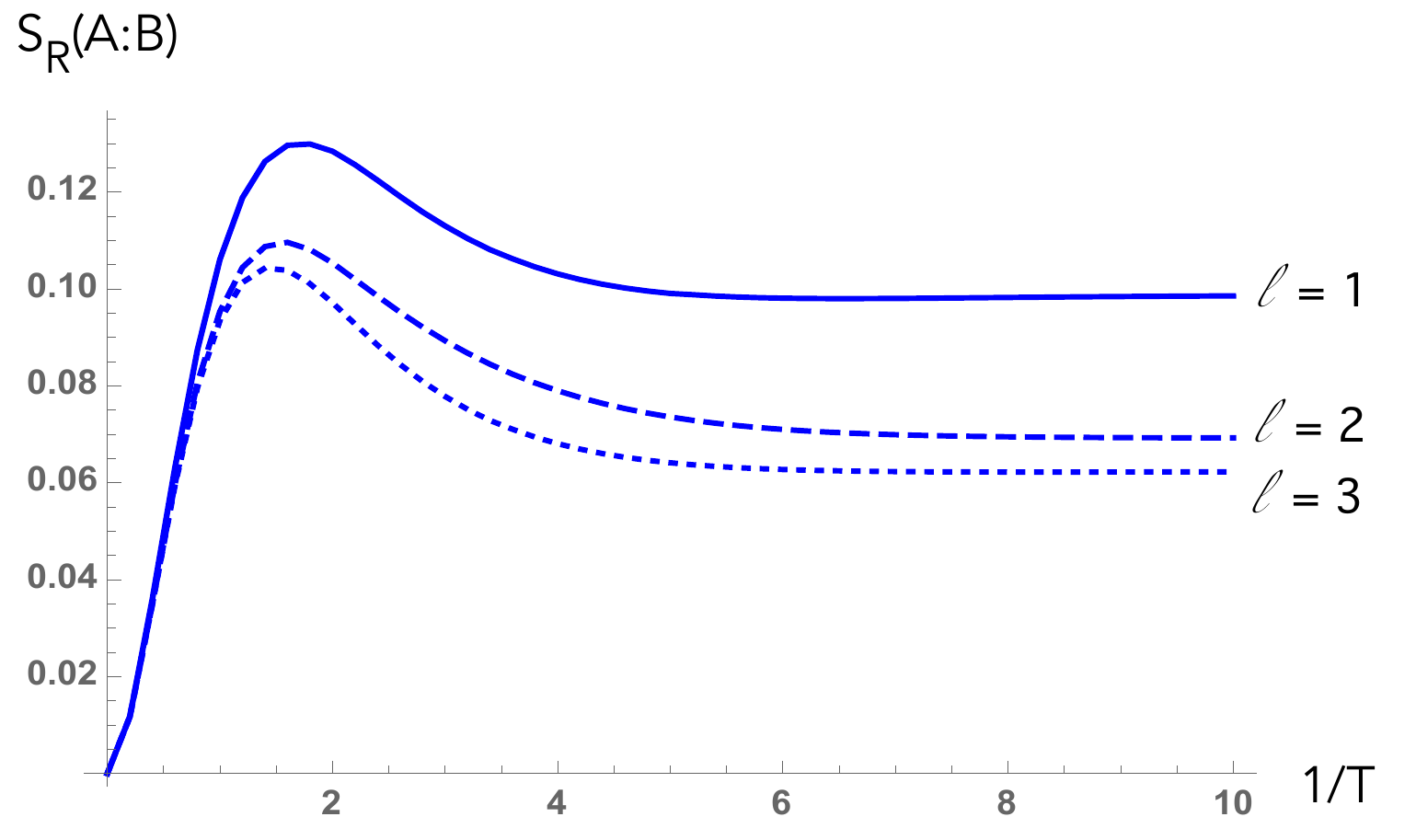}
\caption{ \label{monot-sr} For the XY model, with $\{g=2, N=7, \gamma=1/2\}$, the temperature dependence of the reflected entropy is shown. Under decreasing the size of the $B$ Hilbert space, keeping $\text{dim }\mathcal H_A$ fixed, we show that the reflected entropy reduces monotonically for all $T$.}
\end{figure}

Since the reflected entropy is defined for any bipartite mixed state, we can also pick $\rho_{AB} = \text{Tr}_C \rho_G$ where $\rho_G$ is the Gibbs state at temperature $T$. If the subchain $C$ contains $\ell$ spins, the canonical purification of $\rho_{AB}$ lives in a $2^{2(N-\ell)}$-dimensional Hilbert space. In Figure~\ref{xy-all}, we show the reflected entropy for a chain of length $N=8$ (for $0\leq \ell < N/2$), compared with $I(A:B)$ and $C_R(A:B)$. Among several other features, we notice that $S_R(A:B)$ decreases with increasing $\ell$. This gives an example of the conjectured monotonicity of reflected entropy \eqref{miss2}: $S(A:BB') \geq S(A:B)$, as discussed in Appendix~\ref{app:ineq}. Although we have not managed to prove \eqref{miss2} for the EE our numerics has not turned up a counterexample. Figure~\ref{monot-sr} shows the monotonicity for the XY model, by tracing over $\ell$ spins from a mixed state, while keeping the size of $A$ fixed.

\section{Canonical split inclusions}
\label{sec:split}

Rather than summarize what we have done in the paper, we choose to end with some slightly more speculative ideas for relating the reflected entropy to the split property in the continuum theory. This will then suggest a new(ish) computable regulator for entanglement in QFT.  

\subsection{Relation to the split property}

Take the regions $A$ and $B$ to be spacelike separated by some finite gap - this is the context in which it is natural to \emph{assume} the split property. While there are several equivalent formulations of the split property, we always end up with the following structure \cite{doplicher1984standard}. 
There exists an isomorphism $\phi$ between $\mathcal{A}_A \vee \mathcal{A}_B$ and the spatial tensor product
$\mathcal{A}_A \otimes \mathcal{A}_B$\footnote{For simplicity we will assume that non of these algebras have centers,  although it should be possible to relax this. } which establishes the independence of the two regions $A, B$. This should be compared to a tensor factorization that is sometimes assumed in discussions of entanglement in QFT.
We take a representation of the tensor product algebra that acts on a tensor product of two copies of the QFT Hilbert space $\mathcal{H}_l \otimes \mathcal{H}_r = \mathcal{H} \otimes \mathcal{H}$ that we will refer to as the left and right copies (not to be confused with the LR split used in Section~\ref{sec:refent} which is very different!). This isomorphism is implemented by a unitary:
\begin{equation}
U: \mathcal{H} \rightarrow \mathcal{H}_l  \otimes \mathcal{H}_r
\end{equation}
where $U$ acts as follows on simple operators:
\begin{equation}
U \alpha \beta U^{-1}= \phi(\alpha \beta) \equiv \alpha \otimes \beta \,, \qquad \alpha \in \mathcal{A}_A \,, \beta \in \mathcal{A}_B
\end{equation}
and this extends to an action on the full algebra $U \mathcal{A}_A \vee \mathcal{A}_B U^{-1} = \mathcal{A}_A \otimes \mathcal{A}_B$.  We have the following relations:
\begin{align}
\label{subf}
U \mathcal{A}_A U^{-1} & = \mathcal{A}_A \otimes 1\,, \qquad
U \mathcal{A}_B U^{-1}  = 1\otimes \mathcal{A}_B  \\
U \mathcal{A}_B' U^{-1} & = \mathcal{B}(\mathcal{H}) \otimes \mathcal{A}_B' \,,  \qquad
U \mathcal{N} U^{-1}  = \mathcal{B}(\mathcal{H}) \otimes  1
\label{Nlast}
\end{align}
where $\mathcal{B}(\mathcal{H})$ are all the bounded operators on the QFT Hilbert space which is always a type-I algebra. We can use the last equation in \eqref{Nlast} to define $\mathcal{N}$ given the unitary $U$ or vice-versa.  They satisfy \eqref{split} which we reproduce here:
\begin{equation}
\mathcal{A}_A \subset \mathcal{N} \subset \mathcal{A}_B'
\end{equation}
With this $U$ we can construct factorized states on $\mathcal{H}$ via:
\begin{equation}
\left| S \right> = U^{-1} \left| \psi \right> \otimes \left| \psi \right> \quad \rightarrow  \quad
\left< S \right| \alpha \beta \left| S \right> = \left< \psi \right| \alpha \left| \psi \right> \left< \psi \right| \beta \left| \psi \right>
\end{equation}
which is clearly something we cannot do without the finite spacelike gap between the regions due to the divergent entanglement entropy \cite{Witten:2018lha}. The existence of $S$ demonstrates the independence of the two regions $A,B$ in the QFT and also allows us to construct the mutual information of $\psi$ via the relative entropy:
\begin{equation}
\label{midef}
I(A:B) \equiv S_{\rm rel}\left( \psi | S ; \mathcal{A}_A \vee \mathcal{A}_B \right)
= S_{\rm rel}\left( \xi | \psi \otimes \psi ; \mathcal{A}_A \otimes \mathcal{A}_B \right)
\end{equation}
where one uses the Araki definition \cite{araki1976relative} of relative entropy in this general setting. In the last equality of \eqref{midef} we have defined:
\begin{equation}
\left| \xi \right> \equiv U \left| \psi \right>
\end{equation}
Notice that this later state satisfies:
\begin{equation}
\left<\xi\right| \alpha \otimes \beta \left| \xi\right> =  \left< \psi \right| \alpha \beta \left| \psi \right>
\end{equation}
and similarly for more general operators in $\mathcal{A}_A \vee \mathcal{A}_B$, such that $ \xi $ can be thought of as a different vector representation of the same state $\psi$.

Note that there is not a unique $U$ that implements the structure above. In particular any unitary from $V\in(\mathcal{A}_A \otimes \mathcal{A}_B)'$ can be be used to construct a new $ U \rightarrow V U$. This will then change the intermediate type-I factor $\mathcal{N} \rightarrow V \mathcal{N} V^{-1}$, mixing the operators between those on the two Hilbert spaces in the tensor product. However it is possible to fix a canonical type-I factor given a cyclic and separating state for the algebras above \cite{doplicher1984standard}. Given a state $\psi$ we consider the following vector representative of $\psi$ on the tensor product Hilbert space:
\begin{equation}
\label{uniquexi}
\left| \xi_\psi \right> \in \mathcal{V} 
\, : \left< \xi_\psi \right| \phi( \gamma)  \left| \xi_\psi \right> = \left< \psi \right| \gamma \left| \psi \right> \quad \forall \quad \gamma \in \mathcal{A}_A \vee \mathcal{A}_B
\end{equation}
where $\mathcal{V} \subset \mathcal{H} \otimes \mathcal{H}$ is the natural self-dual cone associated to $\psi \otimes \psi$ and the algebra $\mathcal{A}_A \otimes \mathcal{A}_B$. It is defined as the closure of:
\begin{equation}
 \Delta_{\psi \otimes \psi;  \mathcal{A}_A \otimes \mathcal{A}_B  }^{1/4} \left(\mathcal{A}_A \otimes \mathcal{A}_B \right)^+ \left| \psi \right> \otimes \left| \psi \right>
\end{equation}
where the plus superscript denotes positive elements of the algebra $ \mathcal{A}_A \otimes \mathcal{A}_B $ and $\Delta$ is the modular operator associated to this algebra and the state $\psi \otimes \psi$. There is always a unique state in the natural cone that represents a given state thought of as a linear functional on the algebra \cite{araki1974some}. Thus the state $\xi_\psi$ is uniquely fixed by \eqref{uniquexi}. Note that $\xi$ and $\xi_\psi$ represent the same state on $ \mathcal{A}_A \otimes \mathcal{A}_B $ so there is a unitary $V_\psi \in (\mathcal{A}_A \otimes \mathcal{A}_B)'$ that relates them. Thus we can pick $U \rightarrow U_\psi = V_\psi U$ such that:
\begin{equation}
\left| \xi_\psi \right> = U_\psi \left| \psi \right>
\end{equation}
and $U_\psi : \mathcal{H} \rightarrow \mathcal{H} \otimes \mathcal{H}$ can now be used to \emph{canonically} define the various sub-algebras in (\ref{subf}-\ref{Nlast}) by replacing $U \rightarrow U_\psi$. We will work with this $U_\psi$  and these sub-algebras from now on. 
To make this dependence on the state $\psi$ clear we will also label the resulting type-I factor $\mathcal{N}_\psi$. 
The fact that $\xi_\psi$ is in the natural cone of $\psi \otimes \psi$ implies that the modular conjugation operators associated to the algebra $\mathcal{A}_A \otimes \mathcal{A}_B$ are the same. We will call this $J$ since it will be directly analogous to the finite dimensional $J$ that was defined and used in various places for the reflected entropy, for example \eqref{littlej} and the paragraph below this equation.  From the factorized state $\psi \otimes \psi$ we learn that:
\begin{equation}
\label{defJ2}
J = J_A \otimes J_B \,, \qquad J \left| \xi_\psi \right> = \left| \xi_\psi \right>
\end{equation}
where $J_{A,B}$ are defined for the vector state $\psi$ and the respective algebras $\mathcal{A}_{A,B}$ respectively. The second equation in \eqref{defJ2} is another way to canonically fix this state.  We can fix the type-I factor acting on the original (single QFT) Hilbert space since:
\begin{equation}
\label{typeIdef}
\mathcal{N}_\psi = \left(\mathcal{A}_A \vee \mathcal{A}_A' \right) \otimes 1
=  \left( \mathcal{A}_A \otimes 1 \right) \vee J \left( \mathcal{A}_A \otimes 1 \right) J \quad \implies \quad
U_\psi^{-1} \mathcal{N}_\psi U_\psi = \mathcal{A}_A  \vee  \tilde{J} \mathcal{A}_A \tilde{J}
\end{equation}
where $\tilde{J} = U_\psi^{-1} J U_\psi$ was defined in the introduction below \eqref{split}. 

We would like to develop Conjecture~\ref{conj} which relates the entanglement wedge cross section, defined for QFTs with a holographic dual, to the entropy of the type-I factor or ``splitting entropy'':
\begin{equation}
\label{srvn}
S_R(A:B) = - {\rm Tr}_{\mathcal{H}_r} \rho_\psi \ln \rho_\psi
\end{equation}
where $\rho_\psi$ is the density matrix representation of $\xi_\psi$ on the left Hilbert space:
\begin{equation}
\rho_\psi = {\rm Tr}_{\mathcal{H}_l} \left| \xi_\psi \right> \left< \xi_\psi \right|
\end{equation}
We will give evidence for Conjecture~\ref{conj} by making a correspondence between the finite dimensional discussion of the reflected entropy and the splitting entropy.

The fact that we can even define a density matrix is thanks to the type-I nature of $\mathcal{N}$ that we have represented as bounded operators on a separable Hilbert space. Of course the cross section for holographic theories is finite in the situation at hand since the RT surface never reaches to the boundary. 
The entropy in \eqref{srvn} is not guaranteed to be finite, although there should be a large class of states where it is, where roughly we only expect a finite dimensional subspace to be relevant for the entropy. It would be interesting to find a link between the so called nuclearity bounds \cite{buchholz1986causal} that establish the split property and finiteness of this entropy along the lines of \cite{narnhofer2002entanglement,narnhofer1994entropy,Hollands:2017dov}.

There is more to this conjecture than finiteness. We can develop the following analogy/correspondence to the finite dimensional discussion of the reflected entropy:
\begin{align}
{\rm End} (\mathcal{H}_A) = \mathcal{H}_{AA^\star} &\leftrightarrow \mathcal{H}_l   & \left| \sqrt{\rho_{AB} } \right> &\leftrightarrow \left| \xi_\psi \right> \\ 
{\rm End} (\mathcal{H}_B) =  \mathcal{H}_{BB^\star} &\leftrightarrow \mathcal{H}_r 
 &\left| \sqrt{\rho_A} \otimes \sqrt{\rho_B} \right> &\leftrightarrow \left| \psi \right> \otimes \left| \psi \right>
\end{align}
This correspondence is in addition to the already noted equivalent roles of the modular conjugation operator $J$.
Furthermore the modular conjugation operator $J$ acts on the matrix Hilbert spaces (${\rm End}\mathcal{H}_{A,B}$)  simply via the hermitian conjugation of the matrix. Thus the canonical state $\left| \sqrt{\rho_{AB} } \right>$ is singled out via $J \left| \sqrt{\rho_{AB} } \right> = \left| \sqrt{\rho_{AB} } \right>$ exactly because it is Hermitian. 

Note that the bulk spacetimes constructed in Section~\ref{sec:duality}, using either the $m \rightarrow 1$ limit of the replica trick or the Engelhardt-Wall construction, can now be interpreted as the dual of the state $\left| \xi_\psi \right>$. That is, $\xi_\psi$ is interpreted as a novel kind of \emph{wormhole geometry} that entangles two QFT's that are not interacting with each other and are represented by the doubled Hilbert space $\mathcal{H}_l \otimes \mathcal{H}_r$. The wormhole construction in some sense geometrizes the GNS constructions of the boundary algebras so the boundary of the wormhole is not manifestly living in the standard representation of the QFT Hilbert space that usually arises for the AdS/CFT dictionary. 

For example in the case of a 2d CFT on a circle of circumference $L$ in the vacuum and where $A,B =$ two intervals of length $L_A,L_B$,  the two boundaries of the $\xi_\psi$ wormholes look like they might correspond to circles of radius $2L_A$ and $2 L_B$ respectively. However we note that likely this is not a precise interpretation of the Hilbert space - rather
we would like to think of the two ends of the wormhole more as funny representations of the original CFT on a circle with radius $L$. On the one hand the modular conjugation operators $J_A$ ($J_B$) on the original CFT Hilbert space are determined by simple conformal inversions fixing the intervals $L_A$ ($L_B$). However on the ends of the wormhole they are represented as reflections across the doubled copy.
This action then extends to the bulk of the wormhole in the obvious manner - a geometric manifestation that the chosen state should lie in the same cone as $\Omega \otimes \Omega$. In some sense the bulk solution gives a geometerization of the GNS construction where the algebra of operators in the commutant is simply mirrored.

Considering \eqref{reldef} we can also define reflected entropy via the relative entropy such that:
\begin{equation}
S_R(A:B) = S_{\rm rel} ( \omega_\xi | \omega_\psi \otimes \omega_\psi;  \mathcal{B}(\mathcal{H}) \otimes \mathcal{A}_B ) 
\end{equation}
where here the relative entropy is defined comparing two states thought of as linear functionals on the algebra and $\omega_\xi(\gamma) = \left< \xi_\psi \right| \gamma \left| \xi_\psi \right>$ for $\gamma \in \mathcal{B}(\mathcal{H})  \otimes \mathcal{A}_B 
$  and $\omega_\psi(\alpha) = \left< \psi \right| \alpha \left| \psi \right>$ for $\alpha \in \mathcal{B}(\mathcal{H})$ etc. As usual we can use the Araki definition \cite{araki1976relative}. 
In this form we can prove the bound $2 E_W(A:B)  \geq I(A:B)$ using monotonicity of relative entropy under inclusions $\mathcal{A}_A \otimes \mathcal{A}_B \subset  \mathcal{B}(\mathcal{H})  \otimes \mathcal{A}_B $  and the definition of mutual information in terms of relative entropy.  Thus there is a natural interplay between the mutual information and reflected entropy that persists in the continuum.

Another novelty of our results is that we can now give a replica trick method, using Section~\ref{sec:replica}, for computing quantities associated to the split property and algebraic QFT (AQFT). While the rigor of the replica trick is lacking, it has proven invaluable for giving insight into quantum information aspects of interacting QFT and so likely it will give insight into certain aspects of algebraic QFT for interacting theories. It will be especially useful for the algebraic approach to QFTs that are associated to the AdS/CFT correspondence.

The entanglement of purification can also be given a picture here. It is the minimization over the von Neumann entropy of all possible type-I factors:
\begin{equation}
E_p(A:B) = \inf_{V \in(\mathcal{A}_A \otimes \mathcal{A}_B)'}  S( V \mathcal{N}_\psi V^{-1} )_{\xi_\psi}
\end{equation} 
where $S$ is the von Neumann entropy of the density matrix for $\xi_\psi$ on $V \mathcal{N}_\psi V^{-1}$. 
This is then the analog of \eqref{defep}.  Amusingly both the reflected entropy \cite{schroer2007localization,schroer2010localization} and the entanglement of purification \cite{Narnhofer:2011zz,Otani:2017pmn}, in this algebraic context, have been put forward previously as possible regulators of entanglement entropy. We turn to a discussion of such a regulator now, where the main power we gain here, relative to previous discussions, is its computability in AdS/CFT and some simple CFT cases using the replica trick.

\subsection{A natural regulator for entanglement entropy}

We can use the reflected entropy as a natural regulator for entanglement entropy in QFT. As we have discussed, this is a finite quantity, even in the continuum. Consider a spatial subregion of a time slice $A \subset \Sigma$. Then consider a slightly smaller region $A^-$ contained in $A$, and a slightly larger region $(A^+)^c$, which is the complement (on $\Sigma$) of a region entirely outside of $A$. In this case, we have the inclusions associated to the spacetime regions:
\begin{equation}
\mathcal{D}(A^-) \subset \mathcal{D}(A) \subset \mathcal{D}((A^+)^c)
\end{equation}
which we assume are split.  We parameterize the spatial distance between the boundaries of these various regions via $\delta/2$. We postulate a useful regulator for entanglement entropy in QFT as:
\begin{equation}
S_{EE}^{(\delta)}(A)_\psi \equiv \frac{1}{2} S_R(A^-:A^+) = \frac{1}{2} S_{vN} (\mathcal{N}_\psi)_\psi
\end{equation}
where the type-I factor was constructed around \eqref{typeIdef} with $A \rightarrow A^-$ and $B \rightarrow A^+$. The factor of $1/2$ might seem a bit strange, but it is important for consistency of this proposal.

We note that this regulator works in a similar way to the mutual information regulator \cite{Casini:2015woa}, with the same regions that we defined above. In that case one considers $I(A^-:A^+)/2$ as a regulated version of the entanglement entropy where, roughly speaking, the term $S_{vN}(A^+ A^-)/2$ becomes the entropy associated to a thin disk which can be neglected. This regulator also relies on the split property to define the mutual information, as in \cite{Narnhofer:2011zz}, so we have not made any more assumptions here. 
The advantage that we obtain here is that the regulated quantity is still an entropy. 
Of course, the two regulators satisfy the relation: $S_{EE}^{(\delta)}(A) \geq I(A^+:A^-)/2$. 

This quantity has appeared before in various forms as a regulator of entanglement, although there seems to be one big difference to our proposal. In particular, no previous discussion considered the factor of $1/2$. \cite{Narnhofer:2011zz,Otani:2017pmn} considers the analog to the entanglement of purification after minimization. Indeed, if the original entanglement of purification conjecture \cite{Takayanagi:2017knl} is correct for holographic theories, then this would be a perfectly good regulator of EE, at least in such holographic theories, since it would reproduce the universal terms in the expansion of EE with the cutoff $\delta$. 
 \cite{schroer2007localization,schroer2010localization} considers the canonical type-I factor but also does not include the factor of $1/2$. In this case, this factor is important to reproduce the correct answer for the universal terms in entanglement.
 
For example, we can consider 2d CFTs in vacuum for a single interval, and work out the regulated entanglement entropies for any theory using the OPE discussed in \eqref{opeexp}. In fact, we can also work out the Renyi entropy of the type-I factor:
\begin{equation}
S_n^{(\delta)}(A) \equiv \frac{1}{2} S_n(A^- (A^-)^\star)_{\sqrt{\rho_{A^-A^+}}} = \frac{ c(n+1)}{6n} \ln (2 R/\delta) + \ldots,
\end{equation}
where $R$ is the length of the interval $A$, $\delta$ is the distance between the boundaries of $A^+$ and $A^-$, and the cross ratio of the 4 twist operators in this case is simply $x = \delta^2/R^2$. This reproduces the famous results of \cite{Holzhey:1994we,calabrese2004entanglement}. 

\begin{figure}[h!]
\centering 
\includegraphics[width=.68\textwidth]{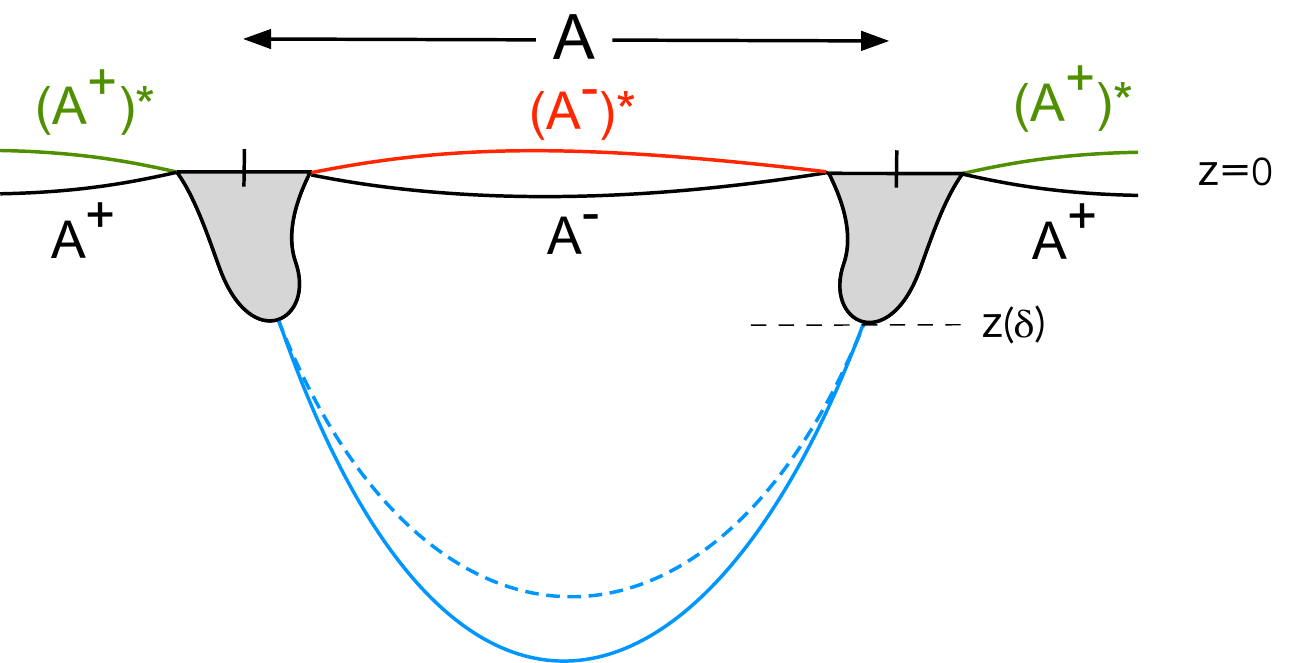} 
\caption{AdS shown in Poincar\'{e} coordinates where $z$ is the radial direction: $A, A^\pm$ are boundary regions, with $\mathcal{D}(A^-) \subset \mathcal{D}(A) \subset \mathcal{D}((A^+)^c)$. The entanglement wedge $rr^\star(AB)$ is the white region in the bulk, within which the reflected minimal surface is shown in blue. $S_R(A^-:A^+)/2$ obtained by this procedure approaches $S_{EE}(A)$ with the natural AdS regulator which cuts the minimal surface off at $z=z(\delta)$.
\label{regulate1}}
\end{figure}

We can also give a picture of how this regulator works in AdS/CFT in Figure~\ref{regulate1}. We note that the regulated EE does approach the EE obtained with a cutoff at some radial coordinate $z=z(\delta)$, away from the boundary at $z=0$. To leading order in small $\delta$, one finds $z(\delta) \propto \delta$, and this is sufficient to guarantee that the universal terms agree with the natural regulator in AdS/CFT \cite{Calabrese:2005in}.

Amusingly we can do even better. We can find a regulator that, in AdS/CFT, is literally a cut-off on the area of the \emph{exact} classical RT surface associated to $A$. We do not attempt to describe this in a continuum language, although one should exist. We will find a regulator that computes the area of the RT surface $m(A)$ integrated up to the point of intersection of this surface with a reflected minimal surface that ends on $m(A)$ near the boundary. To find the dual description of this, we need to apply the canonical purification twice.

Firstly consider the state $\left| \sqrt{\rho_A}\right>$ which, in the bulk, is described by the doubling of the entanglement wedge region $rr^\star(A)$. The boundary is the doubled $AA^\star$ Hilbert space, or a geometerization of the GNS construction. Inside this wedge, we consider two regions $A^- \subset A$ and $(A^-)^\star =$ the mirrored region. We consider the reduced density matrix of the union of these two regions:
\begin{equation}
\rho_{A^- (A^-)^\star} = {\rm Tr}_{ D D^\star} \left| \sqrt{\rho_A} \right> \left< \sqrt{\rho_A} \right|
\end{equation}
where we have split the $A$ Hilbert space as $\mathcal{H}_A = \mathcal{H}_D \otimes \mathcal{H}_{A^-}$
based on the corridor region $D = A \cap (A^-)^c$. We now compute the reflected entropy of this density matrix:
\begin{equation}
S_{EE}^{(\delta), {\rm ver\,\,2.0}}(A)_\psi  \equiv \frac{1}{2} S_R(A^- : (A^-)^\star) = \frac{1}{2} S_{vN} ( A^- (A^-)^{\star_2} )_{\sqrt{\rho_{A^- (A^-)^\star}}}
\end{equation}
where $\big| \sqrt{\rho_{A^- (A^-)^\star}} \big>$ is thought of as a pure state on the twice doubled Hilbert space:
\begin{equation}
\mathcal{H}_{A^- (A^-)^\star (A^- (A^-)^\star )^{\star_2} }.
\end{equation}

 \begin{figure}[h!]
\centering 
\includegraphics[width=0.56\textwidth]{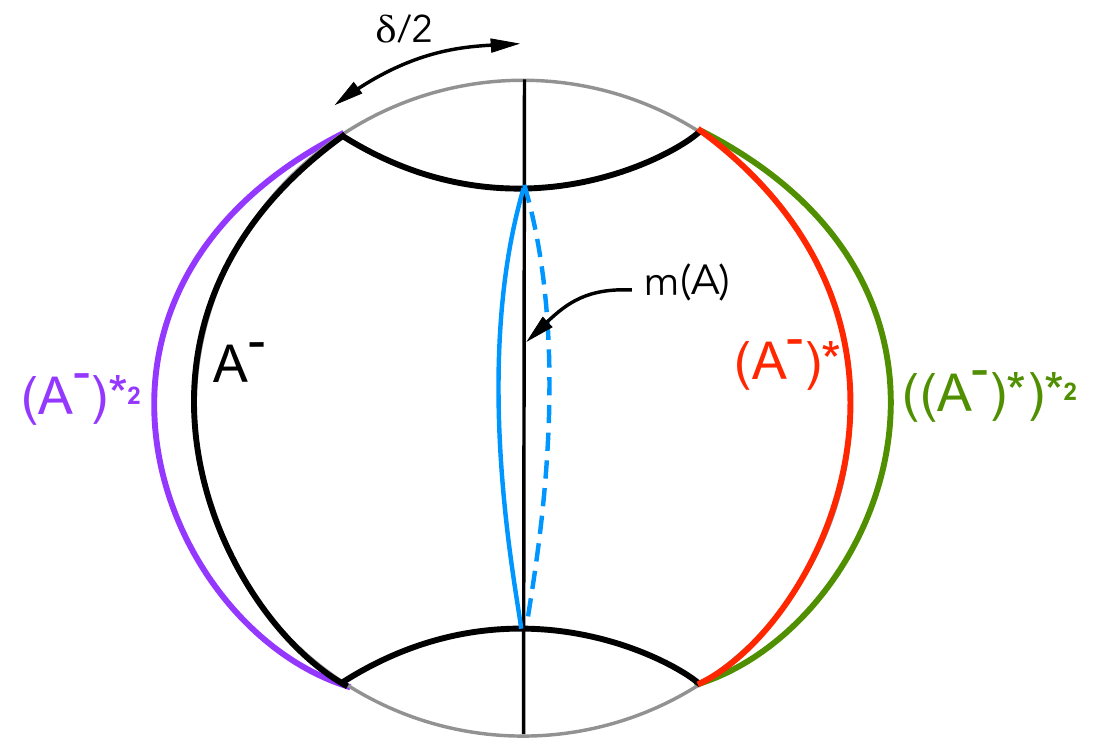} 
\caption{The entire boundary region is $AA^\star$, and the vertical line is the RT surface for $A$ denoted  $m(A)$. $\rho_{A^-(A^-)^\star}$ is the density matrix on $A^-(A^-)^\star$, where $A^- \subset A$, $(A^-)^\star \subset A^\star$, obtained using a corridor region of length $\delta/2$. Its canonical purification, $\left| \sqrt{\rho_{A^-(A^-)^\star}}\right>$, is a state on the $A^-(A^-)^\star(A^-(A^-)^\star)^{\star_2}$ Hilbert space. The reflected minimal surface for $A^-(A^-)^{\star_2}$ (in blue) lives on the twice doubled wedge, such that $S_R(A^-:(A^-)^\star)/2$ approaches $S_{EE}^{(\delta)}(A)$ for small $\delta$, thereby acting as a regulator for $m(A)$.
\label{regulate2}}
\end{figure}

Note that the result is the area of entanglement wedge cross section for $rr^\star(A^-)$  from which
it is not hard to see that, for small $\delta$, this is a cutoff version of the area of $m(A)$. See Figure~\ref{regulate2}.
While this is a natural thing to do holographically it remains to be seen whether this is also a natural thing to do for more general theories. 

\appendix

\section{The GNS construction}
\label{app:gns}

In this appendix we recall the GNS construction for a matrix algebra which defines the same state as the canonical purifications we deal with in this paper.
Consider the matrix algebra $M$ acting on $\mathcal{H}$ with the state,\begin{equation}
\omega: M \rightarrow \mathbb{R} \qquad \omega( \sigma) = {\rm Tr}  \left( \rho \sigma \right)
\end{equation}
which is a state in the sense of a positive linear functional on observables. 
For simplicity we take the density matrix to be full rank to avoid an additional step involving a quotient by an ideal, although the definition of reflected entropy does not rely on this. More specifically we have:
\begin{equation}
\left| \sigma \right>_{GNS} = \left| \sigma \sqrt{\rho} \right>
\end{equation}
with: 
\begin{equation}
\left< \sigma' \right| \left. \sigma \right>_{GNS} = {\rm Tr} \rho (\sigma')^\dagger \sigma 
= \omega ((\sigma')^\dagger \sigma)
\end{equation}
This later equation is usually the starting point for the GNS construction. Note that $\left| 1 \right>_{GNS}$ corresponds to the original state associated to $\rho$. The matrix algebra is represented by the left action:
\begin{equation}
L_\gamma \left| \sigma \right>_{GNS} = \left| \gamma \sigma \right>_{GNS} \qquad \gamma \in M
\end{equation}
and the commutant is given by the right action:
\begin{equation}
\hat{R}_\gamma \left| \sigma \right>_{GNS} =
\big| \sigma \gamma \big>_{GNS}
\end{equation}
We can consider the Tomita-Takesaki modular operators for $\left|1 \right>_{GNS}$. These are:
\begin{equation}
S = J \Delta^{1/2}\,, \qquad \Delta = L_\rho \hat{R}_{\rho^{-1}}\,, \qquad J \left| \sigma \right>_{GNS} = \big| \rho^{1/2} \sigma^\dagger \rho^{-1/2} \big>_{GNS}
\end{equation}
One can check these satisfy:
\begin{equation}
S L_\gamma \left| 1 \right>_{GNS} = L_{\gamma^\dagger} \left| 1 \right>_{GNS}\, \quad  \left( J \left| \sigma' \right>, J \left| \sigma \right> \right)_{GNS} =  \left( \left| \sigma \right>,  \left| \sigma' \right> \right)_{GNS}
\end{equation}
Note that $\hat{R}$ is related to the right action defined in \eqref{lraction} via $R_\gamma =  \hat{R}_{\rho^{1/2} \gamma \rho^{-1/2}}$. The natural cone is defined via the set of states:
\begin{equation}
\label{conegns}
\Delta^{1/4} L_{M^+} \left| 1 \right>_{GNS} = \big| \rho^{1/4} M^+ \rho^{-1/4} \big>_{GNS} = \big| \rho^{1/4} M^+ \rho^{1/4} \big>
\,, \qquad  M^+ \geq 0
\end{equation}
where these states are fixed points of $J$.

The reflected entropy is defined given a tensor factorization of the algebra $M = M_A \otimes M_B$. Then $S_R$ is the von Neumann entropy associated to the state $\left|1 \right>$ and the algebra $M_A \otimes J M_A J$. 
This should be compared to our discussion of the splitting property around \eqref{typeIdef}. Here it is actually more natural to work with the GNS representation for $\rho =\rho_A \otimes \rho_B$ in which case $\rho_{AB}$ lies in the natural cone \eqref{conegns} with 
\begin{equation}
M^+ = \rho_A^{-1/4} \rho_B^{-1/4} \rho_{AB}^{1/2}  \rho_A^{-1/4} \rho_B^{-1/4} \geq 0
 \quad \rightarrow \quad \left| \xi_\psi \right>_{GNS} = \big| \rho_{AB}^{1/2} \rho_{A}^{-1/2} \rho_B^{-1/2} \big>_{GNS} 
\end{equation}

\section{Other inequalities}
\label{app:ineq}

In this Appendix we summarize our efforts to prove some other bounds on the reflected entropy, as inspired by the holographic dual. We will not be able to make general statements about $S_R$, although we will establish some new results for the integer Renyi versions of some of these inequalities.

We can be more explicit about the reduced density matrices involved in the definition of the reflected entropy by writing out the matrix elements:
\begin{equation}
\label{explicit}
\left< \sigma_{ij}^A \right| \rho_{A A^\star} \left| \sigma_{i' j'}^A \right>
= {\rm Tr}_B \left(\left< i_A \right| \rho_{AB}^{1/2} \left| j_A \right>  \left< j'_A \right| \rho_{AB}^{1/2} \left| i'_A \right> \right)
\end{equation}
where we have defined a basis of states denoted $\left| \sigma_{ij} \right>$ where $\sigma_{ij} = \left| i  \right> \left< j \right|$.   We  need to define the ``cross transpose'' for super operators acting on ${\rm End}(\mathcal{H})$:
\begin{equation}
\left< \alpha_{ij} \right| P^{\Gamma} \left| \alpha_{i' j' } \right> \equiv \left< \alpha_{i i'} \right| P \left| \alpha_{jj'} \right> 
\end{equation}
Note that $P^\Gamma$ need not be Hermitian when $P$ is. 

The partial trace $T_B$ is a linear operator from ${\rm End}(\mathcal{H}_{AB})$ 
to ${\rm End} (\mathcal{H}_{A})$:
\begin{equation}
T_B \left| \sigma_{AB} \right> = \left| \sigma_A \right>
\end{equation}
and it is easy to show that $T_B^\dagger \left|\sigma_A \right> = \left| \sigma_A \otimes 1_B \right>$. 
We can also define the isometry:
\begin{equation}
\label{defV}
\mathcal{V}_{A|B} =  R_{\rho_{AB}^{1/2} } T_B^\dagger  R_{\rho_A^{-1/2}}
\end{equation}
which maps $\mathcal{V}_{A|B} : {\rm End} (\mathcal{H}_{A}) \rightarrow  {\rm End} (\mathcal{H}_{AB})$, while commuting with the left action of ${\rm End} (\mathcal{H}_A)$ matrices. Where the $R,L$ are the right and left actions defined in \eqref{lraction}. 
Note that in this section we assume that the relevant density matrices, such as $\rho_A$ in \eqref{defV}, are invertible. The more general case can be studied as a limit of this. One can check the following two defining properties are satisfied by \eqref{defV}:
\begin{align}
\mathcal{V}_{A|B}^\dagger \mathcal{V}_{A|B} & = 1 \\
\label{Vcomm}
\mathcal{V}_{A|B} \left( L_{\sigma_A} \left| \sqrt{\rho_A} \right> \right)
&= L_{\sigma_A \otimes 1_B} \left( \mathcal{V}_{A|B}   \left| \sqrt{\rho_{A} } \right>  \right)
\end{align}
This isometry is crucial in the proofs of monotonicity of relative entropy \cite{petz1986quasi,Nielsen:2004rsv} as reviewed in \cite{Witten:2018lha}. 
With all these new definitions we can introduce an alternative expression for the reduced density matrix on $AA^\star$:
\begin{equation}
\label{newdef}
\rho_{A A^\star} = \left(R_{\sqrt{\rho_A}} Q_{AA^\star}R_{\sqrt{\rho_A}} \right)^\Gamma
\end{equation}
where
\begin{equation}
\label{Qaa}
Q_{AA^\star} = \mathcal{V}_{A|B}^\dagger \Delta_{AB}^{1/2} \mathcal{V}_{A|B}
\end{equation}
and where the modular operator $\Delta$ was defined in \eqref{moddef}. 
The easiest way to show \eqref{newdef}  is to work it out in the basis  $\left| \sigma_{ij} \right>$ and compare to the explicit form \eqref{explicit}. 
Note that the cross transpose has the following property:
\begin{equation}
{\rm Tr} P^\Gamma = \left< 1 \right| P \left| 1 \right>
\end{equation}
where $1$ is the unit matrix. Similarly one can show that:
\begin{equation}
{\rm Tr} (P^\Gamma)^n = \left< 1_A \right|^{\otimes n} \mathbb{S}_n(A)P^{\otimes n}  \left( \mathbb{S}_n(A) \right)^\dagger  \left| 1_A \right>^{\otimes n}
\end{equation}
where we have introduced an $n$-fold tensor product Hilbert space and used the unitary cyclic-swap operator on $A$ defined in \eqref{aswap}.

Applying all this to the density matrix of interest we can write a formula for the integer Renyi entropies: $S_{n}(\rho_{AA^\star})  = - \frac{1}{n-1}\ln Z_n$ where:
\begin{align}
Z_n 
& =  \left< \sqrt{\rho_A}\right|^{\otimes n}  \mathbb{S}_{n}(A)  \left(Q_{AA^\star}\right)^{\otimes n}
\left(\mathbb{S}_{n}(A) \right)^\dagger \left| \sqrt{\rho_A}\right>^{\otimes n}
\label{tocomm}
\end{align}
We can commute the $\mathcal{V}$ factors in $Q_{AA^\star}$ through the twist operators using \eqref{Vcomm}
\begin{equation}
\mathcal{V}_{A|B} \mathbb{S}_n(A)^\dagger =  \left(\mathbb{S}_n(A)^\dagger \otimes 1_{{\rm End} (\mathcal{H}_B}) \right) \mathcal{V}_{A|B}
\end{equation}
since the twist operators consist only of left actions on the doubled Hilbert space.  For simplicity we will often drop the $1_{{\rm End} (\mathcal{H}_B})$ as it should be understood from the labeling in the twist operator. Now using the fact that $ \mathcal{V}_{A|B} \left| \sqrt{\rho_A} \right> = \left| \sqrt{\rho_{AB}} \right>$ we arrive at \eqref{dhalf} with $m=1$ that was derived their in a slightly different way.
 
Having setup some more formalism we would like give an argument for \eqref{miss2}.  We should beware that the density matrix that goes into the two sides of this inequality are not related by a trace. More explicitly we are comparing:
\begin{equation}
S(AA^\star)_{\sqrt{\rho_{ABC}}} \mathop{\geq} S(AA^\star)_{\sqrt{\rho_{AB}}}
\end{equation}
where the subscript specifies the purified state that is used to construct the entropies. 

The hint for establishing this inequality comes from the fact that in the bulk theory this is related to entanglement wedge nesting \cite{Takayanagi:2017knl}. 
From previous work we know that this follows from the positivity properties of modular operators under inclusion \cite{Balakrishnan:2017bjg,Faulkner:2018faa}. Such positivity properties are also used to prove monotonicity of relative entropy \cite{araki1976relative,uhlmann1977relative} and are also linked to the averaged null energy condition \cite{Faulkner:2016mzt}. We will work with finite quantum systems following \cite{Nielsen:2004rsv,Witten:2018lha}.
One notes that the isometry $\mathcal{V}$, defined in \eqref{defV}, has the following action on the modular operator:
\begin{equation}
\mathcal{V}_{AB|C}^\dagger \Delta_{ABC} \mathcal{V}_{AB|C}= \Delta_{AB}
\end{equation}
from which one can prove the inequality \cite{Witten:2018lha}:
\begin{equation}
\Delta_{AB}^{1/2} \geq \mathcal{V}_{AB|C}^\dagger \Delta_{ABC}^{1/2} \mathcal{V}_{AB|C}
\end{equation}
where the ordering $X\geq Y$ means that $X-Y$ is a non-negative operator for $X,Y$ hermitian. 
We also need:
\begin{equation}
\mathcal{V}_{AB|C} \mathcal{V}_{A|B}  = \mathcal{V}_{A|BC}
\end{equation}
which can be checked using the definition in \eqref{defV}. It follows that:
\begin{equation}
\label{qorder}
Q_{AA^\star}^{(A:BC)}  = \left( \mathcal{V}_{AB|C} \mathcal{V}_{A|B} \right)^\dagger  \Delta_{ABC}^{1/2}   \left( \mathcal{V}_{AB|C}  \mathcal{V}_{A|B} \right)
\leq  \mathcal{V}_{A|B}^\dagger \Delta_{AB}^{1/2}  \mathcal{V}_{A|B}  = Q_{AA^\star}^{(A:B)}
\end{equation}
where $Q_{A A^\star}$ was defined in \eqref{Qaa} for the $A:B$ system and this definition generalizes readily to the $A:BC$ system. Expectations of the $n$-fold tensor product of this operator in the state:
\begin{equation}
\left(\mathbb{S}_{n}(A) \right)^\dagger \left| \sqrt{\rho_A}\right>^{\otimes n}
\end{equation}
compute $Z_n$. The state is the same for both $A:B$ and $A:BC$ systems. So the ordering statement \eqref{qorder}, which also applies to the $n$-fold tensor product, becomes the following inequality for the Renyi entropies:
\begin{equation}
S_n(A A^\star)_{\sqrt{\rho_{ABC}}} \geq S_n(A A^\star)_{\sqrt{\rho_{AB}}} \qquad n \in \mathbb{Z} \qquad n \geq 1
\end{equation}
where we used the fact that $\ln x /(1-n) $ is a decreasing function of $x$ for $n > 1$.
So we have managed to show \eqref{miss2} for the integer Renyi entropies. This certainly does not guarantee this is true for $n \approx 1$ which is needed for the von Neumann entropy version. Such a situation is not unfamiliar, and a related but different inequalities was proven for the Renyi entropies in \cite{casini2010entropy} using wedge reflection positivity, which effectively involves the modular operator $\Delta^{1/2}$ for a local modular Hamiltonian.  This inequality is satisfied in holographic states \cite{headrick2014general}, but it was shown to be violated in some simple states. Here we have a similar situation, and a holographic argument can be given since effectively the twist operator correlation function we have setup for integer $m$, continues to be a correlation function to leading order in $G_N$ as $n \rightarrow 1$. The RT surfaces behave like heavy probe co-dimension $2$ operator and so should be expected to be subjected to the same bounds as a regular correlation function. We do not currently have a counter example to \eqref{miss2}, so it remains a possibility that a more sophisticated proof can be found for the von Neumann entropy. 

We next turn to attempts to prove strong super-additivity for $S_R$ which we reproduce here:
\begin{equation}
?? \quad S_R(A_1 A_2 :B_1 B_2) \geq S_R(A_1:B_1) + S_R(A_2:B_2) \qquad ??
\end{equation}
Actually there are again very simple counterexamples to this based on classically correlated states. However we will try to make connections to holographic theories by making some reasonable assumptions about the entropies of holographic like states at leading order in $G_N$. Indeed the analysis below is inspired by the mechanics of the proof of strong superadditivity for the entanglement wedge cross section in holographic states \cite{Takayanagi:2017knl}, although here we work only in the boundary theory. Starting again with the Renyi entropies and using the twist operator expression in \eqref{twistopexp} (for $m=1$):
\begin{equation}
\label{twotwist}
Z_n = \left< \sqrt{\rho_{AB}}\right|^{\otimes n}  \Sigma_{n}(AA^\star) \left| \sqrt{\rho_{AB}} \right>
= \left< \sqrt{\rho_{AB}}\right|^{\otimes n}  \Sigma_{n}(A_1 A_2^\star) \Sigma_{n}(A_1^\star A_2) \left| \sqrt{\rho_{AB}} \right>^{\otimes n}
\end{equation}
where we have set $A=A_1 A_2$ and $B=B_1 B_2$. Now let us insert the following identity between the twist operators in the later expression in \eqref{twotwist}:
\begin{equation}
1= \widetilde{\Delta}^{1/4}_{A_1 A_2^\star B_1 B_2^\star} \widetilde{\Delta}^{1/4}_{A_1^\star A_2 B_1^\star B_2}
\end{equation}
where these are the modular operators for $\left| \sqrt{\rho_{AB}} \right>$ reduced to the respective regions:
\begin{equation}
\widetilde{\Delta}_{A_1 A_2^\star B_1 B_2^\star}  = \rho_{A_1 A_2^\star B_1 B_2^\star} \otimes \rho_{A_1^\star A_2 B_1^\star B_2}^{-1}
\end{equation} 
etc.  Note that since the dimension of the Hilbert space $\mathcal{H}_{A_1 A_2^\star B_1 B_2^\star}$
is the same as the complement Hilbert space $\mathcal{H}_{A_1^\star A_2 B_1^\star B_2}$ the resulting reduced density matrices that go into the definition of $\widetilde{\Delta}$ are generically invertible, and we will assume this is the case here. Again the other cases can be approached with limits. 

Applying the Cauchy-Schwarz inequality:
\begin{align}
\label{cszm}
Z_n^2& \leq \left< \sqrt{\rho_{AB}}\right|^{\otimes n}  \Sigma_{n}(A_1 A_2^\star)\widetilde{\Delta}^{1/2}_{A_1 A_2^\star B_1 B_2^\star} \Sigma_{n}(A_1 A_2^\star)^\dagger  \left| \sqrt{\rho_{AB}} \right>^{\otimes n} \\
 & \qquad \qquad \times \left< \sqrt{\rho_{AB}}\right|^{\otimes n}  \Sigma_{n}(A_1^\star A_2)^\dagger \widetilde{\Delta}^{1/2}_{A_1^\star A_2 B_1^\star B_2}  \Sigma_{n}(A_1^\star A_2) \left| \sqrt{\rho_{AB}} \right>^{\otimes n} \nonumber
\end{align}
Each term on the right hand side has the form:
\begin{equation}
\left< \psi \right| \mathcal{O} \widetilde{\Delta}_{\psi;R}^{1/2} \mathcal{O} \left| \psi \right>
\end{equation}
where $\widetilde{\Delta}_{\psi;R}$ is the modular operator for $\psi$ and the algebra of operators acting on $\mathcal{H}_R$ and where $\mathcal{O}$ is an operator acting in the Hilbert space $\mathcal{H}_R$. This form is invariant under $\left| \psi \right> \rightarrow U_{R^c} \left| \psi \right> $ for some unitary acting in the complement Hilbert space $\mathcal{H}_{R^c}$. Thus we can use this freedom to work with any representative of the state acting on $\mathcal{H}_R$. We choose the canonical (GNS) purification associated to $R$. Concentrating on the first correlator on the right hand side of \eqref{cszm} we can rewrite this as:
\begin{equation}
 \left< \right. \! \sqrt{\rho_{A_1 A_2^\star B_1 B_2^\star}} \! \left. \right|^{\otimes n}  \Sigma_{n}(A_1 A_2^\star)\Delta^{1/2}_{A_1 A_2^\star B_1 B_2^\star} \Sigma_{n}(A_1 A_2^\star)^\dagger  \left| \right. \! \sqrt{\rho_{A_1 A_2^\star B_1 B_2^\star}} \! \left. \right>^{\otimes n}
\end{equation}
where the new modular operator $ \Delta = U_R \widetilde{\Delta} U^\dagger_R $ is defined in the usual way \eqref{moddef} via a left and right action on ${\rm End} (\mathcal{H}_{A_1 A_2^\star B_1 B_2^\star})$.

This now clearly computes the ($n^\text{th}$ Renyi) reflected entropy associated to the state $\rho_{A_1 A_2^\star B_1 B_2^\star}$ with the bipartition $A_1 A_2^\star : B_1 B_2^\star$ denoted:
\begin{equation}
S_R( A_1 A_2^\star : B_1 B_2^\star) \equiv S( A_1 A_2^\star (A_1 A_2^\star)^\star)_{\sqrt{\rho_{A_1 A_2^\star B_1 B_2^\star}}}
\end{equation}
From the CS inequality in \eqref{cszm} we derive that:
\begin{align}
S_n(A_1 A_2 (A_1 A_2)^\star)_{\sqrt{\rho_{A_1 A_2 B_1 B_2}}}
\geq \frac{1}{2}&\left( S_n(A_1 A_2^\star (A_1 A_2^\star)^\star )_{\sqrt{\rho_{A_1 A_2^\star B_1 B_2^\star}}} \right.
 \\ 
& \qquad+   \left. S_n(A_2 A_1^\star (A_2 A_1^\star)^\star)_{\sqrt{\rho_{A_2 A_1^\star B_2 B_1^\star}}} \right)
\end{align}
We have made no assumptions to this point. Although we should recall that the above inequality only works for integer Renyi's for $n \geq 2$. In holography, for the usual reason, we expect this also to apply to the von Neumann entropy limit, which we focus on now.

To proceed we now need to assume the following mutual information vanishes:
\begin{equation}
\label{vanMI}
I(A_1 B_1 : A_2^\star  B_2^\star) = S( A_1 A_2^\star B_1 B_2^\star)- S(A_1 B_1) - S(A_2 B_2 ) = 0
\end{equation}
We showed that such mutual information vanish in holographic states whenever it involved a region $C$ with the mirror region $D^\star$ such that $C$ and $D$ are disjoint.  This rule was discussed directly above \eqref{srcr}. 
Here we have $C = A_1 B_1$ and $D = A_2 B_2$ which indeed are disjoint. Note that \eqref{vanMI} is only true to leading order in $G_N$ and this will lead to an approximately factorized density matrix \footnote{It is possible to bound the trace distance between these two density matrices in terms of the mutual information using the Pinsker inequality. It should then be possible to give a more rigorous discussion for the subsequent approximate equality in \eqref{appadd}, by bounding the differences. Since we are not claiming anything precise here, we do not go through the hassle of this.  } :
\begin{equation}
\label{fact}
\rho_{A_1 A_2^\star B_1 B_2^\star} \approx \rho_{A_1 B_1} \otimes \rho_{A_2^\star  B_2^\star} 
\end{equation}
If such a factorization is exact then the reflected entropy is additive. In our case it should be approximately additive:
\begin{align}
\label{appadd}
S(A_1 A_2^\star (A_1 A_2^\star)^\star )_{\sqrt{\rho_{A_1 A_2^\star B_1 B_2^\star}}}
& \approx S(A_1  (A_1)^\star )_{\sqrt{\rho_{A_1 B_1 }}} + S(A_2^\star  (A_2^\star)^\star )_{\sqrt{\rho_{A_2^\star B_2^\star }}} \\
&= S(A_1  (A_1)^\star )_{\sqrt{\rho_{A_1 B_1 }}}  + S(A_2  (A_2)^\star )_{\sqrt{\rho_{A_2 B_2 }}} 
\end{align}
which, when combined with the analysis of the second factor in \eqref{cszm}, would imply superadditivity:
\begin{equation}
S(A_1 A_2 (A_1 A_2)^\star)_{\sqrt{\rho_{A_1 A_2 B_1 B_2}}}
\gtrapprox S(A_1  (A_1)^\star )_{\sqrt{\rho_{A_1 B_1 }}}  + S(A_2  (A_2)^\star )_{\sqrt{\rho_{A_2 B_2 }}} 
\end{equation}
This at least motivates the strong superaddivity bound in holographic theories. It also gives conditions upon which one might expect this to be true more generally. That is, when applied to the integer Renyi entropies, and where the factorization of \eqref{fact} holds. 

\section{OPE coefficient}
\label{app:ope}

We want to compute the three point function:
\begin{equation}
\left< \sigma_{g_A^{-1}}(z_1) \sigma_{g_B}(z_2)  \sigma_{g_A g_B^{-1}} (z_3)\right>_{CFT^{\otimes nm}(\mathbb{C})}
\end{equation}
where we will keep $z_{1,2,3} \in \mathbb{C}$ arbitrary as a check on our calculation. The branch cut structure implied by these twist operators can be inferred from Figure~\ref{fig:allreplica}. We can un-wrap the $m$-fold branch cut via:
\begin{equation}
\label{wtoz}
w = \frac{ (z-z_1)^{1/m}}{(z-z_2)^{1/m}}
\end{equation}
In the new $w$ space we have removed the two $m$-fold twist operators. However $ \sigma_{g_A g_B^{-1}} $ corresponds now to two twist operators in the $w$ coordinates located at:
\begin{equation}
w_1 =  \frac{ (z_3-z_1)^{1/m}}{(z_3-z_2)^{1/m}}\,, \qquad w_2 = e^{i 2\pi k/m} \frac{ (z_3-z_1)^{1/m}}{(z_3-z_2)^{1/m}}\,, \qquad k = m/2
\end{equation}
These two points are the images of the $(\tau_n^{(0)})^{-1}$ and $\tau_n^{(k)}$ respectively. Where $k$ labels which $m$-sheet the twist operator lives on in the original $z$ coordinates. These two twist operators now have an $n$-fold branch cut that runs between them passing through the point $w=0$ which is the original location of $\sigma_{g_A^{-1}}$. 

The conformal transformation in \eqref{wtoz} does not leave invariant the three point function. Rather there are two effects we must keep track of. There is an anomalous transformation determined by the Liouville action for the Weyl factor:
\begin{equation}
ds^2 = dz d\bar{z} = e^{\phi} dw d\bar{w}\, \qquad \phi  = \ln \frac{\partial z}{\partial w} +\ln \frac{\partial \bar{z}}{\partial \bar{w}} 
\end{equation}
and there is another contribution that comes from the two twist operators on the $w$ plane, again associated to the Weyl factor at the location of these operators. These can be understood as arising from the trace of the CFT stress tensor that has delta function support at the locations of twist operators $w = w_1,w_2$. It can also be simply understood as resulting from the conformal transformation of the primary twist operators.
Together we have:
\begin{align}
\left< \sigma_{g_A^{-1}}(z_1) \sigma_{g_B}(z_2)  \sigma_{g_A g_B^{-1}} (z_3)\right>_{CFT^{\otimes nm}(\mathbb{C})} &= e^{ S_L(\phi)} \left| \frac{\partial w}{\partial z} \right|_{w=w_1}^{2h_n} \left| \frac{\partial w}{\partial z} \right|^{2 h_n}_{w=w_2} \\ &\qquad \, \times \left< \sigma_{(\tau_n^{(0)})^{-1}}(w_1) \sigma_{\tau_n^{(k)}} (w_2) \right>_{CFT^{\otimes n}(\mathbb{C})}
\end{align}
The action $S_L$ needs to be regulated carefully as in \cite{Lunin:2000yv}, whilst also normalizing the twist operators appropriately. Fortunately we actually do not need to work out the details of $S_L(\phi)$.  We know the resulting contributions are  local to the images of the $m$ fold twist operators in the $w$ plane ($w=0$ and $w=\infty$) as well as possibly from the $m$ images of the $z=\infty$ point in the $w$ plane 
($w = e^{2\pi i k/m}, k=0,\ldots n-1$). 
Thus $S_L$ can be computed by applying the same conformal transformation as \eqref{wtoz} but now for the two point function of $m$-fold twist operators at $z_1$ and $z_2$ and with no operator at $z_3$: 
\begin{equation}
e^{S_L(\phi)} = \left(\left< \sigma_{g_B^{-1}}(z_1) \sigma_{g_B}(z_2) \right>_{CFT^{\otimes m}(\mathbb{C})} \right)^n
\end{equation}
We can also just compute:
\begin{equation}
 \left| \frac{\partial w}{\partial z} \right|_{w=w_{1,2}} = \left| \frac{w_{1}}{m} \frac{(z_1-z_2)}{(z_3-z_2)(z_3-z_1)} \right|\,,
 \quad \left< \sigma_{(\tau_n^{(0)})^{-1}}(w_1) \sigma_{\tau_n^{(k)}} (w_2) \right>_{CFT^{\otimes n}(\mathbb{C})}
 = |2w_1|^{-4h_n}
\end{equation}

Putting all this together we have:
\begin{align}
\nonumber \left< \sigma_{g_A^{-1}}(z_1) \sigma_{g_B}(z_2)  \sigma_{g_A g_B^{-1}} (z_3)\right>_{CFT^{\otimes nm}}
&= (2 m)^{-4 h_n}  \left| z_3-z_2 \right|^{-4 h_n}  \left| z_3-z_1 \right|^{-4 h_n}  \\& \qquad \times \left| z_1 - z_2 \right|^{- 4 n h_m+ 4 h_n}
\end{align}
which allows us to read off the OPE coefficient given in \eqref{opec}.

\acknowledgments

It is a pleasure to thank Netta Engelhardt, Udit Gupta, Matt Headrick, Marius Junge, Nima Lashkari, Min Li, Simon Lin, Juan Maldacena, Shinsei Ryu, Aron Wall, Huajia Wang and Tianci Zhou for useful discussions/comments. This work is supported by the Department of Energy contract SC0019183.

\bibliography{ew}

\end{document}